\documentclass[
 reprint,
 superscriptaddress,
showpacs,preprintnumbers,
 nofootinbib,
 amsmath,amssymb,
 aps,
 prd,
floatfix,
]{revtex4-1}

\usepackage{graphicx}
\usepackage{dcolumn}
\usepackage{bm}
\usepackage{hyperref}
\usepackage{color}
\usepackage[justification=justified]{subcaption}

\newcommand{\ga}{g_{\mu\nu}}
\newcommand{\pa}{\partial}

\newcommand{\be}{\begin{equation}}
\newcommand{\ee}{\end{equation}}
\newcommand{\ba}{\begin{eqnarray}}
\newcommand{\ea}{\end{eqnarray}}

\newcommand{\en}{\nonumber\\}

\newcommand{\de}{\delta}

\newcommand{\kk}{\mathbf{k}}
\newcommand{\xx}{\mathbf{x}}

\newcommand{\p}{\mathbf{p}}
\newcommand{\pp}[1]{\mathbf{p}_{#1}}
\newcommand{\qq}{\mathbf{q}}

\definecolor{darkred}{RGB}{175,0,0}
\definecolor{darkblue}{RGB}{0,0,175}

\newcommand{\fn}{f_{\nu}}

\newcommand{\apss}{Astrophysics and Space Science}

\newcommand{\mnras}{Mon. Not. R. Astron. Soc.}

\newcommand{\physrep}{Phys. Rep.}

\newcommand{\jcap}{Journal of Cosmology and Astroparticle Physics}

\begin{document}

\title{The Neutrino Puzzle: Anomalies, Interactions, and Cosmological Tensions}
\author{Christina D. Kreisch}
\email{ckreisch@astro.princeton.edu}
\affiliation{ Department of Astrophysical Sciences, Princeton University, Princeton, New Jersey 08544 USA}

\author{Francis-Yan Cyr-Racine}%
 \email{fcyrraci@physics.harvard.edu}
\affiliation{Department of Physics, Harvard University, Cambridge, Massachusetts 02138, USA}
\affiliation{Department of Physics and Astronomy, University of New Mexico, 1919 Lomas Blvd NE, Albuquerque, New Mexico 87131, USA}

\author{Olivier Dor\'e}
\affiliation{Jet Propulsion Laboratory, California Institute of Technology, Pasadena, California 91109, USA}

\date{\today}

\begin{abstract}

New physics in the neutrino sector might be necessary to address anomalies between different neutrino oscillation experiments. Intriguingly, it also offers a possible solution to the discrepant cosmological measurements of $H_0$ and $\sigma_8$. We show here that delaying the onset of neutrino free-streaming until close to the epoch of matter-radiation equality can naturally accommodate a larger value for the Hubble constant $H_0=72.3 \pm 1.4\,\mathrm{km\, s^{-1} Mpc^{-1}}$ and a lower value of the matter fluctuations $\sigma_8=0.786\pm 0.020$, while not degrading the fit to the cosmic microwave background (CMB) damping tail. We achieve this by introducing neutrino self-interactions in the presence of a non-vanishing sum of neutrino masses. Without explicitly incorporating additional neutrino species, this strongly interacting neutrino cosmology prefers $N_{\rm eff} = 4.02 \pm 0.29$, which has interesting implications for particle model-building and neutrino oscillation anomalies. We show that the absence of the neutrino free-streaming phase shift on the CMB can be compensated by shifting the value of several cosmological parameters, hence providing an important caveat to the detections made in the literature. Due to their impact on the evolution of the gravitational potential at early times, self-interacting neutrinos and their subsequent decoupling leave a rich structure on the matter power spectrum. In particular, we point out the existence of a novel localized feature appearing on scales entering the horizon at the onset of neutrino free-streaming. While the interacting neutrino cosmology provides a better global fit to current cosmological data, we find that traditional Bayesian analyses penalize the model as compared to the standard cosmological scenario due to the relatively narrow range of neutrino interaction strengths that is favored by the data. Our analysis shows that it is possible to find radically different cosmological models that nonetheless provide excellent fits to the data, hence providing an impetus to thoroughly explore alternate cosmological scenarios.   
\end{abstract}


\maketitle


%
\section{Introduction}\label{sec:intro}
The neutrino sector of the Standard Model (SM) of particle physics is a promising area to search for new phenomena that could help pinpoint the Ultraviolet completion of the SM. Indeed, terrestrial neutrino experiments have identified several anomalies that could potentially indicate the presence of new physics in the neutrino sector (see, e.g., Ref.~\cite{Dentler:2018sju} for a recent review). Of particular significance are the $\nu_\mu\rightarrow\nu_{\rm e}$ appearance results from the MiniBooNE \cite{Aguilar-Arevalo:2018gpe} and LSND \cite{Aguilar:2001ty} collaborations which, if interpreted within a neutrino oscillation framework that includes an extra sterile neutrino, would indicate the presence of such a sterile neutrino at very high statistical significance. Within this ``3+1'' neutrino oscillation framework, these results are, however, very difficult to reconcile with the absence of anomalies in the $\nu_\mu\rightarrow\nu_\mu$ disappearance as probed by recent atmospheric \cite{Aartsen:2017bap,TheIceCube:2016oqi} and short-baseline \cite{Adamson:2017zcg,Adamson:2017uda} experiments. If these results are confirmed by future analyses, it is likely that new physics beyond the sterile+active oscillation models would be necessary to resolve the tension between neutrino appearance and disappearance data. 

Astrophysical and cosmological observations provide complementary means of probing the properties of neutrinos. This is perhaps best illustrated by the cosmological constraints on the sum of neutrino masses $\sum m_\nu < 0.12$ eV \cite{Aghanim:2018eyx} obtained by combining cosmic microwave background (CMB) data from the Planck satellite with baryon acoustic oscillation (BAO) measurements. Cosmological observables such as the CMB and large-scale structure (LSS) are also sensitive to the presence of new interactions (see e.g. Refs.~\cite{BialynickaBirula:1964zz,Bardin:1970wq,Gelmini:1980re,Chikashige:1980qk,Barger:1981vd,Raffelt:1987ah,Kolb:1987qy,Konoplich:1988mj,Berkov:1988sd,Belotsky:2001fb, Hannestad:2004qu,Chacko:2004cz,Hannestad:2005ex,Sawyer:2006ju,Mangano:2006mp,Friedland:2007vv,Hooper:2007jr,Serra:2009uu,Aarssen:2012fx,Jeong:2013eza,Laha:2013xua,Archidiacono:2014nda,Ng:2014pca,Cherry:2014xra,Archidiacono:2015oma,Cherry:2016jol,Archidiacono:2016kkh,Dvali:2016uhn,Capozzi:2017auw,Brust:2017nmv,Forastieri:2017bkq,neutrinophilicDM,Lorenz:2018fzb,Choi:2018gho}) in the neutrino sector that would modify their standard free-streaming behavior during the radiation-dominated epoch following their weak decoupling. In the literature, a phenomenological description based on the $c_{\rm eff}$ and $c_{\rm vis}$ parametrization \cite{Hu:1998kj} has often been used to test the free-streaming nature of neutrinos in the early Universe \cite{Trotta:2004ty,Melchiorri:2006xs,DeBernardis:2008ys,Smith:2011es,Archidiacono:2011gq,Archidiacono:2012gv,Gerbino:2013ova,Archidiacono:2013fha,melchiorri14,Sellentin:2014gaa,planckXVI}. While these analyses generally find results consistent with the standard neutrino cosmology, they are difficult to interpret in terms of possible new interactions among neutrinos, as emphasized in Refs.~\cite{Cyr-Racine:2013aa,oldengott15}. Other works \cite{Chacko:2003dt,Beacom:2004yd,Bell:2005dr,Cirelli:2006kt,Basboll:2008fx,Cyr-Racine:2013aa,Archidiacono:2013dua,oldengott15,Forastieri:2015paa,Forastieri:2017oma,lancaster,oldengott17,Barenboim:2019tux} have used more physical parameterizations that make the connection to the underlying particle nature of the neutrino interaction more transparent. 

In particular, Ref.~\cite{oldengott15} has developed a rigorous treatment of the evolution of cosmological neutrino fluctuations in the presence of neutrino self-interactions mediated by either a massive or massless new scalar particle. Using this framework, Ref.~\cite{oldengott17} used CMB data to put constraints on the strength of neutrino self-interactions in the early Universe for the case of a massive mediator. These results largely confirmed earlier constraints from Refs.~\cite{Cyr-Racine:2013aa,Archidiacono:2013dua,lancaster} obtained using an approximate (but nonetheless accurate) form of the neutrino Boltzmann hierarchy. Interestingly, these studies, which focused on four-neutrino interactions parametrized by a Fermi-like coupling constant $G_{\rm eff}$, found a bimodal posterior distribution for this latter parameter. While the first (and statistically dominant) posterior mode is consistent with the onset of neutrino free-streaming being in the very early Universe, the second posterior mode corresponds to a much delayed onset of free-streaming to $z_{\nu, \rm dec}\sim 8300$. In Ref.~\cite{lancaster}, a previously unknown multi-parameter degeneracy involving the amplitude of scalar fluctuations, the scalar spectral index, the Hubble constant, and the neutrino self-interacting strength was identified as being responsible for the existence of this second posterior mode. While intriguing, the neutrino interaction strength favored by this mode is nearly ten orders of magnitude above the standard weak interaction. Taken at face value, this likely constitutes a very serious challenge from a model-building perspective.

Nevertheless, given the current tensions among terrestrial and atmospheric neutrino experiments described above, is the ``interacting'' neutrino mode hinting at the presence of new physics beyond the SM? The simplified interaction models used in Refs.~\cite{Cyr-Racine:2013aa,lancaster,oldengott17,Barenboim:2019tux} are likely capturing parts of a more realistic neutrino interaction scenario, hence leading to a somewhat suboptimal fit to the cosmological data. One aspect that has been neglected in studies of self-interacting neutrinos so far is the presence of neutrino mass. The impact of massive neutrinos on the CMB and matter clustering has been studied extensively in the literature (see e.g.~Refs.~\cite{Kaplinghat:2003bh,lesgourges2006,Lesgourgues:2014zoa,Vagnozzi:2017ovm,Lattanzi:2017ubx}). One of the aims of this paper is to understand how the effects of massive neutrinos on cosmological observables are modified when self-interactions are present in the early Universe. 

Tensions are also growing between different late-time measurements of the Hubble constant $H_0$ \cite{HST,Bernal:2016gxb,Birrer:2018vtm,newH0} and those based on CMB data \cite{Aghanim:2018eyx}. Measurements of the amplitude of matter fluctuations at low redshifts (often parametrized using $\sigma_8$) from weak gravitational lensing and cluster counts are all consistently lower than that inferred from the CMB \cite{Heymans2012,Joudaki:2017zdt,Hikage:2018qbn}. While the statistical significance of the deviation of each individual measurement is less than $3\sigma$, all recent measurements of the amplitude fluctuations in the local universe are below the Planck value. Physics beyond $\Lambda \mathrm{CDM}$ has been proposed to reconcile these tensions, such as early dark energy \cite{earlyDE,DEH0,DEnu,Poulin:2018dzj}, dark matter interactions \cite{DMint1,DMint2}, decaying dark matter \cite{decay,decay2,decays8,Bringmann:2018jpr}, modified gravity \cite{renk,uber}, and new relativistic species \cite{axion}, among others. However, these propositions often struggle to remedy both tensions simultaneously. 

In this paper, we study how the presence of self-interacting massive neutrinos in the early Universe affect cosmological observables such as the CMB, with an eye on how these new effects could help relieve the current tensions among different datasets. In \autoref{sec:nu_int_model}, we describe the simplified neutrino interaction model used in this work. In \autoref{sec:cosmo_pert}, we present the cosmological perturbation equations for massive self-interacting neutrinos. In \autoref{sec:Cosmo_obs}, we describe the physical impacts that massive self-interacting neutrinos have on the CMB and the matter power spectrum. In \autoref{sec:data}, we outline the data and method used in our cosmological analyses of self-interacting. The results from these analyses are presented in \autoref{sec:results} and discussed in \autoref{sec:discuss}. We conclude and highlight future directions in \autoref{sec:conclusions}.

\section{Neutrino Interaction Model}\label{sec:nu_int_model}
In this work, we focus on a simple framework that captures the most important cosmological aspects of realistic neutrino interaction models. We note however that building a successful model of neutrino self-interaction that respects the gauge and flavor structure of the SM likely requires the introduction of a light sterile species which mass-mixes with the active neutrinos and is itself coupled to a massive scalar or vector mediator (see e.g. Refs.~\cite{Bringmann:2013vra,Dasgupta:2013zpn,Hannestad:2013ana,Cherry:2014xra,Tang:2014yla,Kouvaris:2014uoa,Chu:2015ipa,Ghalsasi:2016pcj,Hasenkamp:2016pme}). The presence of these new interactions in the sterile sector suppresses the effective mixing angle between the active and sterile species at early times, ensuring that Big Bang nucleosynthesis (BBN) constraints are respected . At later times  once the active-sterile oscillation rate becomes comparable to the finite temperature effective potential resulting from the new interaction, the mixing angle is no longer suppressed hence allowing the active and sterile sectors to partially thermalize with each other \cite{Archidiacono:2014nda,Archidiacono:2015oma,Archidiacono:2016kkh}. 

Diagonalizing the mass matrix of such a model leads to an effective interaction Lagrangian between the different neutrino mass eigenstates of the generic form
\be\label{eq:int_lagragian}
\mathcal{L}_{\rm int} = g_{ij}\bar{\nu}_i \nu_j \varphi,
\ee
where $g_{ij}$ is a (generally complex) coupling matrix, $\nu_i$ is a left-handed neutrino Majorana spinor, and the indices $i,j$ labeled the neutrino mass eigenstates. Here we have assumed a Yukawa-type interaction with a massive scalar $\varphi$, but note that the results presented in this work also apply if a massive vector is assumed instead. The Lagrangian given in Eq.~\eqref{eq:int_lagragian} could also arise in models where neutrinos couple to a Majoron \cite{Gelmini:1980re,Simpson:2016gph,Berlin:2018ztp}. 

In models where the new interaction arises through active-sterile mixing, the structure of the coupling matrix $g_{ij}$ would generally depend on the flavor content of each mass eigenstate. For instance, a mass eigenstate made of mostly active flavors will couple very weakly to the massive scalar $\varphi$, while an eigenstate being largely composed of the sterile species would couple more strongly to the mediator. In other models of neutrino interaction, the structure of the coupling matrix could be more arbitrary. In all cases though, $g_{ij}$ is subject to important flavor-dependent bounds \cite{Lessa:2007up,Bakhti:2017jhm,Arcadi:2018xdd} arising from meson, tritium, and gauge boson decay kinematics. 

In this work, we consider the simple case of a universal coupling $g_\nu$ between every neutrino mass eigenstate 
\be 
g_{ij} \equiv g_\nu\de_{ij}, 
\ee
where $\de_{ij}$ is the Kronecker delta. While the universal coupling case is likely unrealistic for the reason outlined above, it does provide a simple benchmark to test the sensitivity of cosmological data to new neutrino physics. 

We work in the contact-interaction limit in which the mass of the $\varphi$ mediator is much larger than the typical energy of the scattering event. In this case, one can integrate out this massive mediator and write the interaction as a four-fermion contact interaction. This is an excellent approximation at the energy scale probed by the CMB for $m_\varphi \gtrsim 1$ keV. In this limit, the squared scattering amplitude for a neutrino $\nu_i$ to interact with any other neutrino in the thermal bath is
\begin{align}\label{matrix_elem_majorana}
|\mathcal{M}|_{\nu_i}^2 &=\sum_{\rm spins}|\mathcal{M}|_{\nu_i+\nu_j\rightarrow\nu_k+\nu_l}^2\en 
&=2 G_{\rm eff}^2\left(s^2 +t^2 + u^2\right),
\end{align}
where we have defined the dimensionfull coupling constant $G_{\rm eff}\equiv |g_\nu|^2/m_\varphi^2$. Here, $s$, $t$, and $u$ are the standard Mandelstam variables. While our phenomenological model described by $G_{\rm eff}$ is unlikely to accurately capture all the complexity of novel neutrino interactions, it is nonetheless a useful framework to identify the interesting parameter space, as described in Ref.~\cite{Ng:2014pca}. 

Introducing new neutrino interactions  has an impact beyond cosmology. For a low mass mediator ($< 10$ MeV), SN 1987A \cite{Kolb:1987qy}, Big Bang nucleosynthesis (BBN) \cite{Ahlgren:2013wba,Huang:2017egl}, and the detection of ultra-high energy neutrinos at IceCube \cite{Ng:2014pca,Ioka:2014kca,Cherry:2016jol} provide some of the strongest constraints, with the latter bound having the potential of being the most stringent in the near future. Other limits \cite{Bilenky:1992xn,Bardin:1970wq,Bilenky:1999dn} coming from Z-boson decay do not directly apply at the energy scale probed by the CMB. Also, elastic collisions caused by the new interaction do not affect the time it takes for neutrinos to escape supernovae \cite{Manohar:1987ec,Dicus:1988jh}, although they could lead to interesting phenomena (see e.g.~Refs.~\cite{PhysRevLett.95.191302,PhysRevD.83.117702,PhysRevLett.96.211302,PhysRevD.42.293,Blennow:2008er,Galais:2011jh}). Finally, supernova cooling puts bounds on the coupling of majorons to SM neutrinos \cite{Kachelriess:2000qc,Farzan:2002wx,Zhou:2011rc,Jeong:2018yts}, but the applicability of these likely depends on the details of the exact coupling matrix used.

\section{Cosmological Perturbations}\label{sec:cosmo_pert}
In this section we summarize the key ingredients and simplifications entering our derivation of the Boltzmann equation governing the evolution of massive and self-interacting neutrino fluctuations, at first order in perturbation theory. Our computation uses two main approximations:
\begin{itemize}
\item Based on previous studies \cite{Cyr-Racine:2013aa,lancaster}, we assume that neutrinos decouples while still in the relativistic regime. We thus neglect the presence of the small neutrino mass in the computation of collision integrals. As we shall see, our final results are consistent with this approximation. 

\item We assume that the neutrino distribution function remains exactly thermal throughout the epoch at which neutrinos decouple and start free-streaming. This thermal approximation (also called, relaxation time approximation) implies that the only possible neutrino perturbations are local temperature fluctuations. This approximation was shown to be very accurate in Ref.~\cite{oldengott17} for the type of interaction we consider here.
\end{itemize}

Conformal Newtonian gauge is used throughout this section. 
\subsection{Neutrino distribution function and perturbation variables}
We present a detailed derivation of the left-hand side of the Boltzmann equation for massive neutrino in Appendix \ref{app:Boltz_equns} (see also Ref.~\cite{Ma:1995ey}).
Our starting point is to expand the neutrino distribution function as
\be\label{eq:pert_exp_rev}
f_{\nu}(\xx,\p,\tau)=f_{\nu}^{(0)}(\p,\tau)[1+\Theta_{\nu}(\xx,{\bf p},\tau)],
\ee
where $\xx$ denotes the spatial coordinates, $\tau$ is conformal time, and $\pp{}$ is the proper momentum. The background (spatially uniform) neutrino distribution function is taken to be of a Fermi-Dirac shape
\be\label{eq:background_fnu_rev}
f_{\nu}^{(0)}(\p,\tau) = \frac{1}{e^{p/T_\nu}+1},
\ee
where $p=|\pp{}|$. In the ultra-relativistic regime, for which the thermal approximation implies that the only possible neutrino perturbations are local temperature fluctuations, the perturbation variable $\Theta_\nu$ admits the form
\be
\Theta_\nu(\xx,\pp{},\tau) = - \frac{d \ln{f_\nu^{(0)}}}{d\ln{p}} \frac{\de T_\nu(\xx,\tau)}{\bar{T}_\nu(\tau)},
\ee
where $\bar{T}_\nu$ is the background neutrino temperature, and $\de T_\nu$ is its perturbation. It is therefore convenient to introduce the \emph{temperature fluctuation} variables $\Xi_\nu$
\begin{align}\label{eq:temp_fluct_def}
\Xi_\nu(\xx,\pp{},\tau) &\equiv \frac{-4 \Theta_\nu(\xx,\pp{},\tau)}{\frac{d \ln{f_\nu^{(0)}}}{d\ln{p}}}
\end{align}
which is independent of  $\pp{}$ in the thermal approximation for massless neutrinos. However, the presence of a nonvanishing neutrino mass and the non-negligible momentum transfered in a typical neutrino-neutrino collision would in general introduce some extra $\pp{}$-dependence to $\Xi_\nu$ \cite{Ma:1995ey}. This turns the Boltzmann equation of self-interacting neutrinos into a differentio-integral equation that is particularly difficult to solve exactly \cite{oldengott15}. In practice though, the absence of energy sources or sinks coupled to the neutrino sector implies that the momentum dependence of the right-hand side of Eq.~\eqref{eq:temp_fluct_def} should be vanishingly small at early times when neutrinos form a highly-relativistic tightly-coupled fluid. This allows us to neglect the momentum-dependence of $\Xi_\nu$ in the computation of the collision integrals, an approximation that was found to be accurate in Ref.~\cite{oldengott17}. We do retain, however, the momentum dependence of $\Xi_\nu$ in the left-hand side of the Boltzmann equation. 

In this work, we only consider scalar perturbations and thus expand the angular dependence of the $\tilde{\Xi}_{\nu}$ variable (the Fourier transform on $\Xi_\nu$) in Legendre polynomials $P_l(\mu)$
\be\label{eq:Legendre_exp_rev}
\tilde{\Xi}_{\nu}(\kk,\pp{},\tau)=\sum_{l=0}^{\infty}(-i)^l(2l+1)\nu_l(k,p,\tau)P_l(\mu),
\ee
where $\mu$ is the cosine of the angle between $\kk$ and $\pp{}$. Before presenting the equation of motion for the neutrino multipole moments $\nu_l$, we discuss the structure of the collision integrals.

\subsection{Collision term}
 The details of the collision term calculation for the $
 \nu\nu \rightarrow \nu\nu$ process is given in Appendix \ref{app:coll_int}. As explained above, the main simplification entering this calculation is the use of the thermal approximation in which we neglect the momentum dependence of the $\nu_l$ variables.  Under this assumption, the collision term at first order in perturbation theory $C_\nu^{(1)}$ can be written as
\begin{align}\label{eq:coll_term_final}
C_\nu^{(1)}[\pp{}] & = \frac{G_{\rm eff}^2T_\nu^6}{4}\frac{ \pa \ln f^{(0)}_\nu}{\pa \ln p_1}\\
&\quad\times\sum_{l=0}^\infty(-i)^l(2l+1)\nu_lP_l(\mu)\Bigg(A\left(\frac{p}{T_\nu}\right) \en 
& \qquad\qquad\qquad\qquad + B_l\left(\frac{p}{T_\nu}\right)-2 D_l\left(\frac{p}{T_\nu}\right)\Bigg),\nonumber
\end{align}
where the functions $A(x)$, $B_l(x)$, and $D_l(x)$ are given in Eqs.~\eqref{eq:A_coll}, \eqref{eq:Bl_coll}, and \eqref{eq:Dl_coll}, respectively. Here, we have adopted the notation $T_\nu \equiv \bar{T}_\nu$ to avoid clutter.  
\subsection{Boltzmann equation for self-interacting neutrinos}
Substituting the collision term from Eq.~\eqref{eq:coll_term_final} into Eq.~\eqref{eq:boltz_before_coll} and performing the $\mu$ integral yields the equation of motion for the different neutrino multipoles $\nu_l$. They can be summarized in the following compact form 
\begin{align}\label{eq:raw_comoving_boltz}
&\frac{\pa \nu_l}{\pa \tau}+k\frac{q}{\epsilon}\left(\frac{l+1}{2l+1}\nu_{l+1}-\frac{l}{2l+1}\nu_{l-1}\right) \\
& \qquad\qquad\qquad -4\left[\frac{\pa \phi}{\pa\tau}\de_{l0}+\frac{k}{3}\frac{\epsilon}{q}\psi\de_{l1}\right] \en
& \qquad =- a \frac{G_{\rm eff}^2T_\nu^5\nu_l}{\fn^{(0)}(q)}\left(\frac{T_{\nu,0}}{q}\right)\Bigg(A\left(\frac{q}{T_{\nu,0}}\right) \en 
& \qquad\qquad\qquad\qquad +B_l\left(\frac{q}{T_{\nu,0}}\right)-2D_l\left(\frac{q}{T_{\nu,0}}\right)\Bigg),\nonumber
\end{align}
where we have introduced the comoving momentum $\qq\equiv a \pp{}$, $q=|\qq|$, $\epsilon = \sqrt{q^2+a^2m_\nu^2}$, a is the scale factor normalized to $a=1$ today, $\de_{mn}$ is the Kronecker delta function, and $T_{\nu,0}$ is the current $(a=1)$ temperature of the neutrinos. The fact that the collision term is directly proportional to $\nu_l$ is a consequence of our use of the thermal approximation. We note that energy and momentum conservation ensures that $A+B_0-2D_0=0$ and $A+B_1-2D_1=0$, respectively.

As is standard in analyses of massive neutrino cosmologies, we shall consider our neutrino sector to be composed of a mix of massive and massless neutrinos. In the massless case ($q=\epsilon$), one can integrate Eq.~(\ref{eq:raw_comoving_boltz}) over the comoving momentum to yield a simpler neutrino multipole hierarchy \cite{Cyr-Racine:2013aa,lancaster}
\begin{align}
& \frac{\pa F_l}{\pa \tau}+k\left(\frac{l+1}{2l+1}F_{l+1}-\frac{l}{2l+1}F_{l-1}\right)\\
&\qquad\qquad\qquad -4\left[\frac{\pa \phi}{\pa\tau}\de_{l0}+\frac{k}{3}\psi\de_{l1}\right]  =-aG_{\rm eff}^2T_\nu^5 \alpha_l F_l,\nonumber
\end{align}
where
\begin{align}
\alpha_l &= \frac{120}{7 \pi^4}\int_0^\infty dx\, x^2\Bigg[A\left(x\right) +B_l\left(x\right) -2D_l\left(x\right)\Bigg],
\end{align}
and where we denoted the massless perturbations as $F_l$ to distinguish them from the massive neutrino variables $\nu_l$.

We implement these modified Boltzmann equations in the cosmological code \texttt{CAMB} \cite{camb00}. For computational speed, we precompute the functions $A$, $B_l$ and $D_l$ on a grid of $q/T_{\nu,0}$ values and use an interpolation routine to access them when solving the cosmological perturbation equations. As in standard \texttt{CAMB}, we use a sparse 3-point grid of $q/T_{\nu,0}$ values to evaluate the integrals required to compute the energy density and momentum flux of massive neutrinos. We have checked convergence of our scheme against a 5-point momentum grid and found negligible difference in the CMB and matter power spectrum in the parameter space of interest. We also precompute the coefficient $\alpha_l$ and tabulate them. We emphasize that energy and momentum conservation ensures that $\alpha_0=\alpha_1 = 0$, which we have checked with high accuracy.

For simplicity, we assume throughout this paper that the neutrino sector contains one massive neutrino, with the remaining neutrino species being massless. All neutrinos are assumed to interact with the same coupling strength $G_{\rm eff}$. We find that varying the number of massive neutrinos and number of mass eigenstates, while holding $N_{\mathrm{eff}}$ and $\sum m_\nu$ constant, has a very small impact on the CMB and matter power spectra for all values of $G_{\rm eff}$ consistent with the data used here. It is however possible that future data might be sensitive to the way $\sum m_\nu$ is spread among different mass eigenstates.

\begin{figure*}[t!]
\centering
\includegraphics[width=0.495\linewidth]{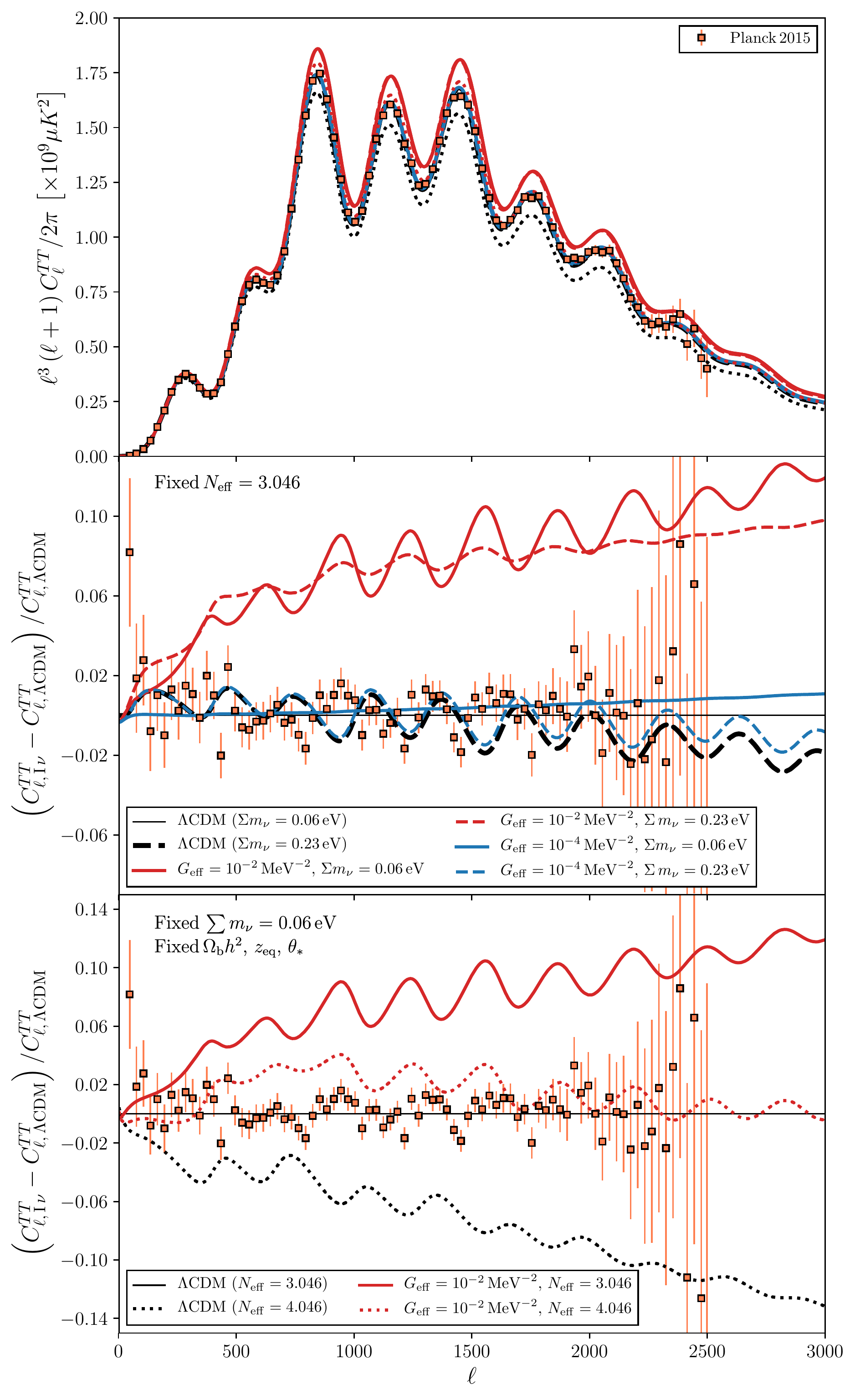}
\includegraphics[width=0.495\linewidth]{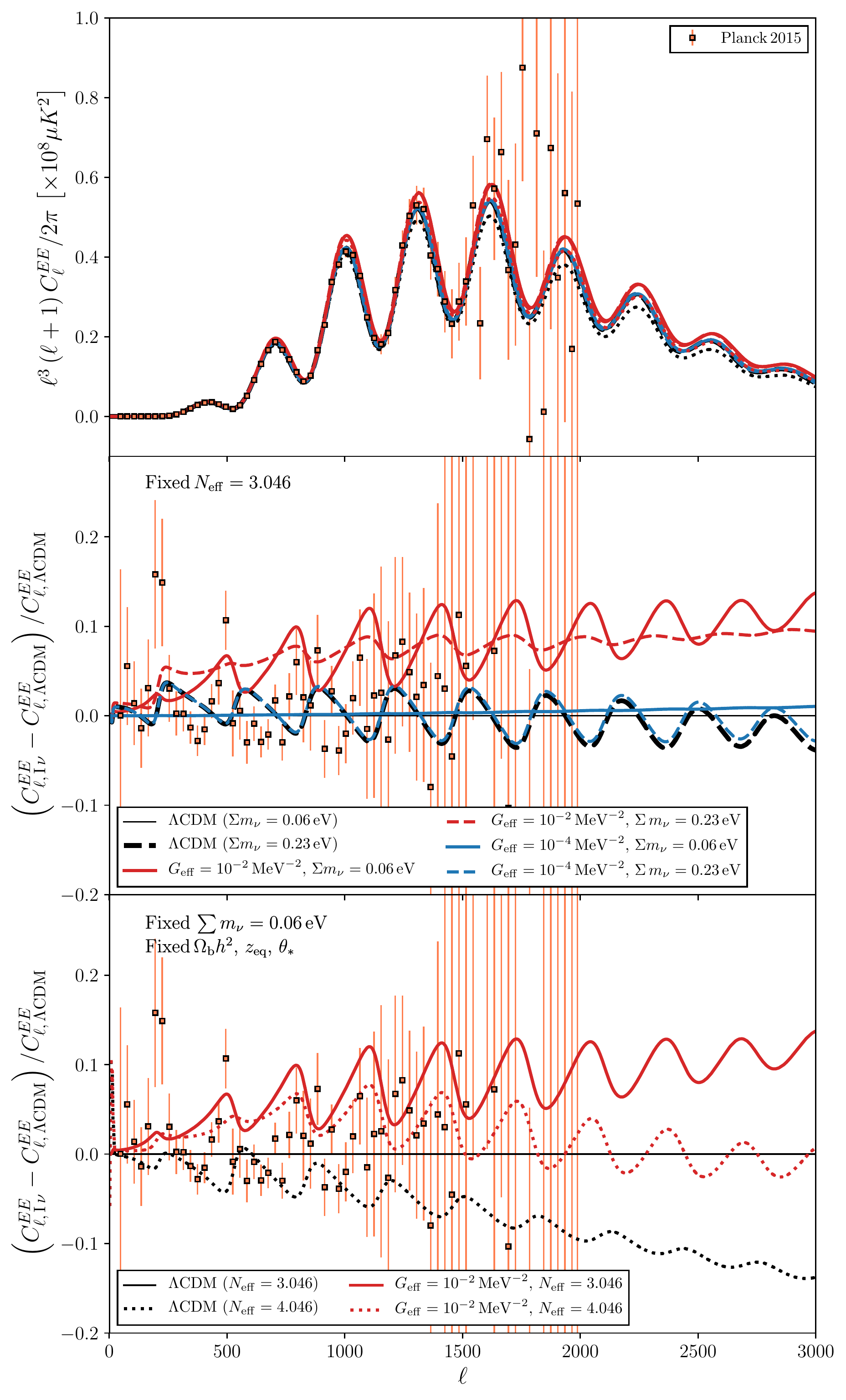}
\caption{Effects of $\sum m_\nu$, $G_\mathrm{eff}$, and $N_{\rm eff}$ on the phase and amplitude of the TT and EE power spectra. Colors denote different values of $G_\mathrm{eff}$. Solid spectra correspond to $\sum m_\nu=0.06\,\mathrm{eV}$ and dashed spectra correspond to $\sum m_\nu=0.23\,\mathrm{eV}$. Measurements from the Planck 2015 data release are included \citep{planckCMB}.}
\label{fig:TT_examples}
\end{figure*}

At early times, the large self-interaction rate of neutrinos renders the equations of motion for multipoles $l\geq2$ extremely stiff. To handle this, we employ a tight-coupling scheme \cite{CyrRacine:2010bk} in which multipole moments with $l\geq2$ are set to zero at early times. Once the neutrino self-interaction rate falls to about a 1000 times the Hubble expansion rate, we turn off this tight-coupling approximation and allows power to flow to the higher multipoles. We have checked that this switch happens early enough as to not affect the accuracy of our results. After neutrino decoupling, once they become non-relativistic, we revert to the standard velocity-integrated truncated Boltzmann hierarchy as described in Ref.~\cite{Lewis:2002nc}. We also modified the adiabatic initial conditions for the cosmological perturbations to take into account the absence of free-streaming neutrinos at early times. Finally, throughout this work, we use the standard BBN predictions to compute the helium abundance given the abundance of relativistic species and the baryon-to-photon ratio. 

\section{Effect on Cosmological Observables}\label{sec:Cosmo_obs}
\subsection{Cosmic microwave background}\label{sec:CMB}

In the standard cosmological paradigm, free-streaming neutrinos travel supersonically through the photon-baryon plasma at early times, hence gravitationally pulling photon-baryon wavefronts slightly ahead of where they would be in the absence of neutrinos \cite{Bashinsky:2003tk,Baumann15,Choi:2018gho}. As a result, the free-streaming neutrinos imprint a net phase shift in the CMB power spectra towards larger scales (smaller $\ell$), as well as a slight suppression of its amplitude. Free-streaming neutrinos thus lead to a physical size of the photon sound horizon at last scattering $r_*$ that is slightly larger than it would otherwise be. This phase shift is thought to be a robust signature of the presence of free-streaming radiation in the early Universe \cite{Follin:2015hya,Baumann:2017lmt,Choi:2018gho}.
 
The neutrino self-interactions mediated by the coupling constant $G_{\mathrm{eff}}$ delay the time at which neutrinos begin to free-stream. Fourier modes entering the causal horizon while neutrinos are still tightly-coupled will not experience the gravitational tug of supersonic neutrinos and will therefore not receive the associated phase shift and amplitude reduction. Compared to the standard $\Lambda$CDM model, neutrino self-interactions thus shift the CMB power spectra peaks towards smaller scales (larger $\ell$) and boost their fluctuation amplitude. This leads to a net reduction of the physical size of the photon sound horizon at last scattering $r_*$. As we shall see, this is the key feature of our model that helps reconcile CMB and late-time measurements of the Hubble constant $H_0$.

The left panels of \autoref{fig:TT_examples} show the temperature CMB power spectra and their relative difference to a $\Lambda \mathrm{CDM}$ model for different values of $G_{\mathrm{eff}}$, $\sum m_\nu$, and $N_{\rm eff}$ to illustrate the effects of neutrino self-scattering in the presence of a non-vanishing mass term. Here, we keep $\Omega_{\rm m}$ fixed as $\sum m_\nu $ changes, and use the best-fit Planck TT+lowP+lensing $\Lambda\mathrm{CDM}$ values as our fiducial cosmology \citep{planck2015}. The middle left panel of \autoref{fig:TT_examples} displays the combined effect of changing both $G_{\rm eff}$ and $\sum m_\nu$. For the minimal sum of neutrinos masses $\sum m_\nu = 0.06\,\mathrm{eV}$, an interaction strength of $G_{\mathrm{eff}}=10^{-4}\,\mathrm{MeV}^{-2}$ (solid blue line)  has for only effect a slight increase of power at large multipoles. On the other hand, increasing the neutrino coupling strength to $G_{\mathrm{eff}}=10^{-2}\,\mathrm{MeV}^{-2}$ (solid red line) significantly boosts the amplitude of the TT spectrum and introduces a clear phase shift (identifiable from the oscillatory pattern of the residuals), which are the two telltale signatures of self-scattering neutrinos as described above. 

Increasing the sum of neutrinos masses to $\sum m_\nu = 0.23\,\mathrm{eV}$ (at fixed $\Omega_{\rm m}$) delays the time of matter-radiation equality. The delay slightly increases the amplitude of the TT spectrum near the first few acoustic peaks and dampens the spectrum at smaller scales (see dashed black line in \autoref{fig:TT_examples}). The resulting changes to the photon-baryon sound horizon at recombination and to the angular diameter distance to the surface of last scattering create a net phase shift towards low $\ell$ \citep{lesgourges2006}, that is, in the opposite direction to that caused by increasing $G_{\mathrm{eff}}$. This opens the door for possible cancellations between the relative phase shift (as compared to $\Lambda$CDM) caused by neutrino self-scattering and that resulting from a large sum of neutrino masses. Such cancellation partially occurs in the middle left panel of \autoref{fig:TT_examples} for $G_{\mathrm{eff}}=10^{-2}\,\mathrm{MeV}^{-2}$ as $\sum m_\nu$ is increased from $0.06$ to $0.23$ eV (dashed red line).  Similarly, the boost in amplitude from $G_{\mathrm{eff}}$ can also compensate for the damping effects of increasing $\sum m_\nu$ at small scales (see e.g. the dashed blue line). Overall, we see that the effect of massive neutrinos and increased interaction strength are nearly \emph{additive}\footnote{Indeed, combining the spectrum for \{$G_{\mathrm{eff}}=10^{-2}\,\mathrm{MeV}^{-2}$, $\Sigma m_\nu=0.06\,\mathrm{eV}$\} (solid red line) with that of the $\Sigma m_\nu=0.23\,\mathrm{eV}$ $\Lambda \mathrm{CDM}$ model (dashed black line) yields a spectrum similar to the model with \{$G_{\mathrm{eff}}=10^{-2}\,\mathrm{MeV}^{-2}$, $\Sigma m_\nu=0.23\,\mathrm{eV}$\} (dashed red line).}, reflecting the fact that the physical processes associated with each of these properties take place at different times in the cosmological evolution. 

The lower left panel of \autoref{fig:TT_examples} displays the impact of increasing the energy density of the neutrino fluid, which we parametrized here through the standard parameter $N_{\rm eff}$, defined via the relation
\begin{align}
  \label{eq:rhoR}
  \rho_{\rm R} = \left[1+N_{\mathrm{eff}}\frac{7}{8}\left(\frac{4}{11}\right)^{4/3}\right]\rho_\gamma,
\end{align}
where $\rho_{\rm R}$ and $\rho_\gamma$ are the total energy density in radiation and in photons, respectively. The effects on the CMB of increasing $N_{\rm eff}$ have been well-studied in the literature (see e.g.~Ref.~\cite{Hou:2011ec}) for the case of free-streaming neutrinos. For fixed values of the angular scale of the sound horizon, the epoch of matter-radiation equality, and the physical baryon abundance, it was found that the most important net impact of increasing $N_{\rm eff}$ was to damp the high-$\ell$ tail of the TT spectrum and to induce a phase shift towards larger scales (low-$\ell$). Interestingly, self-interacting neutrinos can partially compensate for these effects, hence pointing to a possible degeneracy between $G_{\rm eff}$ and $N_{\rm eff}$. An example of this can be seen in the dotted red line in the lower left panel of \autoref{fig:TT_examples}, where the excess of damping caused by $N_{\rm eff}=4.046$ (dotted black line) is compensated by suppressing neutrino free-streaming with $G_{\rm eff}=10^{-2}$ MeV$^{-2}$. 

$G_{\mathrm{eff}}$ affects the EE polarization power spectrum in a similar manner as the temperature spectrum. The right panel of \autoref{fig:TT_examples} shows that the phase shift between the standard $\Lambda$CDM model and that with self-interacting neutrinos is more visible in this case due to the sharp, well defined peaks of the polarization spectrum \citep{Baumann15}. This allows to directly see in which direction the spectrum is shifted compared to $\Lambda$CDM since the oscillations in the residuals lean in the direction of the phase shift, that is, there is a sharper drop off in the residuals in the direction that the spectrum is shifted. Once again, we clearly see that the absence of phase shift caused by a large value of $G_{\mathrm{eff}}$ can be partially canceled by increasing $\sum m_\nu$, in a nearly additive fashion. For the EE polarization spectrum, suppressing neutrino free-streaming can somewhat compensate the extra damping caused by a large $N_{\rm eff}$ (at fixed $\theta_*$, $z_{\rm eq}$, and $\Omega_{\rm b}h^2$; see lower right panel of 
\autoref{fig:TT_examples}).
\subsection{Matter power spectrum}\label{sec:matter}
\begin{figure*}[t!]
\centering
\includegraphics[width=0.498\linewidth]{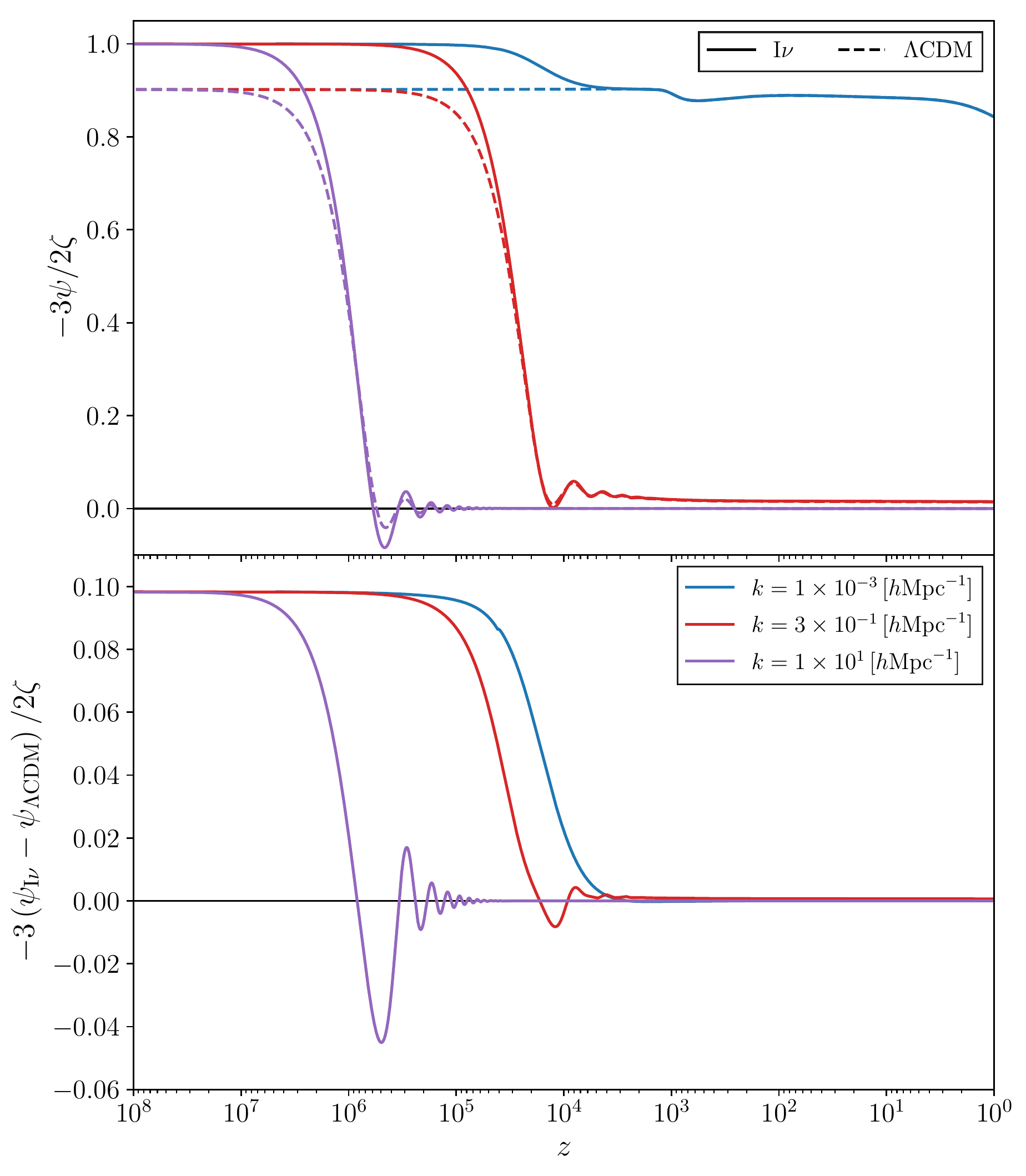}
\includegraphics[width=0.487\linewidth]{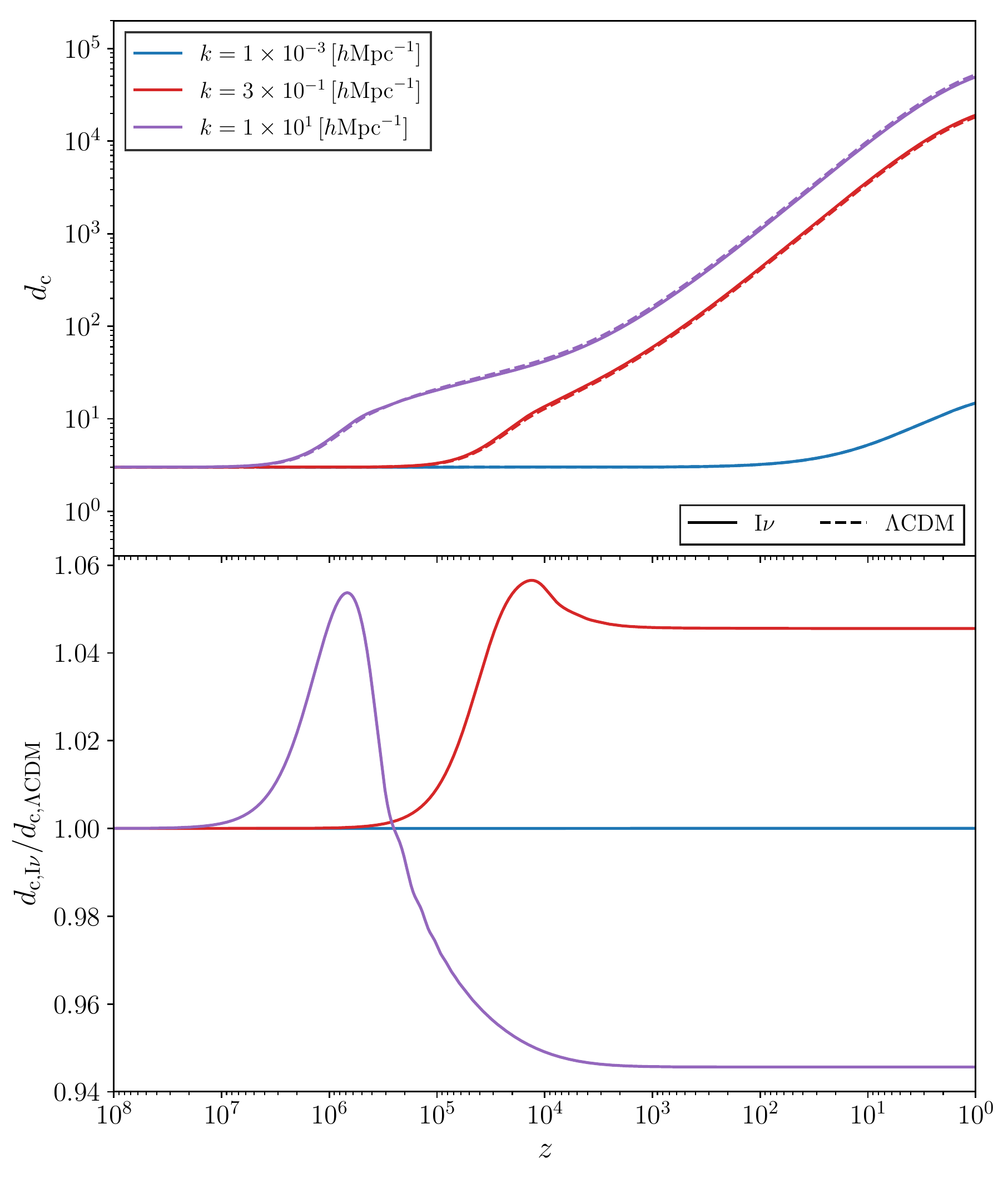}
\caption{The evolution of the $\psi$ gravitational potential (left) and of the gauge invariant dark matter density contrast $d_{\rm c}$ (right) for different $k$-modes as a function of redshift. Solid lines correspond to the interacting neutrino case with $G_{\rm eff} = 10^{-2}$ MeV$^{-2}$, $N_{\rm eff}= 3.046$, and $\sum m_\nu=0.06$ eV, whereas dashed lines correspond to the $\Lambda\mathrm{CDM}$ case. On the left, we plot $-3\psi/(2\zeta)$, where $\zeta$ is the gauge-invariant curvature perturbation. The lower left panel shows the normalized difference between the interacting neutrino and $\Lambda$CDM $\psi$ potential, while the lower right panel shows the ratio of the dark matter fluctuations in the two models. The onset of neutrino free-streaming for the interacting neutrino model shown here occurs at $z_{\rm dec,\nu}\simeq 10^4$. Dark matter fluctuations entering the horizon while neutrinos are still tightly coupled decay and appear damped at present relative to $\Lambda \mathrm{CDM}$, while those entering the horizon during neutrino decoupling receive a net boost that persists until the present epoch.}
\label{fig:potentials}
\end{figure*}

The growth of matter fluctuations is sensitive to the presence of self-interacting neutrinos through the neutrinos' impact on the two gravitational potentials $\phi$ and $\psi$. Indeed, neutrino self-interactions suppress the anisotropic stress of the universe, leading to $\phi-\psi=0$ before the onset of neutrino free-streaming. This contrasts with the $\Lambda$CDM case for which $\phi = (1 + 2R_\nu/5)\psi$ on large scales at early times for the adiabatic mode \cite{Ma:1995ey}, where $R_\nu$ is the radiation free-streaming fraction. This difference in the evolution of the potentials modifies the gravitational source term driving the growth of matter fluctuations. The equation describing the evolution of dark matter fluctuations can be written in Fourier space as \cite{Bashinsky:2003tk}
\be\label{eq:dm_growth}
\ddot{d}_{\rm c} + \frac{\dot{a}}{a}\dot{d}_{\rm c} = -k^2\psi,
\ee
where
\be 
d_{\rm c} \equiv \de_{\rm c} - 3\phi,
\ee
and where $\de_{\rm c} = \de\rho_{\rm c}/\rho_{\rm c}$ is the standard dark matter energy density contrast in Newtonian gauge. Here, an overhead dot denotes a derivative with respect to conformal time $\tau$. The gauge-invariant variable $d_{\rm c}$ represents the fractional dark matter number density perturbation by unit coordinate volume. At late times, $d_{\rm c}$ is nearly equal to $\de_{\rm c}$ and it is thus a useful quantity to understand the structure of the matter power spectrum at $z=0$. In the radiation-dominated epoch where $\dot{a}/a = \tau^{-1}$, the solution to Eq.~\eqref{eq:dm_growth} can be written \cite{Dodelson-Cosmology-2003}
\be\label{eq:sol_growth_cdm} 
d_{\rm c}(k,\tau) = -\frac{9}{2}\phi_{\rm p} + k^2\int_0^\tau d\tau' \tau'\psi(k,\tau')\ln{(\tau'/\tau)},
\ee
where $\phi_{\rm p}$ is the primordial value of $\phi$ on large scales. The integral appearing in Eq.~\eqref{eq:sol_growth_cdm} obtains most of its contribution when $k\tau\sim1$. The changes to the growth of dark matter fluctuations can thus be understood by examining the behavior of the $\psi$ potential at horizon entry.

We compare the evolution of $\psi$ in the presence of self-interacting neutrinos with $G_{\rm eff}=10^{-2}$ MeV$^{-2}$ to that of standard $\Lambda$CDM in the left panel of \autoref{fig:potentials}. There, we track the evolution of three different Fourier modes: $k=10 \, h/\mathrm{Mpc}$ which enters the horizon during the radiation dominated era while neutrinos are still tightly-coupled to each other, $k=0.3 \, h/\mathrm{Mpc}$ which roughly corresponds to the scale entering the horizon when neutrinos begin to free-stream, and $k=10^{-3}\, h/\mathrm{Mpc}$ which does not enter the horizon until far after neutrino decoupling. We use here the same cosmological parameters as in \autoref{fig:TT_examples}. The resulting evolution of dark matter fluctuations for these three modes is shown in the right panel of \autoref{fig:potentials}.

When modes enter the horizon during the radiation-dominated era, the gravitational potential $\psi$ decays in an oscillatory fashion \cite{Dodelson-Cosmology-2003}. The absence of anisotropic stress implies that $\psi$ starts its oscillatory decaying behavior from a larger amplitude. This boosts the amplitude of the envelope of the decaying oscillations as compared to $\Lambda$CDM, leading to an overall slower decay. While this at first increases the amplitude of dark matter fluctuations at horizon entry as compared to $\Lambda$CDM (see bottom right panel of \autoref{fig:potentials}), the subsequent oscillations of the integrand appearing in Eq.~\eqref{eq:sol_growth_cdm} lead to a net \emph{damping} of the dark matter perturbation amplitude. Another way to think about this is that the slower decay of the potential $\psi$ in the presence of self-interacting neutrinos reduces the horizon-entry boost that dark matter fluctuations experience as compared to $\Lambda$CDM.

For modes entering the horizon at the time of neutrino decoupling, the potential $\psi$ begins decaying from its larger value with $R_\nu=0$ but rapidly locks into its standard $\Lambda$CDM evolution due to the onset of neutrino free-streaming. This case thus displays the quickest damping of the $\psi$ potential after horizon entry, which leads to a net boost of dark matter fluctuations as compared to $\Lambda$CDM. Indeed, these modes receive a positive contribution near horizon entry from the integral in Eq.~\eqref{eq:sol_growth_cdm}, but without the subsequent extra damping due to the $\psi$ potential quickly converging to its $\Lambda\mathrm{CDM}$ behavior. The evolution of the $k=0.3 \, h/\mathrm{Mpc}$ mode in \autoref{fig:potentials} displays this behavior. 

Finally, modes entering the horizon well-after the onset of neutrino free-streaming behave exactly like their $\Lambda\mathrm{CDM}$ counterparts, as illustrated by the $k=10^{-3}\, h^{-1}\mathrm{Mpc}$ mode in \autoref{fig:potentials}. Taking together the evolution of the different Fourier modes entering before, during, and after neutrino decoupling, we expect the matter power spectrum to have the following properties (at fixed neutrino mass). For large wavenumbers entering the horizon while neutrinos are tightly coupled, we expect the matter power spectrum to be suppressed compared to $\Lambda$CDM. As we go to larger scales and approach modes entering the horizon at the onset of free-streaming, we expect a ``bump''-like feature displaying an excess of power as compared to $\Lambda$CDM. As we go to even larger scales, the matter power spectrum is expected to asymptote to its standard $\Lambda$CDM value.  

\begin{figure}
\centering
\includegraphics[width=\linewidth]{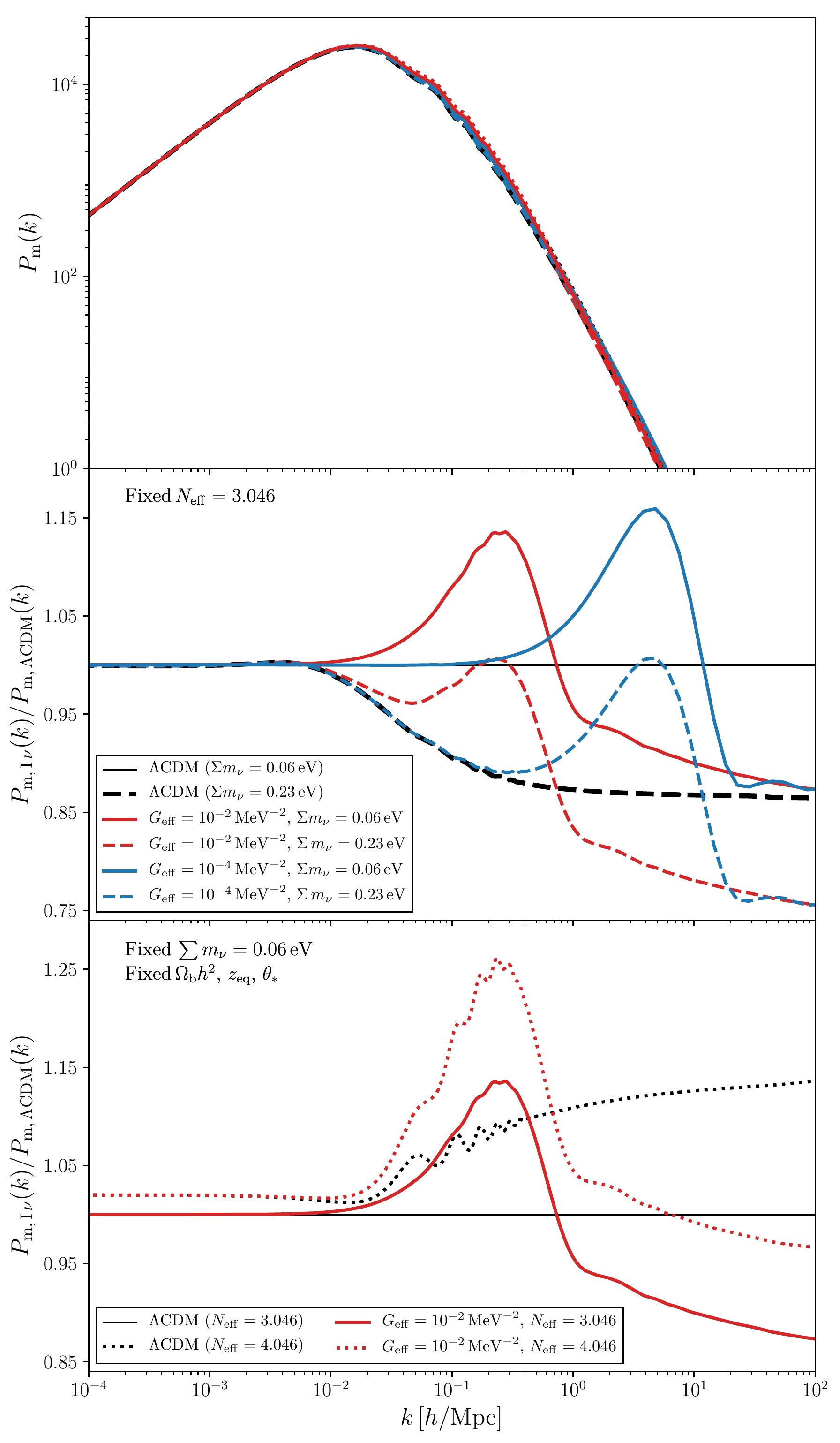}
\caption{Effects of $G_\mathrm{eff}$, $\sum m_\nu$, and $N_{\rm eff}$ on the matter power spectrum. Colors denote different values of $G_\mathrm{eff}$. Solid spectra correspond to $\sum m_\nu=0.06\,\mathrm{eV}$ and dashed spectra correspond to $\sum m_\nu=0.23\,\mathrm{eV}$. Dotted lines in the bottom panel have $N_{\rm eff}=4.046$. Note the localized increase in amplitude at the scales entering the horizon at the onset of neutrino free-streaming.}
\label{fig:matter_examples}
\end{figure}

These expectations are indeed realized as shown in \autoref{fig:matter_examples}. The middle panel shows the power spectrum ratios between the interacting neutrino models and $\Lambda$CDM. Focusing for the moment on the cases with $\sum m_\nu=0.06$ eV, we see that the matter power spectrum is damped at large wavenumbers and then displays a broad peak-like feature with an excess of power as compared to $\Lambda$CDM. The shape of this power excess is determined by the neutrino visibility function \cite{Cyr-Racine:2013aa} encoding the details of neutrino decoupling. Increasing the sum of neutrino masses (at fixed $\Omega_{\rm m}$) leads to a damping of the matter power spectrum on small scales \citep{lesgourges2006,Lattanzi:2017ubx}. This standard reduction of power is shown for $\Lambda$CDM as the thick black dashed line in \autoref{fig:matter_examples}. Interestingly, this small-scale suppression is also present for self-interacting neutrinos and occurs \emph{in addition} to that caused by the slower decay of the gravitational potential $\psi$ discussed above. Thus, the matter power spectrum for massive self-interacting neutrinos is even \emph{more suppressed} at large $k$ than in the standard $\Lambda$CDM case with massive neutrinos.  

This fact might seem counterintuitive at first since the reduction of small-scale power from massive neutrinos is often refereed to as ``free-streaming'' damping. We see this moniker is somewhat of a misnomer since the damping is present whether or not neutrinos are actually free-streaming. Instead, the small-scale reduction of power is simply caused by the large pressure term that prohibits neutrino clustering on these scales. This pressure term is always there as long as neutrinos are relativistic, even when neutrinos are self-scattering. As was the case for the CMB, the effects of a non-vanishing sum of neutrino masses and large $G_{\rm eff}$ are largely additive. Comparing the $\sum m_\nu =0.23$ eV cases to that with $\sum m_\nu =0.06$ eV in \autoref{fig:matter_examples} for both interacting neutrino models shown illustrates this well. Again, this near additivity reflects the fact that part of the effect comes from the behavior of dark matter fluctuations at horizon entry, while the rest is caused by the large pressure term of relativistic neutrinos on small scales. 

The lowest panel of \autoref{fig:matter_examples} shows the effect of increasing $N_{\rm eff}$ (at fixed $\theta_*$, $z_{\rm eq}$, and $\Omega_{\rm b}h^2$) on the matter power spectrum. For $\Lambda$CDM, the main impact is to increase the amplitude of Fourier modes that enter the causal horizon during radiation domination. This results from the larger radiation density and free-streaming fraction $R_\nu$ \cite{Bashinsky:2003tk} at early times. Suppressing neutrino free-streaming for $N_{\rm eff}=4.046$ (dotted red line) nullifies this increase of power on small-scales, even leading to a net damping compared to $\Lambda$CDM for $k>10h$/Mpc. However, as the neutrinos start to decouple from one another, the larger radiation density leads to a higher amplitude feature on scales entering the horizon at that time. 

We thus see that taken together, the joint effect of $G_\mathrm{eff}$, $\sum m_\nu$, and $N_{\rm eff}$ can lead to matter power spectra having a significantly different structure and shape than the standard $\Lambda$CDM paradigm.  

\section{Data \& Methodology}\label{sec:data}
We use our modified versions of {\tt CAMB} \cite{camb00} and {\tt CosmoMC + Multinest} \cite{Lewis:2002ah,feroz08} to place constraints on $G_{\mathrm{eff}}$, $N_{\mathrm{eff}}$, and $\sum m_\nu$, as well as the standard cosmological parameters. We use nested sampling \cite{skilling2006} to ensure that we properly sample our posterior, which we expect to be multi-modal as in previous cosmological studies of self-interacting neutrinos \cite{Cyr-Racine:2013aa,lancaster,oldengott17}.

We use a combination of CMB and low-redshift data sets in our analysis:
\begin{itemize}
  \item {\bf TT}: low-$\ell$ and high-$\ell$ CMB temperature power spectrum from the Planck 2015 release\footnote{Explicitly, we use the likelihood {\tt plik\_lite\_v18\_TT} for high-$\ell$ and {\tt commander\_rc2\_v1.1\_l2\_29\_B} at low-$\ell$.} \citep{planckCMB}.
  \item {\bf EE, TE}: low-$\ell$ and high-$\ell$ CMB E-mode polarization and their temperature cross-correlation from the Planck 2015 data release\footnote{Explicitly, we use the likelihood {\tt plik\_lite\_v18\_TTTEEE} for high-$\ell$ and {\tt lowl\_SMW\_70\_dx11d\_2014\_10\_03\_v5c\_Ap} at low-$\ell$.} \citep{planckCMB}. The 2015 polarization data is known to have residual systematics and results drawn using this dataset should be interpreted with caution. While our main conclusions will not make use of this dataset, we nonetheless present results including this dataset for completeness. 
  \item {\bf lens}: CMB lensing data from the Planck 2015 data release \citep{plancklens}.
  \item {\bf BAO}: Baryon Acoustic Oscillation (BAO) measurements from the 6dF Galaxy Survey constraining $D_V$ at $z=0.106$ \citep{BAO1}, Sloan Digital Sky Survey (SDSS-III) Baryon Oscillation Spectroscopic Survey (BOSS) data release 11 low-$z$ data measuring $D_V$ at $z=0.32$ and CMASS data measuring $D_V$ at $z=0.57$ \citep{BAO2}, and data from the SDSS Main Galaxy Sample measuring $D_V$ at $z=0.15$ \citep{BAO3}
  \item ${\bf H_0}$: Local measurement\footnote{We note that the mean value of $H_0$ used in our analysis is slightly lower ($\sim 0.14\sigma$) than the value quoted in the published version of Ref.~\cite{HST} (ours corresponds to the value found in an earlier version of their manuscript). We do not expect this very small difference to impact our results in any way.} of the Hubble parameter $H_0=73.0 \pm 1.75\,\mathrm{km\,s}^{-1}\mathrm{Mpc}^{-1}$ at $z=0.04$ from Ref.~\citep{HST}.
\end{itemize}

We use the lite high-$\ell$ likelihood, which marginalizes over nuisance parameters, to reduce the number of free parameters in our analysis. We use the following data set combinations for our nested sampling analysis: `TT+lens+BAO', `TT+lens+BAO+$H_0$', `TT,TE,EE', and `TT,TE,EE+lens+$H_0$'.

In \autoref{tab:priors} we list our adopted prior ranges. We place uniform priors on all these parameters, except for the Planck calibration parameter $y_{\rm cal}$, for which we use a Gaussian prior $y_{\rm cal} = 1.0000 \pm 0.0025$. For the analyses using the Planck polarization data, we also include a Gaussian prior on the optical depth to reionization given by $\tau=0.058 \pm 0.012$ from Ref.~\cite{Adam:2016hgk}. 

We use 2000 live points in our nested sampling runs, setting the target sampling efficiency to 0.3. We impose an accuracy threshold on the log Bayesian evidence of $20\%$, which ensures that our confidence intervals are highly accurate. We use the mode-separation feature of {\tt Multinest} to isolate each posterior mode and compute their respective summary statistics.  
\begin{table}[tbp]
\caption{Adopted prior ranges\label{tab:priors}}
\center
\begin{tabular}{cc}
\hline
 Parameter & Prior \\
\hline \\[-1.5ex]
$\rm{log}_{10}(G_{\rm{eff}}{\rm MeV}^2)$ & $\left[-5.5, -0.000001\right]$ \\
$\sum m_\nu$\,[eV] & $\left[0.0001, 1.5\right]$  \\
$N_\mathrm{eff}$ & $\left[2.0, 5.0\right]$ \\[+1.5ex]
\hline \\[-1.5ex]
{$\Omega_{\rm b} h^2$} & $\left[0.01, 0.04\right]$ \\
{$\Omega_{\rm c} h^2$} & $\left[0.08, 0.16\right]$ \\
{$100\theta_\mathrm{MC}$} & $\left[1.03, 1.05\right]$ \\
{$\tau$} & $\left[0.01, 0.25\right]$ \\
{${\rm{ln}}(10^{10} A_s)$} & $\left[2, 4\right]$ \\
{$n_s$} & $\left[0.85, 1.1 \right]$ \\ 
 \hline \\[-1.5ex]
{$y_{\rm cal}$} & $\left[0.9, 1.1 \right]$ \\
\hline
\end{tabular}
\end{table}

\section{Results}\label{sec:results}
In this section, we first present the main highlights of our analysis, before discussing the physical properties of the two categories of interacting neutrino models that are favored by the data. We end this section with a brief discussion about which properties of interacting neutrino models help alleviate current tensions in cosmological data. Throughout this section, we quote and analyze results for the $\mathrm{TT+lens+BAO+}H_0$ data set combination unless otherwise specified. We discuss the impact of other data (including CMB polarization) in \autoref{sec:discuss} and list parameter constraints for these other data set combinations in \autoref{tab:SI_nu_params} and \autoref{tab:LCDM_nu_params}, in Appendix \ref{app:all_res}.
\subsection{Highlights}\label{sec:highlights}

\begin{figure*}
\centering
  \includegraphics[width=0.49\linewidth]{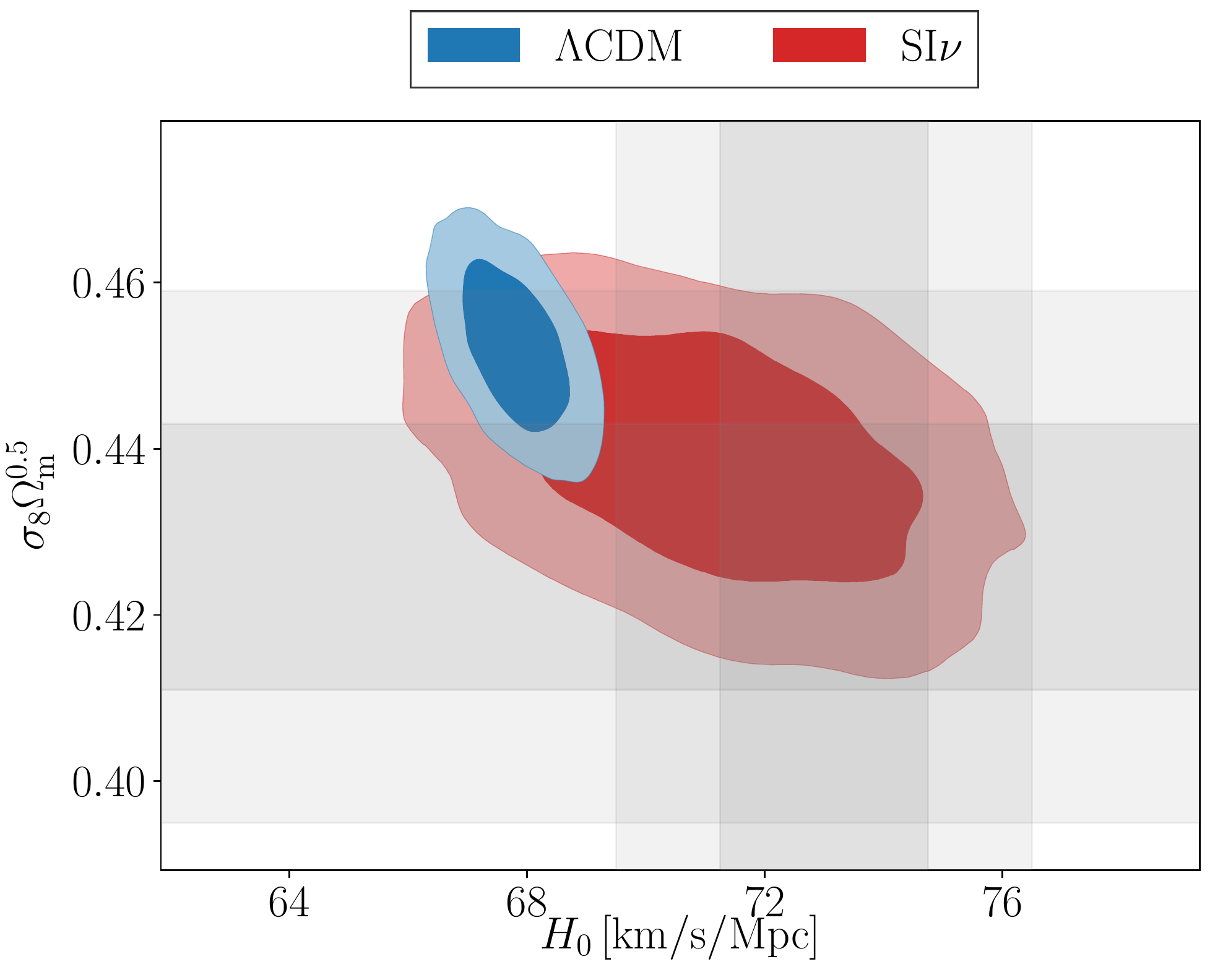}
  \includegraphics[width=0.49\linewidth]{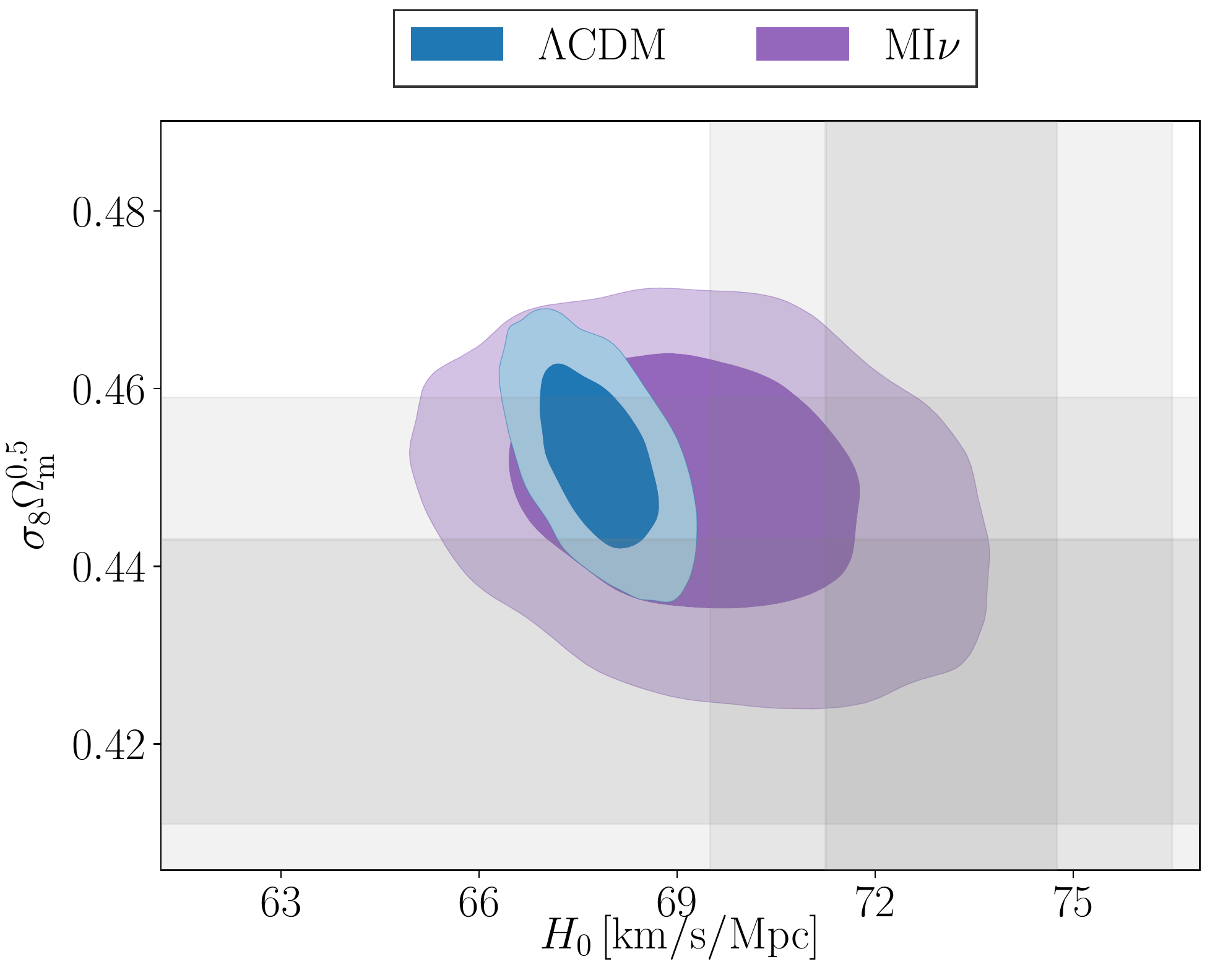}
  \caption{2D posteriors for $S_8$ and $H_0$ illustrating how neutrino self-interactions can remedy cosmological tensions. We compare the Planck $\mathrm{TT+lens+BAO}$ $\Lambda \mathrm{CDM}$ posterior to the $\mathrm{SI}\nu$ and $\mathrm{MI}\nu$ posteriors for $\mathrm{TT+lens+BAO}$. We overlay $2\sigma$ bands for the measurements $S_8 = 0.427 \pm 0.016$ \cite{Hikage:2018qbn} and $H_0 = 73 \pm 1.75$ km/s/Mpc \cite{HST}.}
  \label{fig:money_SInu}
\end{figure*}
 \begin{table*}
  \caption{TT + lens + BAO + $H_0$ Constraints: Parameter $68\%$ Confidence Limits \label{tab:params}}
  \begin{ruledtabular}
  \begin{tabular}{ccc}
   Parameter & Strongly Interacting Neutrino Mode & Moderately Interacting Neutrino Mode \\
  \hline
  {$\Omega_\mathrm{b} h^2$} & $0.02245^{+0.00029}_{-0.00033}$ & $0.02282\pm 0.00030$\\
  {$\Omega_\mathrm{c} h^2$} & $0.1348^{+0.0056}_{-0.0049}$ & $0.1256^{+0.0035}_{-0.0039}$\\
  {$100\theta_{\rm MC} $} & $1.04637\pm 0.00056 $ & $1.04062^{+0.00049}_{-0.00056}$\\
  {$\tau$} & $0.080\pm 0.031$ & $0.127^{+0.034}_{-0.029}$\\
  {$\sum m_\nu$}\,[eV] & $0.42^{+0.17}_{-0.20}$ & $0.40^{+0.17}_{-0.23}$\\
  {$N_\mathrm{eff}$} &  $4.02\pm 0.29$	& $3.79\pm 0.28$\\
  {$\rm{log}_{10}(G_{\rm{eff}}{\rm MeV}^2)$} & $-1.35^{+0.12}_{-0.066}$	& $-3.90^{+1.0}_{-0.93}$\\
  {${\rm{ln}}(10^{10} A_s)$} & $3.035\pm 0.060$	& $3.194^{+0.068}_{-0.056}$\\
  {$n_\mathrm{s}$} & $0.9499\pm 0.0098$	& $0.993^{+0.013}_{-0.012}$\\
  \hline \\[-1.5ex]
  $H_0$\,[km/s/Mpc] & $72.3\pm 1.4$	& $71.2\pm 1.3$\\
  $\Omega_\mathrm{m}$ & $0.3094 \pm 0.0083$	& $0.3010 \pm 0.0080$\\
  $\sigma_8$ & $0.786\pm 0.020$	& $0.813^{+0.023}_{-0.020}$\\
  $10^9 A_{\rm s}$ & $2.08^{+0.11}_{-0.13}$ & $2.44\pm 0.15$\\
  $10^9 A_{\rm s} e^{-2\tau}$ & $1.771\pm 0.016$ & $1.892^{+0.019}_{-0.017}$\\
  $r_*$\,[Mpc] & $136.3\pm 2.4$	& $139.1\pm 2.3$\\
  $100\theta_*$ & $1.04604\pm 0.00056$ & $1.04041^{+0.00058}_{-0.00064}$\\
  $D_{\rm{A}}$ \,[Gpc]	& $13.03\pm 0.23$ & $13.37\pm 0.21$\\
  $r_{\rm drag}$\,[Mpc] & $138.8 \pm 2.5$ & $141.6 \pm 2.3$\\
  \end{tabular}
  \end{ruledtabular}
 \end{table*}
 
 \begin{table*}{}
  \caption{Comparison to $\Lambda\mathrm{CDM}$ for TT + lens + BAO + $H_0$ \label{tab:chi2lcdm_plain}}
  \begin{ruledtabular}
  \begin{tabular}{ccccc}
  Parameter & Strongly Interacting Neutrino Mode & Moderately Interacting Neutrino Mode \\
  \hline \\ [-2ex]
  $\Delta \chi^2_{\mathrm{low\,}\ell}$ 	& $0.66$	& $-0.75$ \\
  $\Delta \chi^2_{\mathrm{high\,}\ell}$ 	& $-1.15$	& $1.08$ \\
  $\Delta \chi^2_\mathrm{lens}$ 	& $0.06$	& $-0.24$ \\
  $\Delta \chi^2_{H_0}$ 	& $-6.68$	& $-6.12$ \\
  $\Delta \chi^2_\mathrm{BAO}$ 	& $-0.81$	& $-0.36$ \\[0.5ex]
  \hline \\[-2ex]
  $\Delta \chi^2_\mathrm{Total}$ 	& $-7.91$	& $-6.39$ \\
  $\Delta \mathrm{AIC}$ 	& $-1.91$	& $-0.39$ 
  \end{tabular}
\end{ruledtabular}
\end{table*}
 
Similarly to previous works \cite{Cyr-Racine:2013aa,lancaster,oldengott17},
we find two unique neutrino cosmologies preferred by the data: a strongly interacting neutrino cosmology (hereafter denoted $\mathrm{SI}\nu$ model) characterized by $\mathrm{log}\left(G_{\mathrm{eff}}\,\mathrm{MeV}^2\right)=-1.35^{+0.12}_{-0.07}$ for the TT+lens+BAO+$H_0$ combination, and a moderately interacting neutrino cosmology (hereafter, $\mathrm{MI}\nu$ model) characterized by $\mathrm{log}\left(G_{\mathrm{eff}}\,\mathrm{MeV}^2\right)=-3.90^{+1.00}_{-0.93}$ for the same data set. Values of $G_{\rm eff}$ between these two modes are strongly disfavored by the data since they either prefer to have a phase shift that is largely consistent with free-streaming neutrinos, or no phase shift at all. We present constraints on cosmological parameters for the $\mathrm{SI}\nu$ and $\mathrm{MI}\nu$ modes in \autoref{tab:params}. While the MI$\nu$ cosmology was nearly indistinguishable from the $\Lambda$CDM scenario with massless neutrinos in previous work \cite{lancaster}, the addition of neutrino mass and $N_{\rm eff}$ in combination with the $H_0$ measurement from Ref.~\cite{HST} leads to a slight preference for a delayed onset of neutrino free-streaming. We expand more on this new development in \autoref{sec:MInu} below.  

Cosmological parameters in the SI$\nu$ cosmology admit values that are significantly different from $\Lambda\mathrm{CDM}$:
\begin{enumerate}
\item The angular scale of the baryon-photon sound horizon at last scattering $100\theta_*=1.04604 \pm 0.00056$ ($68\%$ C.L.) takes a value that is radically different ($>5\sigma$ away) than in the $\Lambda$CDM scenario, reflecting the absence of the free-streaming neutrino phase shift. 
\item The large $N_\mathrm{eff}$ value $4.02 \pm 0.29$ ($68\%$ C.L.) suggests the presence of an additional neutrino species, which might help reduce tensions between different neutrino oscillation experiments.
\item A smaller value of the baryon drag scale $r_\mathrm{drag} = 138.8 \pm 2.5\,\mathrm{Mpc}$ ($68\%$ C.L.) helps reconcile BAO with local Hubble constant measurements, leading to $H_0=72.3\pm1.4\,\mathrm{km\,s}^{-1}\,\mathrm{Mpc}^{-1}$ ($68\%$ C.L.). 
\item The impact of self-interacting neutrinos on the growth of dark matter perturbations and a preferred suppressed spectrum of primordial scalar fluctuations lead to $\sigma_8 = 0.786 \pm 0.020$ ($68\%$ C.L.).
\end{enumerate}
 
To illustrate the ability of neutrino self-interactions to help resolve current cosmological tensions, we compare the $S_8 \equiv \sigma_8 \Omega_{\rm m}^{0.5}$ and $H_0$ 2D posteriors for the $\mathrm{SI}\nu$ model and $\mathrm{MI}\nu$ model with the base $\Lambda \mathrm{CDM}$ model in \autoref{fig:money_SInu}. We overlay bands for HSC constraints on $S_8$ \cite{Hikage:2018qbn} and local measurements of $H_0$ \cite{HST}. In order for our analysis to be independent from these measurements, we show posteriors for the TT+lens+BAO constraints for both the neutrino self-interaction models and $\Lambda \mathrm{CDM}$. Intriguingly, the strong neutrino self-interactions in the $\mathrm{SI}\nu$ model are able to independently produce the preferred values for $S_8$ and $H_0$, \emph{even without using these measurements in our analysis}. The base $\Lambda \mathrm{CDM}$ model is unable to achieve these values, and the weak neutrino interactions of the $\mathrm{MI}\nu$ model can only achieve such values with weak significance.

In \autoref{tab:chi2lcdm_plain} we compute the $\Delta \chi^2$ values between the two neutrino self-interaction models and $\Lambda \mathrm{CDM}$. The data favor the strongly interacting neutrino model over $\Lambda \mathrm{CDM}$ with $\Delta \chi^2_\mathrm{Total} = -7.91$. This is a significant difference, even after accounting for the three extra parameters in the SI$\nu$ model (see \autoref{sec:LCDM_significance} for further discussion about this point). The preference for the self-interacting neutrinos comes from the local measurements of $H_0$, the high-$\ell$ TT data, and the BAO data. 

\begin{figure*}
\centering
  \includegraphics[width=\linewidth]{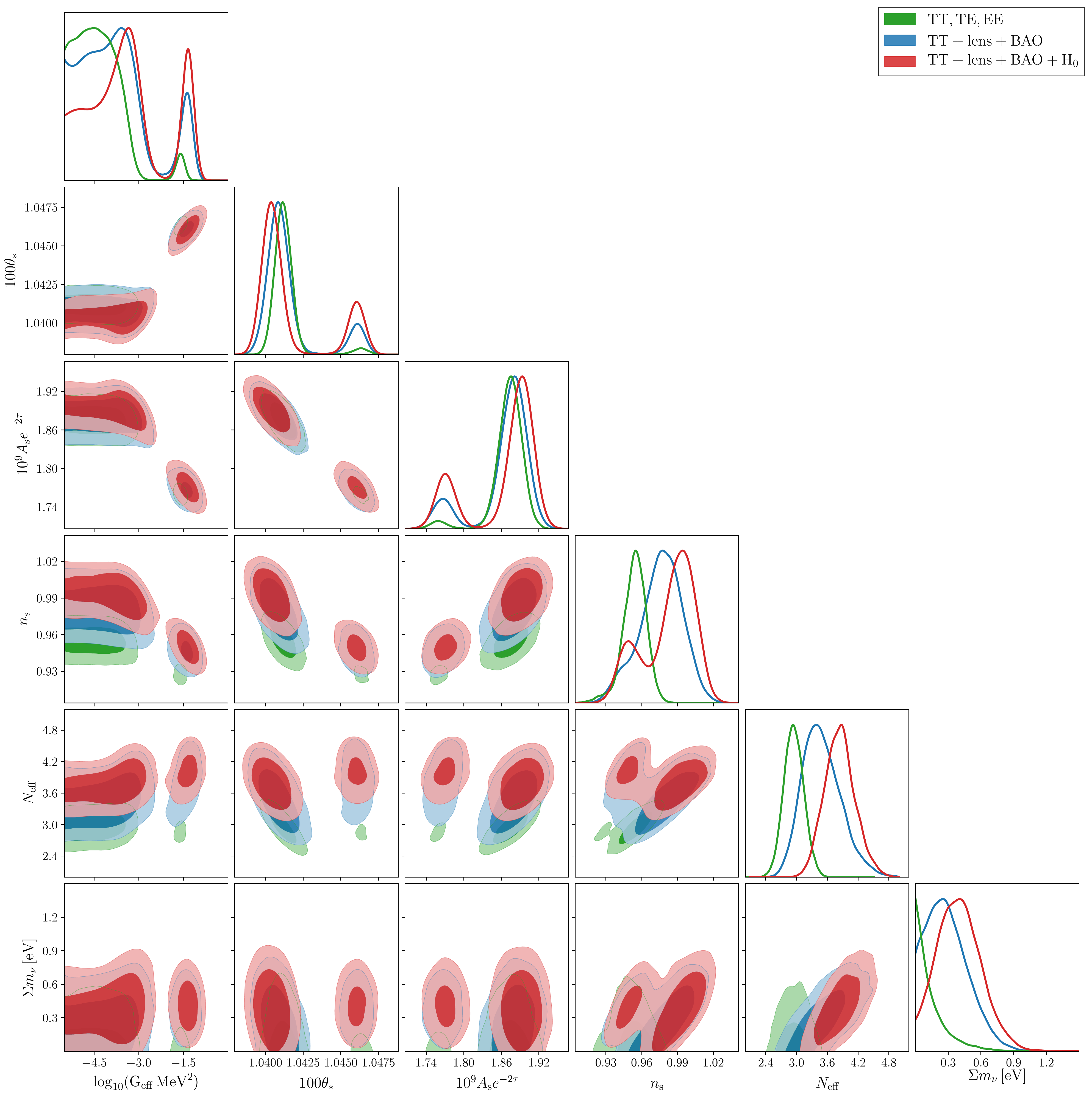}
  \caption{Marginalized posterior distributions for select parameters for three of the data set combinations used in this work. Here, we focus on parameters illustrating the difference between the two modes in relation to sound horizon, the amplitude of the spectrum, and the neutrino properties. Posterior for $H_0$ and $\sigma_8$ are shown in \autoref{fig:controversy}.}
  \label{fig:bimodal_contours}
\end{figure*}

In \autoref{fig:LCDM_interact_standard_contours}, we separate the posterior modes and plot their separate statistical distribution for the most salient parameters. For comparison, we also show the marginalized posteriors for the standard $\Lambda$CDM paradigm, as well as for its $N_{\rm eff}+\sum m_\nu$ two-parameter extension. In \autoref{fig:bimodal_contours}, we show the different covariances between the most relevant model parameters for three of the dataset combinations used in this work\footnote{Due to the presence of two posterior modes with different width, it is difficult to choose a smoothing scale that faithfully captures the intrinsic shape of the whole posterior while removing sampling noise. This particularly affects the $\mathrm{SI}\nu$ mode and results in a significantly reduced height which appears to visually suppress its statistical significance. See \autoref{fig:bimodal_post} in Appendix \ref{app:all_res} for a figure with a smoothing scale more appropriate for the SI$\nu$ mode.}.

\subsection{Strongly interacting neutrino mode}\label{sec:SInu}

\begin{figure*}
\centering
  \includegraphics[width=0.9\linewidth]{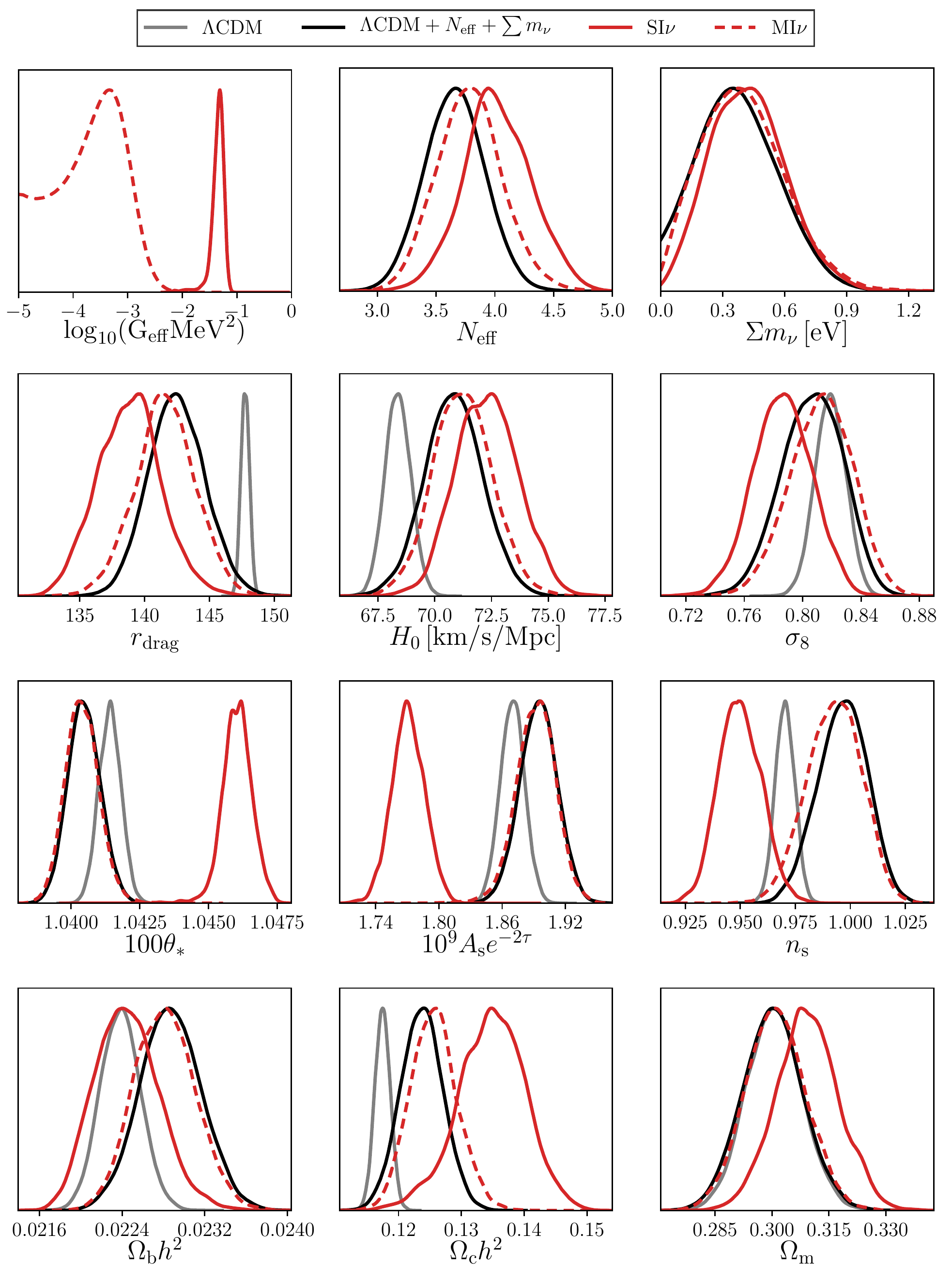}
  \caption{1D posteriors for the TT+lens+BAO+$H_0$ data combination after separating the $\mathrm{SI}\nu$ and $\mathrm{MI}\nu$ modes and plotting them independently. For this reason, the peak locations and posterior shapes are of physical interest rather than the relative heights of the peaks.}
  \label{fig:LCDM_interact_standard_contours}
\end{figure*}

The existence of the SI$\nu$ mode was first pointed out in Ref.~\cite{Cyr-Racine:2013aa}, and further studied in Refs.~\cite{lancaster,oldengott17}. As discussed there, the SI$\nu$ cosmology arises due to a multi-parameter degeneracy that opens up in CMB data when the onset of neutrino free-streaming is delayed until redshift $z\sim 8000$. This approximately coincides with the epoch when Fourier modes corresponding to multipole $\ell \approx 400$ enters the causal horizon \citep{lancaster}, which lies somewhere between the first and second peak of the CMB temperature spectrum. 
We review below the properties of this alternate cosmology, emphasizing its differences with the standard $\Lambda$CDM model. 

\noindent{\bf Sound horizon} One of the most striking features of the SI$\nu$ model is the significantly larger value of the angular size of the sound horizon $\theta_*$. This is probably the most confusing aspect of our results since the angular size of the CMB sound horizon at last scattering is thought to be the best measured quantity in all of cosmology. To understand this apparent discrepancy, it is important to realize that the  angular sound horizon is \emph{defined} as $\theta_* \equiv r_*/D_{\rm A}$, where
\be 
r_* = \int_0^{a_*}\frac{c_{\rm s}(a)}{a^2 H(a)}  da, \qquad D_{\rm A} = \int_{a_*}^1\frac{da}{a^2 H(a)},  
\ee
where $c_{\rm s}$ is the baryon-photon sound speed, $H$ is the Hubble rate, and $a_*$ is the scale factor at last scattering. We thus see that $\theta_*$ is purely defined in terms of \emph{background} quantities, independent of the behavior of cosmological perturbations. In particular, it is independent of the gravitational tug that neutrinos exert on the photons. 

Of course, when fitting CMB data we use the full temperature and polarization spectra computed from the Boltzmann equation which includes the effect of neutrinos. For the SI$\nu$ model, the absence of free-streaming neutrinos means that the CMB spectra do not receive the standard phase shift, and thus appear slightly displaced toward larger $\ell$ as compared to the corresponding $\Lambda$CDM spectra. In order to fit the data, we must compensate for this shift by \emph{increasing} the value of $\theta_*$. Thus, the difference between the values of $\theta_*$ in the SI$\nu$ and $\Lambda$CDM models directly reflects the absence of the free-streaming neutrino phase shift in the former. 

We note that it was a priori far from obvious that such a dramatic change in  the angular size of the sound horizon was possible without introducing other artifacts that would significantly worsen the fit to CMB and BAO data. Our analysis shows that the larger value of $\theta_*$ is achieved by increasing $H_0$ and $\Omega_{\rm c}h^2$ above their $\Lambda$CDM values. 

\noindent{\bf Primordial spectrum} In addition to removing the CMB phase shift, suppressing neutrino free-streaming also increases the amplitude of the temperature and polarization spectra, as discussed in \autoref{sec:CMB}. In the SI$\nu$ model, these changes are reabsorbed by modifying the primordial spectrum of scalar fluctuations parametrized by the amplitude $A_{\rm s}$ and spectral index $n_{\rm s}$. As was found in Refs.~\cite{Cyr-Racine:2013aa,lancaster}, lower values of both $A_{\rm s}$ and $n_{\rm s}$ are required to fit the temperature data in the SI$\nu$ mode. The difference between this alternative cosmology and $\Lambda$CDM is even more apparent if we compare the values of the parameter $A_{\rm s}e^{-2\tau}$ which directly determines the amplitude of the CMB temperature spectrum. As shown in \autoref{fig:LCDM_interact_standard_contours}, this amplitude parameter admits values that are radically ($> 5\sigma$) different than in $\Lambda$CDM, again reflecting the large impact that suppressing neutrino free-streaming has on the CMB. 

\noindent{\bf Neutrino properties} The $\mathrm{SI}\nu$ model is consistent with having an entire additional neutrino species ($N_{\mathrm{eff}}=4.02\pm0.29$, see \autoref{fig:LCDM_interact_standard_contours}), which has interesting implications for neutrino oscillation experiments. By comparing the SI$\nu$ cosmology with a more standard $\Lambda$CDM + $N_{\rm eff}$ + $\sum m_\nu$ model, we can understand how much of this preference is driven by the neutrino self-interaction. As shown in \autoref{fig:LCDM_interact_standard_contours}, the two-parameter extension of the $\Lambda$CDM cosmology already favors a larger $N_{\rm eff}$, but the introduction of strong neutrino self-interactions shifts the posterior to even larger values. To a certain extent, this shift is driven by the need to fit the large value of the local Hubble rate from Ref.~\cite{HST} by reducing the size of the sound horizon at the baryon drag epoch (see e.g.~Ref.~\cite{Aylor:2018drw}). However, in the presence of free-streaming neutrinos, increasing $N_{\rm eff}$ also leads to a larger phase shift toward low $\ell$ which puts a limit on how much extra free-streaming radiation can be added before severely degrading the fit to CMB data. For the SI$\nu$ model, the absence of this phase shift allows for larger $N_{\rm eff}$, which leads to a smaller value of $r_{\rm drag}$ and, in turn, a larger Hubble constant. This is the key feature of the SI$\nu$ model that allows it to severely reduce the Hubble rate tension between CMB and late-time measurements, as we shall discuss in \autoref{sec:H0_tension}.  

The SI$\nu$ model also statistically prefers a nonzero value for the sum of neutrino masses. This preference was however already present at a less significant level ($<2\sigma$) in the $N_{\rm eff}$ + $\sum m_\nu$ extension of the $\Lambda$CDM scenario. In this latter case, nonzero neutrino masses arise from the need to suppress the amplitude of matter of fluctuations at late times (as measured here through CMB lensing) in the presence of a larger $N_{\rm eff}$ and $\Omega_{\rm m}$. For the SI$\nu$ model, the preference for a nonzero sum of neutrino masses is increased ($>2\sigma$) due to the even larger $N_{\rm eff}$ and $\Omega_{\rm m}$ values favored by this scenario.

We note that, in our analysis, the primordial helium abundance $Y_P$ is highly correlated with $N_{\mathrm{eff}}$ due to our use of the BBN consistency condition. Allowing $Y_P$ to take a different set of values in the SI$\nu$ scenario could lead to an even better fit to cosmological data. 

\noindent{\bf Matter clustering} Several competing effects act to set the amplitude of late-time matter fluctuations (as captured by the parameter $\sigma_8$) in the SI$\nu$ model. First, the large values of $N_{\rm eff}$ and $\Omega_{\rm m}$ (the latter necessary to keep the epoch of matter-radiation equality fixed) tend to boost the amplitude of matter fluctuation as discussed in \autoref{sec:matter}. On scales entering the horizon before the onset of neutrino free-streaming, this increase is counteracted by both a nonzero sum of neutrino masses and the reduction of the horizon entry boost for dark matter fluctuations in the presence of self-interacting neutrinos.  Dark matter fluctuations entering the horizon during neutrino decoupling, which for the SI$\nu$ model are coincidentally those primarily contributing to $\sigma_8$, are however enhanced by the rapid decay of the gravitational potential on these scales. Finally, the lower amplitude and spectral index of the primordial scalar spectrum in the SI$\nu$ model tend to suppress power on scales probed by $\sigma_8$. Putting all of these effects together leads to a net \emph{lower} value of $\sigma_8$, which, as discussed in the previous section, might be favored by some probes of late-time matter clustering. The overall shape of the matter power spectrum in the SI$\nu$ model will be further discussed in \autoref{sec:discuss}. 

\subsection{Moderately interacting neutrino mode}\label{sec:MInu}

Within the $\mathrm{MI}\nu$ mode, the onset of neutrino free-streaming occurs before most Fourier modes probed by the Planck high-$\ell$ data enter the causal horizon. As such, the cosmological parameter values preferred by this mode are very similar to those from the $N_{\rm eff}$ + $\sum m_\nu$ extension of the $\Lambda$CDM scenario (see \autoref{fig:LCDM_interact_standard_contours}). The main difference here is that high-$\ell$ CMB modes do not receive the full amplitude suppression associated with free-streaming neutrinos due to the finite width of the neutrino visibility function. In other words, even though these high-$\ell$ modes enter the horizon after most neutrinos have started to free-stream, residual scattering in the neutrino sector still influences the amplitude of the CMB damping tail (see, e.g., the model with $G_{\rm eff}=10^{-4}$ MeV$^{-2}$ and $\sum m_\nu=0.06$ eV in \autoref{fig:TT_examples}). This increased small-scale power allows for a larger $N_{\rm eff}$, which, by reducing the baryon drag scale, leads to slightly larger Hubble constant. This shift is however quite small. 

A surprising fact about the MI$\nu$ mode (also pointed out in Ref.~\cite{oldengott17}) is that it shows a slight statistical preference for a nonzero value of $G_{\rm eff}$. As we can see in \autoref{fig:bimodal_contours}, this preference is nearly entirely driven by the local Hubble constant measurement of Ref.~\cite{HST}. Indeed, removing this dataset from our analysis (blue contours) eliminates most of the preference for a nonzero value of $G_{\rm eff}$. 

\subsection{Mediating Controversy: Effects on \texorpdfstring{$H_0$}{H0} and \texorpdfstring{$\sigma_8$}{sigma8}}\label{sec:H0_tension}

\begin{figure*}[t!]
\begin{subfigure}[t]{\textwidth}
\includegraphics[width=\linewidth]{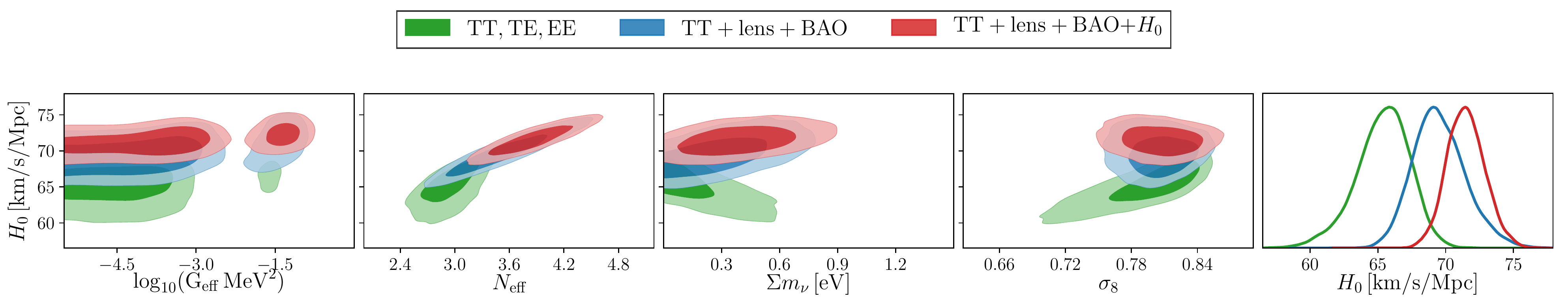}
        \caption{$H_0$ correlations.}\label{fig:H0_contours}
\end{subfigure}
\begin{subfigure}[b]{0.8\textwidth}
\includegraphics[width=\linewidth]{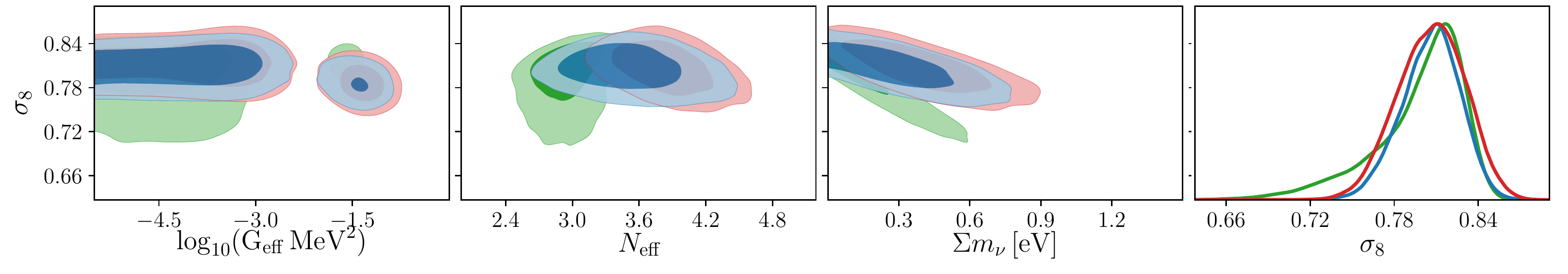}
        \caption{$\sigma_8$ correlations.}\label{fig:sigma8_contours}
\end{subfigure}
\caption{Correlations between $H_0$ and $\sigma_8$ with neutrino properties.\label{fig:controversy}}
\end{figure*}

We show in \autoref{fig:H0_contours} the impact of $G_{\rm eff}$, $N_{\rm eff}$ and $\sum m_\nu$ on the inferred value of the Hubble parameter. As described above, the large values of $N_{\rm eff}$ allowed in the presence of neutrino self-interactions reduce the size of the baryon drag scale, which allows a larger value of $H_0$ without damaging the fit to the BAO scale \cite{Aylor:2018drw} and without introducing extra damping at large multipoles (see \autoref{fig:TT_examples} for an illustration of this latter effect). In the SI$\nu$ model, this effect is compounded by the larger value of $\theta_*$ necessary to compensate for the absence of the free-streaming neutrino phase shift. This further slightly increases the value of $H_0$ necessary to fit the data, as can be seen by comparing the two modes in the left-most panel of \autoref{fig:H0_contours}.

It is worth noting that when $N_{\mathrm{eff}}$ is fixed at $3.046$, $H_0$ and $\sum m_\nu$ are usually negatively correlated (see e.g.~Ref.~\cite{planck2015}). If both $N_{\mathrm{eff}}$ and $\sum m_\nu$ are allowed to vary, there is not a strong correlation between $N_{\mathrm{eff}}$ and $\sum m_\nu$ for CMB data alone. However, when $H_0$ or BAO data are added, $N_{\mathrm{eff}}$ and $\sum m_\nu$ become positively correlated \citep{hou2014} as shown in \autoref{fig:bimodal_contours}. The tight correlation between $N_{\mathrm{eff}}$ and $H_0$ then permits a positive correlation between $H_0$ and $\sum m_\nu$, seen in the third panel from the left of \autoref{fig:H0_contours}. Thus, instead of larger $\sum m_\nu$ being correlated with a smaller Hubble constant, here a larger sum of neutrino masses corresponds to a slightly larger $H_0$. Allowing the neutrinos to self-interact does not dramatically change the direction of this degeneracy, but does allow it to stretch to larger $H_0$ values. 

$G_\mathrm{eff}$'s direct effects on matter clustering are scale dependent. As discussed in \autoref{sec:matter}, dark matter fluctuations that enter the horizon during neutrino decoupling receive a boost, while fluctuations that enter the horizon before neutrino decoupling are damped. For the $\mathrm{SI}\nu$ mode, neutrino decoupling is coincident with the modes entering the horizon that contribute most to $\sigma_8$, giving them a gravitational boost. However, low values of $A_{\rm s}$ and $n_{\rm s}$ must accompany a large $G_\mathrm{eff}$ for the $\mathrm{SI}\nu$, as discussed in \autoref{sec:SInu}, which consequently damp these same scales. The combination of $G_\mathrm{eff}$'s effects thus leads to an overall decrease in matter clustering at scales probed by $\sigma_8$, seen in the $\mathrm{SI}\nu$ island in the left panel of \autoref{fig:sigma8_contours}. 

The sum of neutrino masses is negatively correlated with $\sigma_8$ since massive neutrinos do not contribute to matter clustering for small scales where their pressure term is large (see third panel in \autoref{fig:sigma8_contours}). Typically large $N_\mathrm{eff}$ boosts the dark matter fluctuations upon horizon entry, leading to a positive correlation between $N_\mathrm{eff}$ and $\sigma_8$. However, the positive correlation between $N_\mathrm{eff}$ and $\sum m_\nu$ when including BAO data causes $N_\mathrm{eff}$ and $\sigma_8$ to be negatively correlated (see second panel in \autoref{fig:sigma8_contours}). This allows the interacting neutrino model to both be compatible with a large value of the Hubble constant and not overpredict the amplitude of matter fluctuations at late times. 

\subsection{Impact of CMB polarization data} 
As can be seen in \autoref{fig:bimodal_contours} and \autoref{fig:controversy}, the addition of EE polarization data tends to significantly reduce the statistical significance of the SI$\nu$ cosmology. This is in contrast with Ref.~\cite{lancaster} which found that polarization data slightly increased the significance of the SI$\nu$ mode (see also Ref.~\cite{oldengott17}). The degradation of the fit for the SI$\nu$ model in our case is the result of (i) our use of the reionization optical depth prior from Ref.~\cite{Adam:2016hgk} whenever we use polarization data, and (ii) our use of BBN calculations to predict the helium abundance for a given value of $N_{\rm eff}$. As we discuss in \autoref{sec:conclusions}, it is likely that the fit could improve significantly by replacing this strong prior with the actual low-$\ell$ polarization data used to obtain it, and by letting the helium fraction float freely in the fit. As might be expected, the addition of the local Hubble constant measurement increases the statistical significance of the SI$\nu$ mode, as can be seen from TT,TE,EE+lens+$H_0$ data set in \autoref{fig:bimodal_post} in Appendix \ref{app:all_res}.

\section{Statistical Significance}\label{sec:significance}
In this section, we quantify the relative statistical significance of the two modes of the posterior, and compare the maximum likelihood values between our interacting neutrino models and standard extensions of the $\Lambda$CDM paradigm.
\subsection{Mode Comparison}\label{sec:mode_significance}

To determine the statistical significance of the $\mathrm{SI}\nu$ mode relative to the $\mathrm{MI}\nu$ mode, we can compare their relative Bayesian evidence. It is defined as the parameter-averaged likelihood of the data
\begin{align}
  \mathcal{Z} \equiv \mathrm{Pr}\left(\mathbf{d}|\mathcal{M}\right) = \int_{\Omega_{\theta}}\mathrm{Pr}\left(\mathbf{d}|\mathbf{\theta},\mathcal{M}\right)\mathrm{Pr}\left(\mathbf{\theta}|\mathcal{M}\right)d\mathbf{\theta},
\end{align}
where $\textbf{d}$ is the data, $\mathcal{M}$ is the cosmological model, $\mathbf{\theta}$ are the parameters in model $\mathcal{M}$, and $\Omega_{\theta}$ is the domain of the model parameters. We use {\tt Multinest}'s \citep{feroz13} mode separation algorithm to compute the Bayesian evidence for each mode. In practice, this mode separation occurs near a neutrino coupling value of $\log_{10}\left(G_{\mathrm{eff}}\,\mathrm{MeV}^{2}\right) \approx -2.2$. This separation in parameter space defines $\Omega_{\theta}$ for each mode.

To compare the SI$\nu$ to the MI$\nu$ mode, we compute the following Bayes factor:
\begin{align}
  \mathcal{B}_{\mathrm{SI}\nu} \equiv \frac{\mathrm{Pr}\left(\mathcal{M}_{\mathrm{SI}\nu}|\mathbf{d}\right)}{\mathrm{Pr}\left(\mathcal{M}_{\mathrm{MI}\nu}|\mathbf{d}\right)} = \frac{\mathcal{Z}_{\mathrm{SI}\nu}}{\mathcal{Z}_{\mathrm{MI}\nu}}\frac{\mathrm{Pr}\left(\mathcal{M}_{\mathrm{SI}\nu}\right)}{\mathrm{Pr}\left(\mathcal{M}_{\mathrm{MI}\nu}\right)}.
\end{align}
We place a uniform prior on $\mathrm{log}_{10}\left(G_{\mathrm{eff}}\,\mathrm{MeV}^{2}\right)$ rather than a uniform prior on $G_{\mathrm{eff}}$ to avoid introducing a preferred energy scale. With our choice of prior, small values of $G_{\mathrm{eff}}$ can be thoroughly explored, which is particularly important since the actual Fermi constant governing neutrino interaction in the Standard Model takes the value $G_\mathrm{F} \sim \mathcal{O}\left(10^{-11}\,\mathrm{MeV}^{-2}\right)$. Taking a uniform prior on $G_{\mathrm{eff}}$ would greatly increase the statistical significance of the interacting mode (see Ref.~\cite{Cyr-Racine:2013aa}). We thus consider it conservative to adopt a uniform prior on $\mathrm{log}_{10}\left(G_{\mathrm{eff}}\,\mathrm{MeV}^{2}\right)$, but note that the statistical significance of the SI$\nu$ mode could be greatly enhanced by a different choice of prior. 

The probability of the prior is equivalent for each mode (or cosmological model), so $\mathrm{Pr}\left(\mathcal{M}_{\mathrm{SI}\nu}\right) / \mathrm{Pr}\left(\mathcal{M}_{\mathrm{MI}\nu}\right)=1$. In \autoref{tab:ratios} we show the Bayes factor for each data set combination we consider in this work.  A Bayes factor less than unity indicates the data prefer the $\mathrm{MI}\nu$ mode for the specified parameter space. All values are below unity, indicating the data, on average, do not prefer the $\mathrm{SI}\nu$ mode. As expected though, incorporating the local Hubble rate measurement does increase the significance of the $\mathrm{SI}\nu$ mode.

A useful method to understand if the $\mathrm{SI}\nu$ mode is ever preferred and to further investigate the significance's dependence on LSS data is to compare the maximum-likelihood value of each model:
\begin{align}
  \mathcal{R}_{\mathrm{SI}\nu} = \frac{\mathrm{max}\left[\mathcal{L}\left(\theta_{\mathrm{SI}\nu}|\mathbf{d}\right)\right]}{\mathrm{max}\left[\mathcal{L}\left(\theta_{\mathrm{MI}\nu}|\mathbf{d}\right)\right]}.
\end{align}
In \autoref{tab:ratios} we show the maximum-likelihood ratios for the data set combinations in our analysis. Again, adding $H_0$ and CMB lensing data increases the likelihood of the strongly interacting mode. Intriguingly, the $\mathrm{SI}\nu$ mode has a larger maximum-likelihood (by a factor larger than 2) than the $\mathrm{MI}\nu$ mode for TT+lens+BAO+$H_0$ (see the unsmoothed posteriors in \autoref{fig:bimodal_post}). It is reasonable that the Bayes factor for TT+lens+BAO+$H_0$ is below unity while the maximum-likelihood ratio is above unity since the former is a global, parameter-averaged statistic while the latter is based on a single set of best-case scenario parameters. This indicates that the parameter space for which strong neutrino interactions are preferred has a small volume.

It is also informative to look at the individual $\chi^2$ values for the different data sets. To compare the two modes, we list the $\Delta \chi^2 = \chi^2_{\mathrm{SI}\nu} - \chi^2_{\mathrm{MI}\nu}$ values in \autoref{tab:ratios}. A positive $\Delta \chi^2$ value thus means that the MI$\nu$ mode is preferred, and vice versa. The $H_0$ and high-$\ell$ TT,TE,EE data show preference for the $\mathrm{SI}\nu$ mode for TT,TE,EE+lens+$H_0$, but this is compensated by a poorer fit to low-$\ell$ and CMB lensing data. For the TT+lens+BAO data combinations, the BAO and high-$\ell$ TT data display a slight preference for the $\mathrm{SI}\nu$ mode, which is again overshadowed by the low-$\ell$ data. We see that the slight preference for the SI$\nu$ mode with the TT+lens+BAO+$H_0$ data combination is largely  due to improvement of the BAO and high-$\ell$ likelihoods. 

\begin{table*}{}
  \caption{Mode Comparison. Here, $\mathcal{B}_{\mathrm{SI}\nu}$ is the Bayes factor between the SI$\nu$ and the MI$\nu$ mode, $\mathcal{R}_{\mathrm{SI}\nu}$ is theie maximum likelihood ratio, and $\Delta \chi^2 = \chi^2_{\mathrm{SI}\nu} - \chi^2_{\mathrm{MI}\nu}$. The low-$\ell$ dataset refers to low-$\ell$ TEB if polarization was included and low-$\ell$ TT if only temperature was used. Similarly, the high-$\ell$ dataset refers to high-$\ell$ TT,TE,EE if polarization was included and high-$\ell$ TT if only temperature was used.  \label{tab:ratios}}
  \begin{ruledtabular}
  \begin{tabular}{ccccc}
  Parameter & TT,TE,EE & TT,TE,EE + lens + $H_0$ & TT + lens + BAO & TT + lens + BAO + $H_0$ \\ 
  \hline \\ [-2ex]
  $\mathcal{B}_{\mathrm{SI}\nu}$ 	& $0.03 \pm 0.01$	& $0.10 \pm 0.04$	& $0.13 \pm 0.04$ & $0.37 \pm 0.10$ \\
  $\mathcal{R}_{\mathrm{SI}\nu}$ 	& $0.26$	& $0.63$	& $0.81$ & $2.14$ \\ [0.5ex]
  \hline \\[-2ex]
  $\Delta \chi^2_{\mathrm{low\,}\ell}$ 	& $2.47$	& $2.18$	& $2.00$ & $1.41$ \\
  $\Delta \chi^2_{\mathrm{high\,}\ell}$ 	& $0.22$	& $-0.16$	& $-1.53$ & $-2.23$ \\
  $\Delta \chi^2_\mathrm{lens}$ 	& --	& $1.34$	& $0.16$ & $0.30$ \\
  $\Delta \chi^2_{H_0}$ 	& --	& $-2.12$	& -- & $-0.56$ \\
  $\Delta \chi^2_\mathrm{BAO}$ 	& --	& --	& $-0.20$ & $-0.44$ \\
  $\Delta \chi^2_\mathrm{Total}$ 	& $2.69$	& $0.92$	& $0.43$ & $-1.52$ 
  \end{tabular}
\end{ruledtabular}
\end{table*}

\subsection{Comparison to \texorpdfstring{$\Lambda\mathrm{CDM}$}{LambdaCDM} and its extensions}
\label{sec:LCDM_significance}

Comparing how well each mode fits the data relative to $\Lambda\mathrm{CDM}$ and its common extensions tells us if these neutrino self-interaction models offer a viable improvement to current cosmological theory. For the TT+lens+BAO+$H_0$ data set, we list the $\Delta \chi^2 = \chi^2_{\mathrm{SI}\nu} - \chi^2_{\Lambda \mathrm{CDM+ext}}$ values and the $\Delta \chi^2 = \chi^2_{\mathrm{MI}\nu} - \chi^2_{\Lambda \mathrm{CDM+ext}}$ values for each observable in \autoref{tab:chi2lcdm}. Here, $\Lambda \mathrm{CDM+ext}$ refers to the $N_{\rm eff} + \sum m_\nu$ two-parameter extension of the $\Lambda \mathrm{CDM}$ cosmology. Comparison to plain $\Lambda$CDM was given in \autoref{tab:chi2lcdm_plain} above. For all data sets except the low-$\ell$ TT data, both modes offer a better fit to the data than $\Lambda\mathrm{CDM}+\mathrm{ext}$. In fact, the $\mathrm{SI}\nu$ mode has a total $\Delta \chi^2$ of $-3.33$, a significant difference. The improvement of the high-$\ell$ CMB data is notable since jointly fitting CMB and local $H_0$ data usually results in a worse fit to the CMB damping tail. For the SI$\nu$ model, this is somewhat compensated by a degradation of the low-$\ell$ fit.

What if the strong improvement in fit over $\Lambda \mathrm{CDM}$ is due to overfitting from the extra parameter we have added? To take this into account we compute the Akaike information criterion (AIC) \citep{AIC}. The AIC takes into account how well the model fits the data and penalizes extra parameters, thereby discouraging overfitting. The AIC is defined as
\begin{align}
\mathrm{AIC} = -2 \ln \left(\mathcal{L}\right) + 2k = \chi_{\mathrm{Total}}^2 + 2 k,
\end{align}
where $\chi^2_{\mathrm{Total}} = \chi^2_{\mathrm{low\,}\ell} + \chi^2_{\mathrm{high\,}\ell} + \chi^2_{\mathrm{lens}} + \chi^2_{H_0} + \chi^2_{\mathrm{BAO}}$, $\mathcal{L}$ is the maximum-likelihood, and $k$ is the number of fit parameters. Then we can write 
\begin{align}
\Delta \mathrm{AIC} = \mathrm{AIC}_{\mathrm{I}\nu} - \mathrm{AIC}_{\Lambda\mathrm{CDM}} = \Delta \chi^2 + 2\Delta k,
\end{align}
where $\Delta k$ is the difference in the number of parameters between the two models. The lower AIC between two models corresponds to the preferred model. Thus, for us, a negative $\Delta \mathrm{AIC}$ value indicates the data prefer the specified $\mathrm{I}\nu$ model over $\Lambda\mathrm{CDM}$, while a positive $\Delta \mathrm{AIC}$ value indicates the data prefer $\Lambda\mathrm{CDM}$ over the $\mathrm{I}\nu$ model.

We list the $\Delta \mathrm{AIC}$ values relative to $\Lambda \mathrm{CDM} + N_{\rm eff} + \sum m_\nu$ in \autoref{tab:chi2lcdm}. Here $\Delta k =1$, and the SI$\nu$ mode has a negative $\Delta \mathrm{AIC} = -1.33$, indicating a genuine statistical preference for the suppression of neutrino free-streaming in the early Universe for the TT+lens+BAO+$H_0$ data set. On the other hand, $\Delta \mathrm{AIC} = 0.19$ for the $\mathrm{MI}\nu$ mode, indicating that the neutrino self-interactions do not add value to the fit beyond what is already provided by the $N_{\rm eff} + \sum m_\nu$ extension. Values of $\Delta \mathrm{AIC}$ between the I$\nu$ models and standard $\Lambda$CDM ($\Delta k =3$) are also given in \autoref{tab:chi2lcdm_plain}. The fact that $\Delta \mathrm{AIC}$ values for the SI$\nu$ cosmology are similar ($-1.91$ versus $-1.33$) when comparing it to plain $\Lambda$CDM and $\Lambda$CDM $+ N_{\rm eff} + \sum m_\nu$ means that \emph{suppressing neutrino free-streaming is the true driving factor behind the improvement of the fit}. Thus, even after penalizing the self-interacting neutrino models for incorporating additional parameters, the TT+lens+BAO+$H_0$ data still significantly prefer the strongly interacting neutrino cosmology over $\Lambda \mathrm{CDM}$. 

\begin{table*}{}
  \caption{Comparison of the interacting neutrino cosmology to $\Lambda\mathrm{CDM} + N_\mathrm{eff} + \sum m_\nu$ for TT + lens + BAO + $H_0$ \label{tab:chi2lcdm}}
  \begin{ruledtabular}
  \begin{tabular}{ccccc}
  Parameter & Strongly Interacting Neutrino Mode & Moderately Interacting Neutrino Mode \\
  \hline \\ [-2ex]
  $\Delta \chi^2_{\mathrm{low\,}\ell}$ 	& $2.40$	& $0.99$ \\
  $\Delta \chi^2_{\mathrm{high\,}\ell}$ 	& $-3.40$	& $-1.17$ \\
  $\Delta \chi^2_\mathrm{lens}$ 	& $-0.20$	& $-0.50$ \\
  $\Delta \chi^2_{H_0}$ 	& $-1.32$	& $-0.76$ \\
  $\Delta \chi^2_\mathrm{BAO}$ 	& $-0.81$	& $-0.36$ \\[0.5ex]
  \hline \\[-2ex]
  $\Delta \chi^2_\mathrm{Total}$ 	& $-3.33$	& $-1.81$ \\
  $\Delta \mathrm{AIC}$ 	& $-1.33$	& $0.19$ 
  \end{tabular}
\end{ruledtabular}
\end{table*}

\section{Discussion}\label{sec:discuss}

\begin{figure*}
\centering
\includegraphics[width=0.49\linewidth]{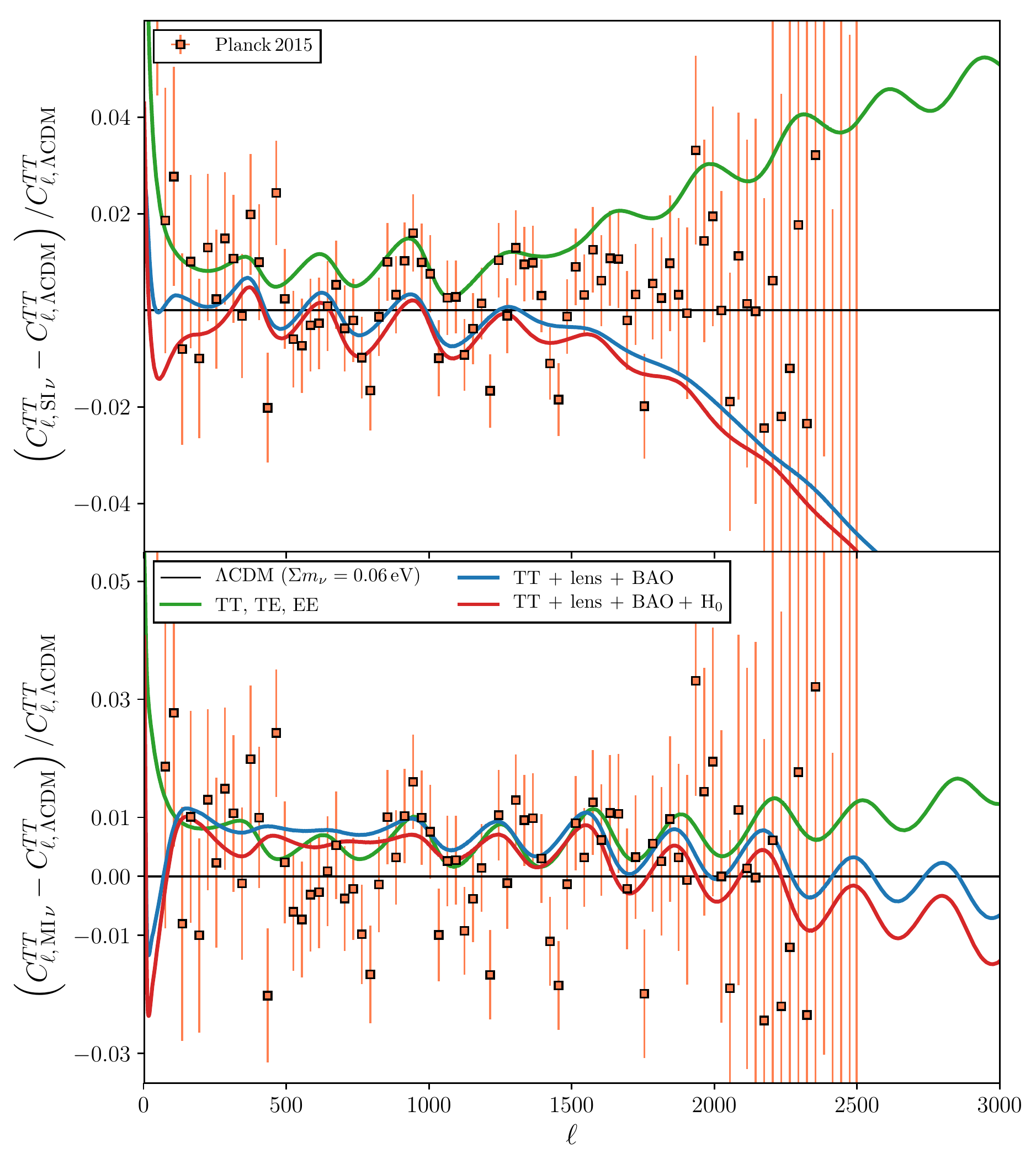}
\includegraphics[width=0.49\linewidth]{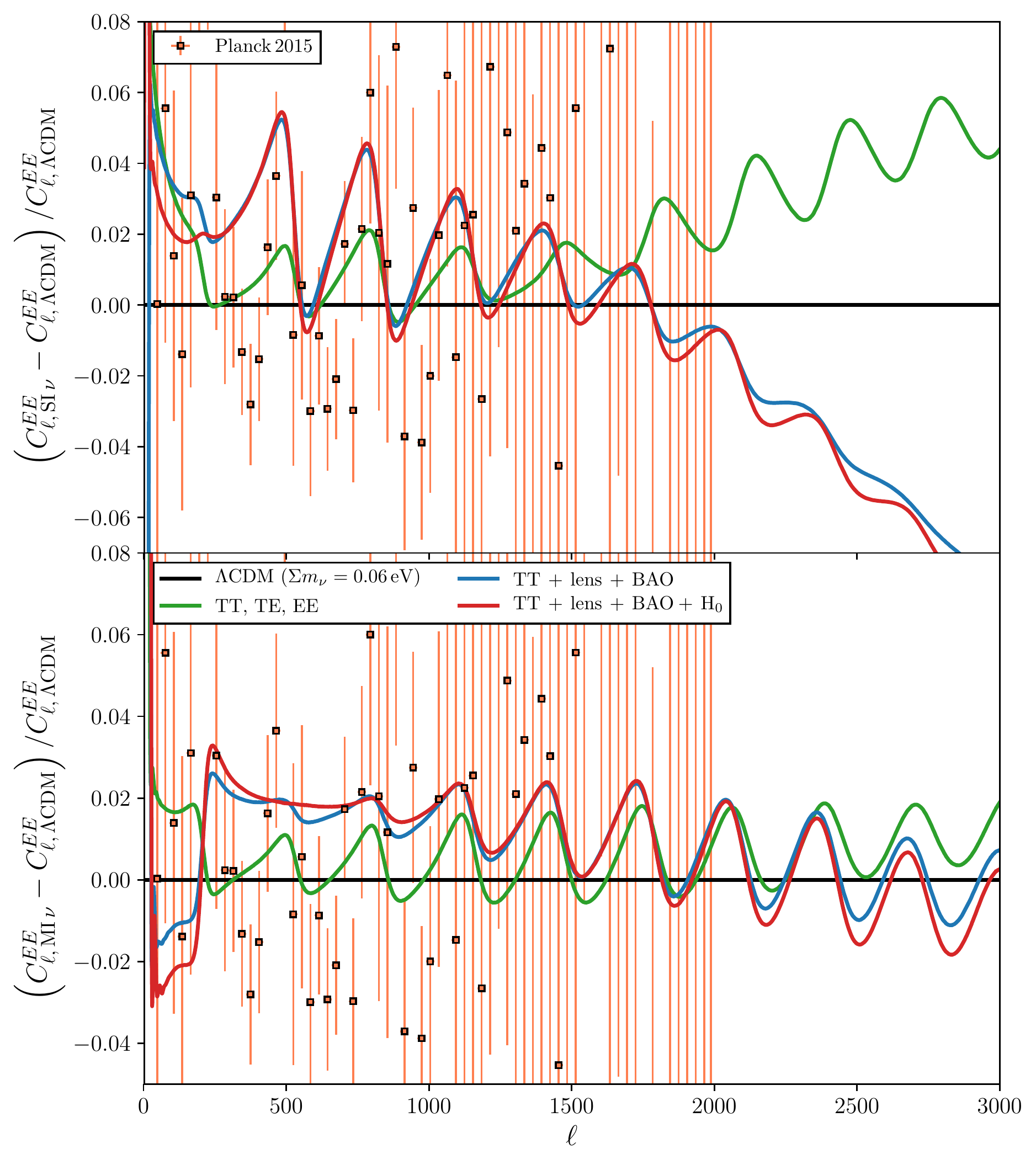}
\caption{Relative difference between the $\mathrm{SI}\nu$ mode (upper panels) or $\mathrm{MI}\nu$ mode (lower panels) and $\Lambda\mathrm{CDM}$ for the high-$\ell$ TT (left) and EE (right) power spectra. The $\mathrm{SI}\nu$ mode and $\mathrm{MI}\nu$ mode spectra are produced using the maximum likelihood parameter values for each respective mode. Colors denote the data set combination used. Measurements from the Planck 2015 data release are included \citep{planckCMB}.}
\label{fig:TT_max-likes}
\end{figure*}

\begin{figure}
\centering
\includegraphics[width=\linewidth]{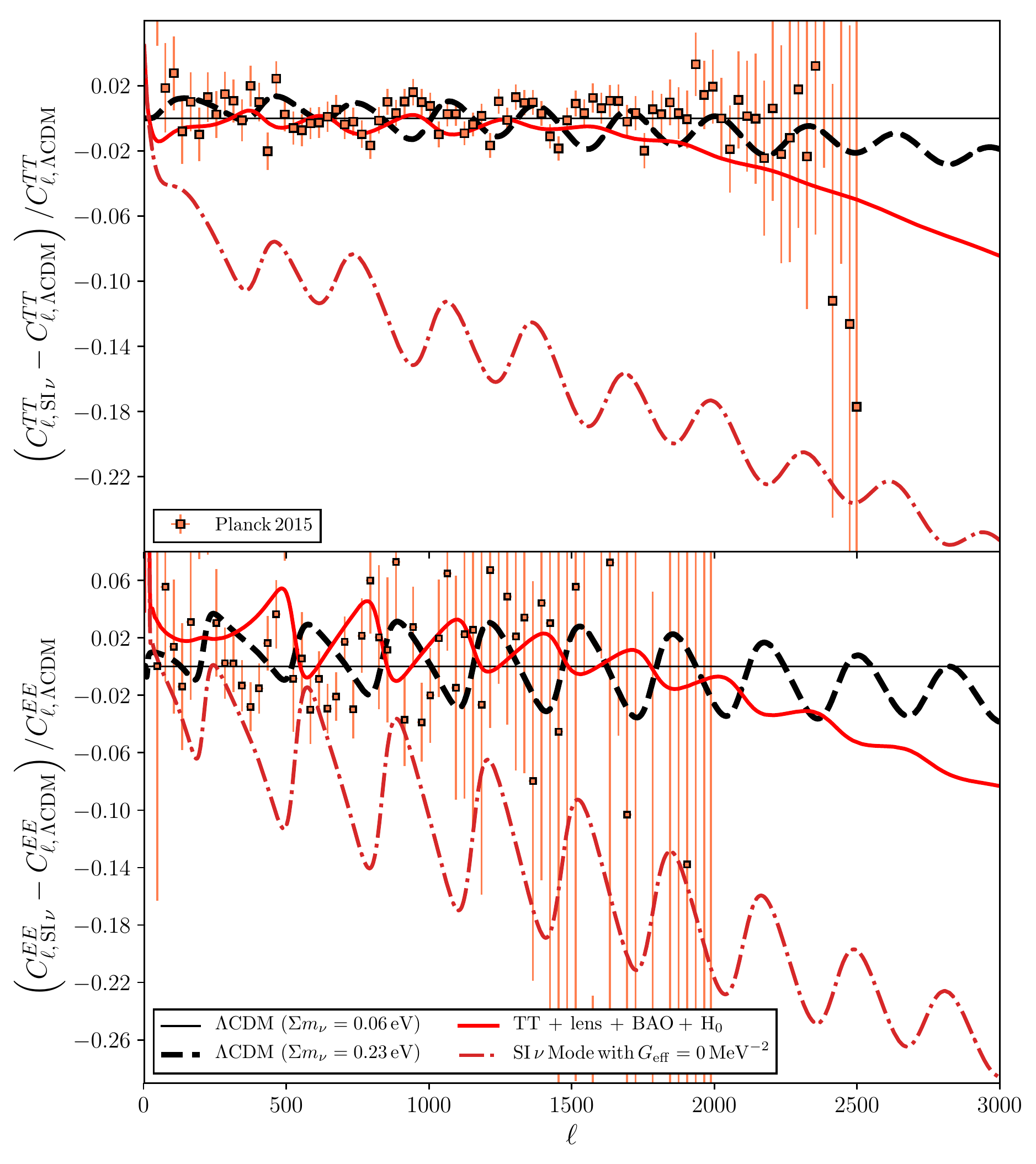}
\caption{Illustration of the importance of the neutrino self-interaction to the fit to CMB data for the SI$\nu$ cosmology. The red solid spectra corresponds to the best-fit $\mathrm{SI}\nu$ model, while the red dashed-dot spectra use the same best-fit cosmological parameters but allows neutrino free-streaming by setting $G_\mathrm{eff}=0$. }
\label{fig:intGeff0}
\end{figure}

\begin{figure*}
\centering
\begin{subfigure}[t]{0.445\textwidth}
\centering
\includegraphics[width=\linewidth]{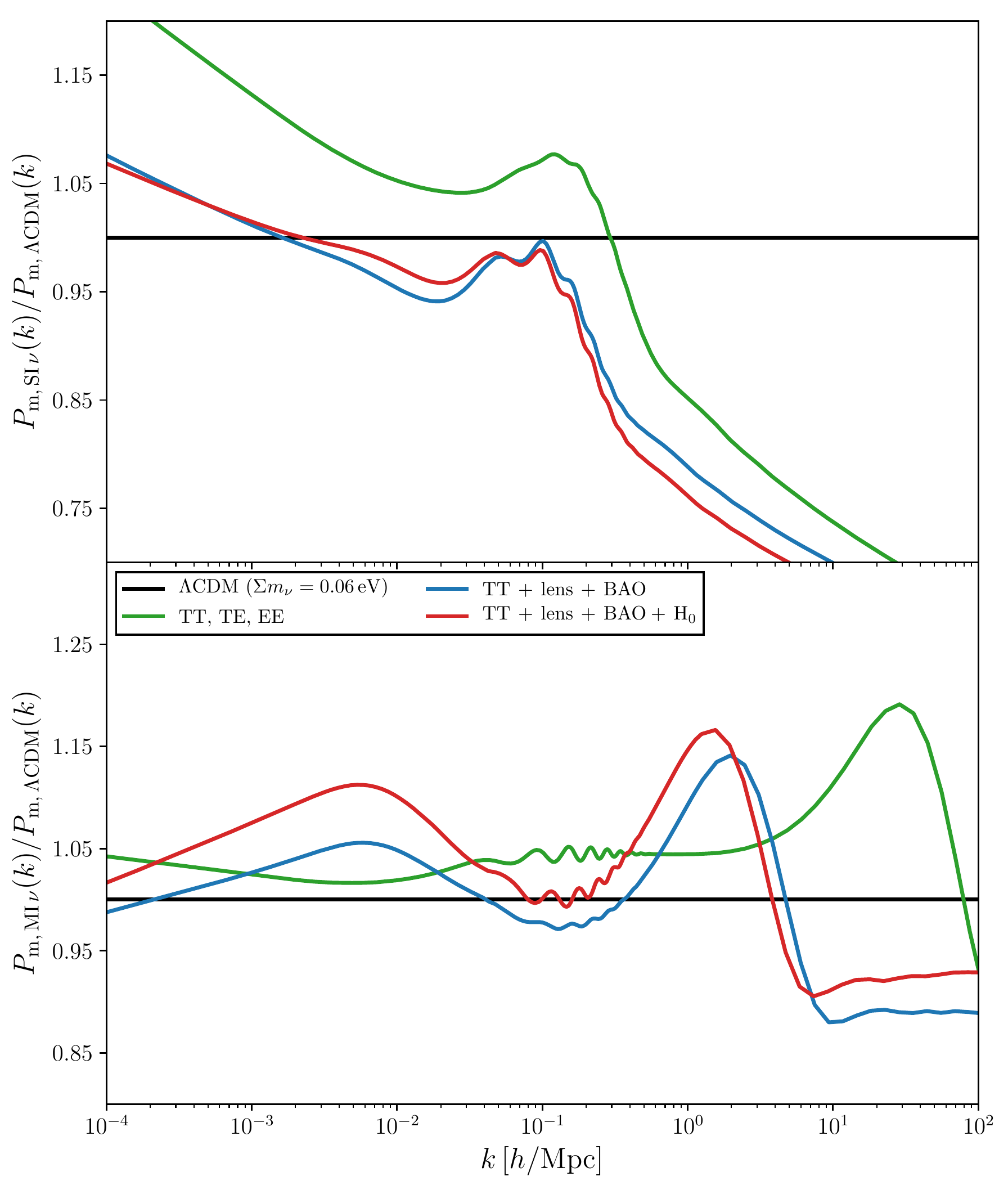}
\caption{Linear matter power spectrum}
\label{fig:matter_max-likes}
\end{subfigure}
\begin{subfigure}[t]{0.455\textwidth}
\centering
\includegraphics[width=\linewidth]{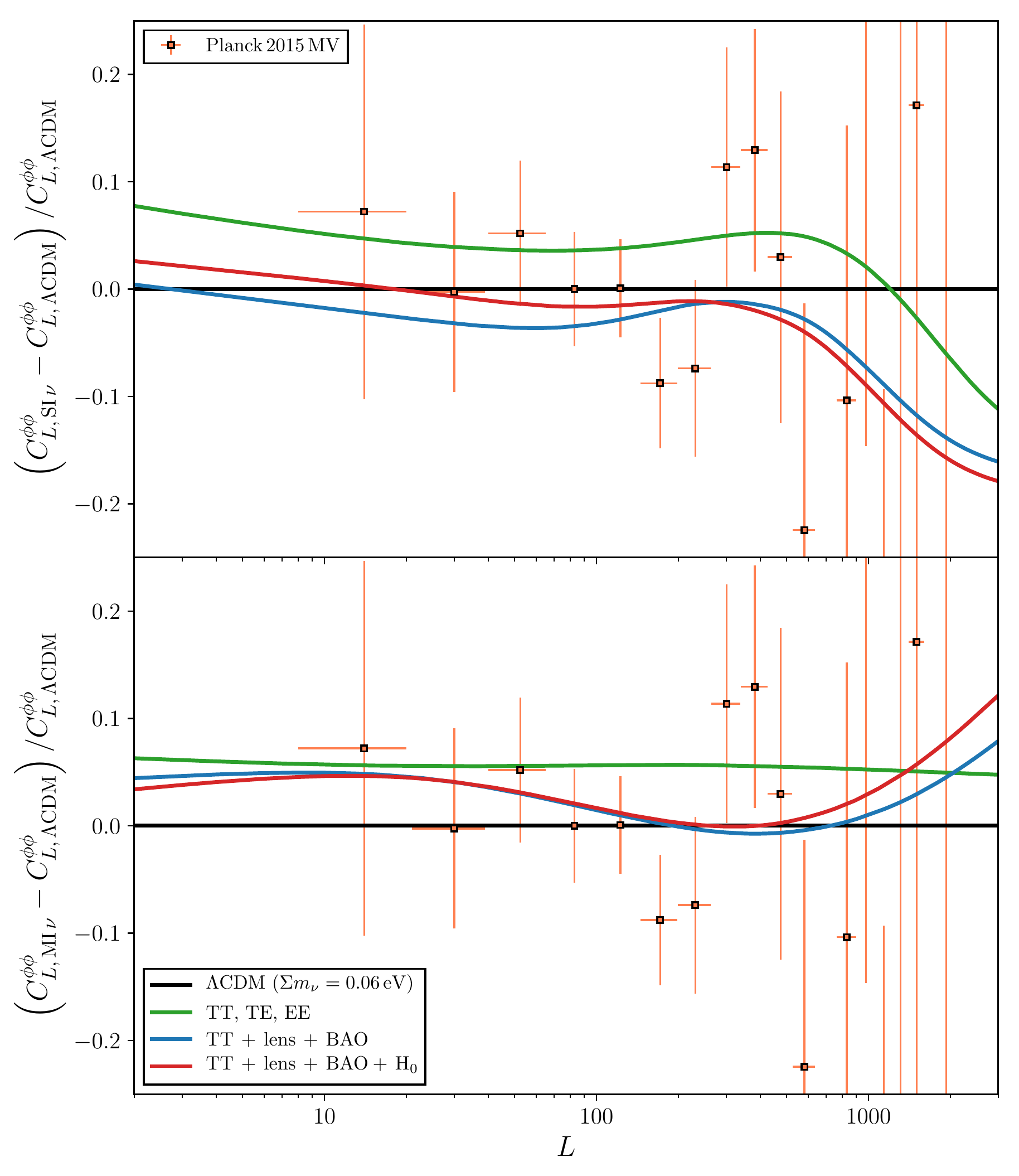}
\caption{CMB Lensing power spectrum}
\label{fig:lens_max-likes}
\end{subfigure}
\caption{ Relative difference between the $\mathrm{SI}\nu$ mode (upper panels) or $\mathrm{MI}\nu$ mode (lower panels) and $\Lambda\mathrm{CDM}$ for the linear matter power spectrum (left) and the CMB lensing power spectrum (right). The $\mathrm{SI}\nu$ and $\mathrm{MI}\nu$ spectra are produced using the maximum likelihood parameter values for each respective mode. Measurements from the Planck 2015 data release \citep{planckCMB} are included in the right panel.}
\end{figure*}

\subsection{Cosmic microwave background}

In \autoref{fig:TT_max-likes}, we plot the high-$\ell$ TT and EE power spectra residuals between the maximum likelihood parameters for each data set combination used and the best-fit Planck $\Lambda$CDM model. For the $\mathrm{SI}\nu$ mode (upper panels), the most striking feature of the residuals is the deficit of power at high multipoles ($\ell>1500$) as compared to $\Lambda$CDM for the TT+lens+BAO and TT+lens+BAO+$H_0$ data combinations. This is caused by the large value of $N_{\rm eff}$ and the resulting high helium abundance\footnote{We remind the reader that we use the standard BBN predictions to compute the helium abundance for given $N_{\rm eff}$ and $\Omega_{\rm b}h^2$ values.} $Y_P$ for this category of models. This implies that the multi-parameter degeneracy that allows the SI$\nu$ cosmology to provide a decent fit to CMB temperature data at $\ell < 1500$ could be broken by the addition of high-resolution CMB data (see e.g.~Refs.~\cite{Louis:2016ahn,Henning:2017nuy}). However, it is reasonable to assume that the BBN helium abundance is modified in the presence of the new neutrino physics we explore here, and that the deficit of power at large multipoles could be compensated by a smaller value of $Y_P$ \cite{Hou:2011ec}. We leave the study of the impact of a free helium fraction on interacting neutrino cosmologies to future works. 

The EE polarization residuals shown in the right panel of \autoref{fig:TT_max-likes} for the TT+lens+BAO and TT+lens+BAO+$H_0$ data combinations also display strong oscillations for the SI$\nu$ mode (upper panel). This implies that the shift in $\theta_*$ (and other parameters, see \autoref{sec:SInu}) that was required to compensate for the absence of the free-streaming neutrino phase shift in the temperature spectrum does not fully realign the peaks of the polarization spectrum with the data. This is a consequence of the polarization data being more sensitive to the phase of the acoustic peaks \cite{Baumann15}. With the current size of the Planck error bars, this does not constitute an overwhelmingly strong constraint on the absence of a neutrino-induced phase shift, but it is possible that future CMB polarization data could entirely rule out this possibility. 

The TT+TE+EE CMB-only data combination in the upper panels of \autoref{fig:TT_max-likes} display an excess of power as compared to $\Lambda$CDM at nearly all scales, resulting in an overall poorer fit to the CMB data. At large multipoles, this is of course in contrast with the deficit of power that the TT+lens+BAO and TT+lens+BAO+$H_0$ fits display. Our use of the polarization-driven prior on the reionization optical depth from Ref.~\cite{Adam:2016hgk} is largely responsible for this excess of power as compared to $\Lambda$CDM for the SI$\nu$ mode with the TT+TE+EE data set. Again, this shows that polarization data could in principle break the multi-parameter degeneracy that allows the SI$\nu$ cosmology to exist. 

All dataset combinations we consider display an excess of power at $\ell<50$ for the SI$\nu$ mode. This is largely caused by the lower value of the scalar spectral index $n_{\rm s}$ which adds power on large scales. While error bars are large in this regime due to cosmic variance, the dip in power around $\ell\sim 20$-$30$ in the CMB temperature data tends to penalize any model displaying more low-$\ell$ power than $\Lambda$CDM. If this dip were to be explained by some other physics (from the inflationary epoch, say), then it is possible that the fit to the data using the SI$\nu$ cosmology could significantly improve. 

It is important to emphasize how the suppression of neutrino free-streaming plays a very important role in the existence of the SI$\nu$ cosmology. To illustrate this, we plot in \autoref{fig:intGeff0} the spectra corresponding to the best-fit $\mathrm{SI}\nu$ parameters for the TT+lens+BAO+$H_0$ data set but allow neutrino to free-stream at all times by setting $G_{\mathrm{eff}}=0$ (red dashed-dot line), along with the original TT+lens+BAO+$H_0$ best-fit $\mathrm{SI}\nu$ model (solid red) and a $\Lambda \mathrm{CDM}$ model with $\sum m_\nu=0.23\,\mathrm{eV}$ (black dashed), for reference. Here, the difference between the dashed-dot and solid red lines is entirely driven by the streaming property of neutrinos. Figure \ref{fig:intGeff0} reinforces our discussion from \autoref{sec:LCDM_significance} that the $G_{\mathrm{eff}}$ parameter plays a statistically significant role in improving the fit to the CMB data, beyond what is already provided by the $N_{\rm eff} +\sum m_\nu$ two-parameter extension of $\Lambda$CDM.

\subsection{Matter clustering}

We show in \autoref{fig:matter_max-likes} the linear matter power spectrum residuals between the best-fit SI$\nu$ (and MI$\nu$) models and the corresponding $\Lambda$CDM models. The most striking feature for the SI$\nu$ mode is the overall red tilt of the matter power spectrum residuals for all data combinations shown. This tilt is due to the low preferred value of $n_\mathrm{s}$ for this mode. Despite this global shape difference with $\Lambda$CDM, the enhancement of matter fluctuations on scales entering the horizon at the onset of neutrino free-streaming discussed in \autoref{sec:matter} causes the matter power spectrum to only slightly deviate from the CDM prediction on scales contributing the most to $\sigma_8$ ($0.02h\,\mathrm{Mpc}^{-1} \lesssim k\lesssim 0.2h\,\mathrm{Mpc}^{-1}$). For the TT+lens+BAO and TT+lens+BAO+$H_0$ data combinations, this difference is less than $5\%$ on these scales and results in a $\sigma_8$ value that is slightly lower than in $\Lambda$CDM, potentially bringing low-redshift measurements of the amplitude of matter fluctuations in agreement with CMB data as discussed in \autoref{sec:H0_tension}.

Nevertheless, it is important to note that the localized feature in the matter power spectrum caused by the late onset of neutrino free-streaming nearly coincides with the BAO scale, that is, it is on scales where we have a large amount of data from, for example, spectroscopic galaxy surveys (see e.g.~Ref.~\cite{BAO2}). While an analysis that takes into account the full shape of the measured galaxy power spectrum at these scales is beyond the scope of this work, we note that both the SI$\nu$ and MI$\nu$ cosmologies only mildly deviate from the $\Lambda$CDM model near the BAO scale. On smaller scales, the $\mathrm{SI}\nu$ mode displays a net suppression of power which has implications for probes of small-scale structure such as the Lyman-$\alpha$ forest \cite{2017PhRvD..96b3522I} and the satellite galaxy count surrounding the Milky Way \cite{Kim:2017iwr}. It is an interesting possibility that the SI$\nu$ cosmology could help alleviate the small-scale structure problems \cite{Bullock:2017xww} without introducing a nongravitational coupling between neutrinos and dark matter. 

The MI$\nu$ residuals (lower panel) in \autoref{fig:matter_max-likes} display an even richer structure than those shown in \autoref{fig:matter_examples}. Indeed, even in the case of relatively weak neutrino interactions, their impact on the matter power spectrum is significant, and potentially provide a different channel to constrain new physics in the neutrino sector. Since the dominant constraining power of the data used here comes from $k\sim0.1\,h\,\mathrm{Mpc}^{-1}$, we observe that the $\mathrm{MI}\nu$ power spectra have values similar to $\Lambda\mathrm{CDM}$ near this scale. Outside the scales probed by $\sigma_8$, the linear matter power spectra deviate more significantly (up to $\sim20\%$) from $\Lambda$CDM.

The lensing potential power spectrum in \autoref{fig:lens_max-likes} shows a similar pattern to the matter power spectrum for the different best-fit models, as expected. The current large error bars of the Planck lensing measurements allow substantial freedom to the SI$\nu$ and MI$\nu$ cosmologies. As shown in \autoref{tab:ratios}, the lensing data prefer the MI$\nu$ mode for all data-set combinations, but we note that the $\mathrm{SI}\nu$ modes are typically within the error bars of the lensing data.

\section{Conclusions}\label{sec:conclusions}

The presence of yet-unknown neutrino interactions taking place in the early Universe could delay the onset of neutrino free-streaming, imprinting the CMB and probes of matter clustering with distinct features. We have performed a detailed study of the impact of neutrino self interactions with a rate scaling as $\Gamma_\nu\sim G_{\rm eff}^2 T_\nu^5$ on the CMB and the matter power spectrum, taking into account the presence of nonvanishing neutrino masses and of a nonstandard neutrino thermal history. Using recent measurements of the BAO scale, the local Hubble rate, and of the CMB, we find that a cosmological scenario (originally pointed out in Ref.~\cite{Cyr-Racine:2013aa}) in which the onset of neutrino free-streaming is delayed until close to the epoch of matter-radiation equality can provide a good fit to CMB temperature data while also being consistent with the Hubble constant inferred from the local distance ladder \cite{HST}.  

This strongly interacting neutrino cosmology has the following properties:
\begin{itemize}
  \item Using the data combination TT+lens+BAO+$H_0$, it displays a strong preference ($>3\sigma$) for an addition neutrino species ($N_{\mathrm{eff}}=4.02\pm0.29$, $68\%$ C.L.). This can have important implications given the current anomalies in neutrino oscillation experiments. It also prefers a nonvanishing value of the sum of neutrino masses $\sum m_\nu = 0.42^{+0.17}_{-0.20}$ eV ($68\%$ C.L.). 
  \item  It can easily accommodate a larger value of $H_0$ and smaller $\sigma_8$, hence possibly alleviating tensions between current measurements. Quantitatively, the data combination TT+lens+BAO+$H_0$ favors $H_0=72.3\pm1.4\,\mathrm{km\,s}^{-1}\,\mathrm{Mpc}^{-1}$ and $\sigma_8=0.786\pm0.020$ at $68\%$ C.L.
\end{itemize}

It is remarkable that a cosmological model admitting parameter values that are so different (see \autoref{fig:LCDM_interact_standard_contours}) than in the standard $\Lambda$CDM paradigm can provide a better fit to the data at a statistically-significant level ($\Delta$AIC$=-1.91$). We believe that this is the most important lesson to be drawn from our work: While most analyses have focused on mild deformation from the standard $\Lambda$CDM scenario in trying to reconcile the current cosmological datasets, it is important to entertain the possibility that a radically different scenario (i.e. statistically disjoint in cosmological parameter space) could provide a better global fit to the data.

Despite the success of the strongly interacting neutrino cosmology in addressing tensions between certain cosmological data sets, there are several important obstacles that still tilt the balance towards the standard $\Lambda$CDM cosmology. First, the addition of polarization data seems to degrade the quality of the fit for the strongly interacting neutrino cosmology. We have traced this deterioration of the fit to our use of a Gaussian prior on the reionization optical depth from Ref.~\cite{Adam:2016hgk}. This prior was utilized as a way to capture the constraint on the optical depth from low-$\ell$ HFI Planck polarization data before the full likelihood is made available. It it likely that the Gaussian form of the prior leads to constraints that are too strong as compared to what the full likelihood will provide. Only a complete analysis with the legacy Planck data, once available, will allow us to determine whether this is the case. An important fact to keep in mind is that Ref.~\cite{lancaster} found that the addition of CMB polarization data (without an additional $\tau$ prior) tends to increase the statistical significance of the strongly interacting neutrino cosmology. 

Second, the low values of the Bayes factor (see \autoref{tab:ratios}) consistently favor either very weakly interacting neutrinos or no interaction at all. This reflects the fact that strongly interacting neutrinos can only fit the data better for a narrow window of interaction strengths, while $\Lambda$CDM provides a decent (but overall less good) fit over a broader part of the parameter space. This is a fundamental feature of Bayesian statistics and it is unlikely to change in future analyses. This highlights the need to consider a portfolio of statistical measures to assess the quality of a given cosmological model. 

Third, it might be difficult from a particle model-building perspective to generate neutrino self interactions with the strength required by the strongly interacting neutrino cosmology while not running afoul of other constraints on neutrino physics. A viable model might look similar to that presented in Ref.~\cite{Cherry:2014xra}, but it remains to be seen whether the necessary large interaction strength can be generated while evading current constraints \cite{Lessa:2007up} on new scalar particles coupling to Standard Model neutrinos. It is also possible that a successful self-interacting neutrino model could have a different temperature dependence than that considered in this work ($\Gamma_\nu\propto T_\nu^5$). This would change the shape of the neutrino visibility function (see Refs.~\cite{Cyr-Racine:2013aa,lancaster}) and potentially improve the global fit to the data. We leave the study of different temperature scalings of the neutrino interacting rate to future works.

Our analysis could be improved in a few different ways. Given the computational resources we had at our disposal and the need to obtain accurate values of the Bayesian evidence, we used the ``lite'' version of the Planck high-$\ell$ likelihoods in our analysis. Since some of the assumptions that went into generating these likelihoods \cite{planck15nuisance} might not apply to the interacting neutrino cosmologies, it would be interesting, given sufficient computing power, to reanalyze these models with the complete version of the likelihoods that include all the nuisance parameters. In particular, it is possible that some of the foreground nuisance parameters might be degenerate with the effect of self-interacting neutrinos. For simplicity, we have also assumed that the helium fraction is determined by the standard big-bang nucleosynthesis calculation throughout our analysis. Given the new physics and the resulting modified thermal history of the neutrino sector for the type of models we explore here, it reasonable to assume that the helium fraction would in general be different than in $\Lambda$CDM. While the details of the helium production within any interacting neutrino model are likely model-dependent, a sensible way to take these effects into account would be to let the helium fraction float freely in the fit to CMB data. We leave such analysis to future works. 

Given the structure of the residuals between the best-fit interacting neutrino cosmologies and the $\Lambda$CDM model presented in \autoref{sec:discuss}, it is clear that future high-$\ell$ CMB polarization and matter clustering measurements will play an important role in constraining or ruling out these models \citep[see e.g.][]{minsu}. In particular, the overall red tilt of the matter power spectrum in the strongly interacting neutrino cosmology could have important consequences on both large and small scales. Since current anomalies in terrestrial neutrino experiments \cite{Aguilar-Arevalo:2018gpe,Aguilar:2001ty} may indicate the presence of new physics in the neutrino sector, it is especially timely to use the complementary nature of cosmological probes to look for possible clues about physics beyond the Standard Model.

\acknowledgments
We thank Kris Sigurdson and Roland de Putter for collaboration at early stages of this work. We also thank Jo Dunkley, David Spergel, and Lyman Page for comments on an early version of this manuscript, and Prateek Agrawal and David Pinner for useful conversations. C.~D.~K. acknowledges the support of the National Science Foundation award number DGE1656466 at Princeton University and of the Minority Undergraduate Research Fellowship at the Jet Propulsion Laboratory. F.-Y. C.-R.~acknowledges the support of the National Aeronautical and Space Administration (NASA) ATP grant NNX16AI12G at Harvard University. This work was performed in part at the California Institute of Technology for the Keck Institute for Space Studies, which is funded by the W. M. Keck Foundation. Part of the research described in this paper was carried out at the Jet Propulsion Laboratory, California Institute of Technology, under a contract with NASA.  The computations in this paper were run on the Odyssey cluster supported by the FAS Division of Science, Research Computing Group at Harvard University.
\\
\appendix
%
\section{Results for all data sets}\label{app:all_res}
\begin{figure*}
\centering
  \includegraphics[width=0.6\linewidth]{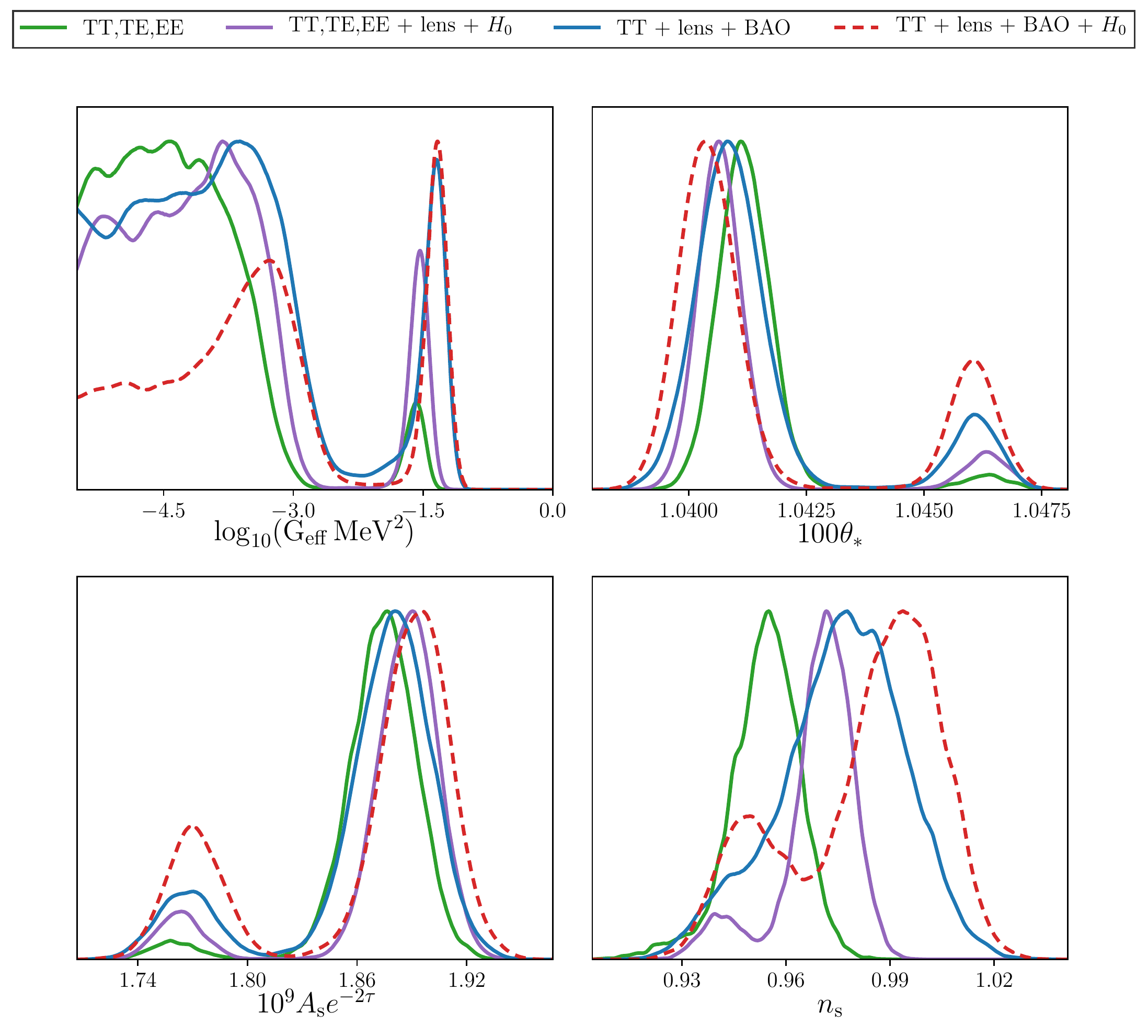}
  \caption{1D posteriors for bimodal parameters with low smoothing.}
  \label{fig:bimodal_post}
\end{figure*}

\begin{figure*}
\centering
  \includegraphics[width=0.6\linewidth]{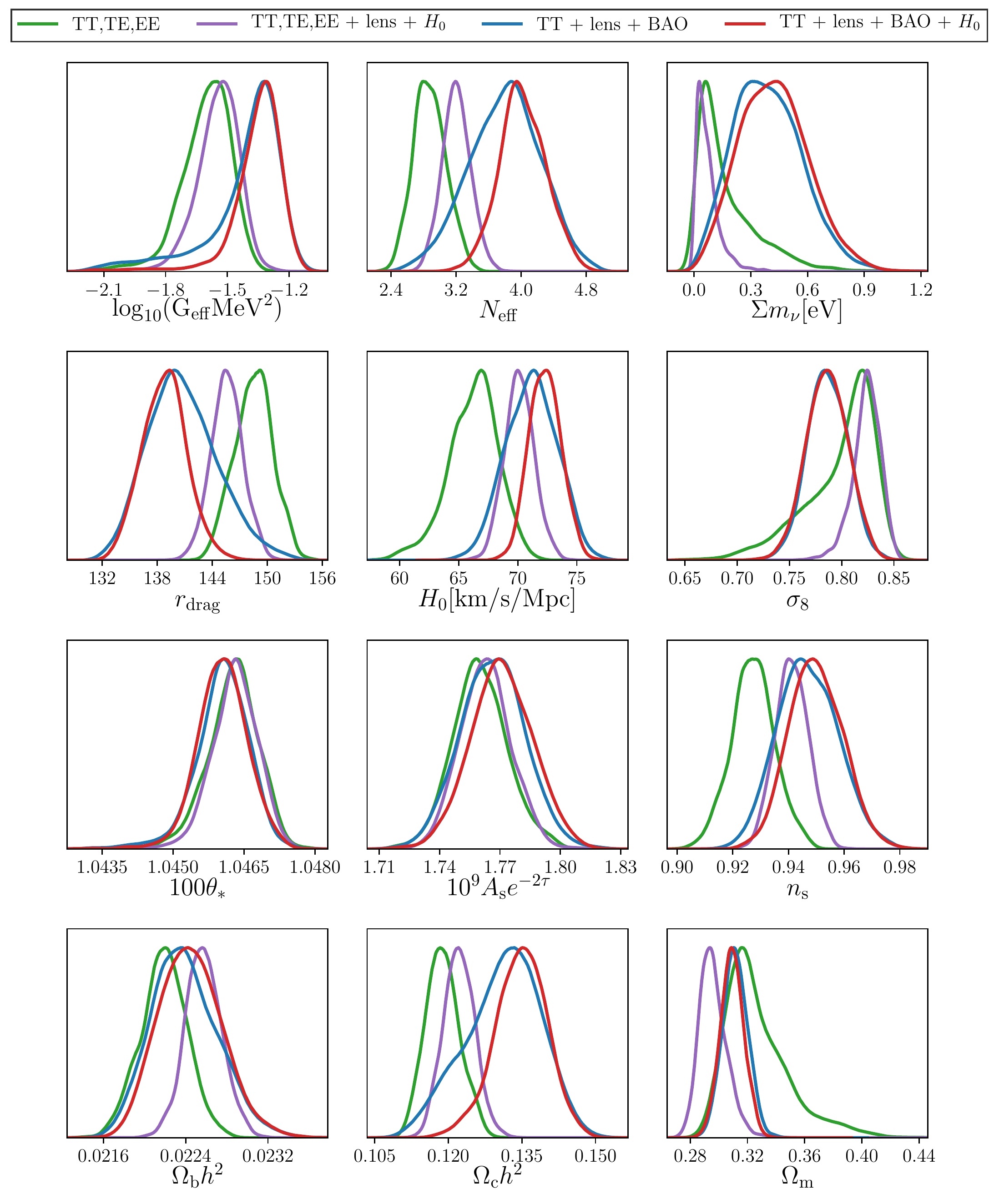}
  \caption{$\mathrm{SI}\nu$ mode posteriors for all data set combinations.}
  \label{fig:LCDM_interact_standard_contours_TTTEEE}
\end{figure*}

We display in \autoref{tab:SI_nu_params} and \autoref{tab:LCDM_nu_params} the $68\%$ confidence limits for the strongly interacting and moderately interacting neutrino modes, respectively. In \autoref{fig:bimodal_post}, we show the marginalized posteriors for key cosmological parameters for a choice of smoothing kernel that represents more accurately the shape of the SI$\nu$ mode. In \autoref{fig:LCDM_interact_standard_contours_TTTEEE}, we compare the marginalized posterior distribution of the SI$\nu$ mode for the four data set combinations considered in this work.

 \begin{table*}
  \caption{Strongly interacting neutrino cosmology: Parameter $68\%$ Confidence Limits \label{tab:SI_nu_params}}
  \begin{ruledtabular}
  \begin{tabular}{ccccc}
   Parameter & TT,TE,EE & TT,TE,EE + lens + $H_0$ & TT + lens + BAO & TT + lens + BAO + $H_0$ \\
  \hline
  {$\Omega_\mathrm{b} h^2   $} 	& $0.02219^{+0.00024}_{-0.00022}$	& $0.02257\pm 0.00018        $	& $0.02239^{+0.00029}_{-0.00036}$	& $0.02245^{+0.00029}_{-0.00033}$\\
  {$\Omega_\mathrm{c} h^2   $} 	& $0.1189^{+0.0032}_{-0.0038}$	& $0.1222\pm 0.0032          $	& $0.1311^{+0.0090}_{-0.0065}$	& $0.1348^{+0.0056}_{-0.0049}$\\
  {$100\theta_{\rm MC} $} 	& $1.04590^{+0.00061}_{-0.00045}$	& $1.04622^{+0.00054}_{-0.00042}$	& $1.04623^{+0.00067}_{-0.00044}$	& $1.04637\pm 0.00056        $\\
  {$\tau           $} 	& $0.0633^{+0.0089}_{-0.0066}$	& $0.0624^{+0.0089}_{-0.0074}$	& $0.082^{+0.028}_{-0.036}   $	& $0.080\pm 0.031            $\\
  {$\sum m_\nu$} [eV] 	& $0.166^{+0.064}_{-0.18}    $	& $0.069^{+0.027}_{-0.066}   $	& $0.39^{+0.16}_{-0.20}      $	& $0.42^{+0.17}_{-0.20}      $\\
  {$N_\mathrm{eff}        $} 	& $2.88^{+0.19}_{-0.22}      $	& $3.20\pm 0.18              $	& $3.80\pm 0.45              $	& $4.02\pm 0.29              $\\
  {$\rm{log}_{10}(G_{\rm{eff}}{\rm MeV}^2)$} 	& $-1.60^{+0.14}_{-0.089}    $	& $-1.55^{+0.12}_{-0.080}    $	& $-1.41^{+0.20}_{-0.066}    $	& $-1.35^{+0.12}_{-0.066}    $\\
  {${\rm{ln}}(10^{10} A_s)$} 	& $2.995^{+0.019}_{-0.015}   $	& $2.994\pm 0.017            $	& $3.036^{+0.054}_{-0.071}   $	& $3.035\pm 0.060            $\\
  {$n_\mathrm{s}            $} 	& $0.9273\pm 0.0080          $	& $0.9412\pm 0.0061          $	& $0.947\pm 0.011            $	& $0.9499\pm 0.0098          $\\
  \hline \\[-1.5ex]
  $H_0$ [km/s/Mpc] 	& $66.2^{+2.3}_{-1.9}        $	& $70.1\pm 1.3               $	& $71.1\pm 2.2               $	& $72.3\pm 1.4               $\\
   $\Omega_{\rm m}                  $ 	& $0.327^{+0.013}_{-0.026}   $	& $0.2961^{+0.0075}_{-0.011} $	& $0.3115\pm 0.0090          $	& $0.3094\pm 0.0083          $\\
  $\sigma_8                  $ 	& $0.799^{+0.041}_{-0.017}   $	& $0.824^{+0.015}_{-0.010}   $	& $0.786\pm 0.020            $	& $0.786\pm 0.020            $\\
  $10^9 A_{\rm s}                  $ 	& $1.998^{+0.039}_{-0.030}   $	& $1.998\pm 0.034            $	& $2.09^{+0.10}_{-0.15}      $	& $2.08^{+0.11}_{-0.13}      $\\
  $10^9 A_{\rm s} e^{-2\tau}       $ 	& $1.760\pm 0.014            $	& $1.763\pm 0.013            $	& $1.766\pm 0.016            $	& $1.771\pm 0.016            $\\
  $Y_P                       $ 	& $0.2430\pm 0.0029          $	& $0.2476\pm 0.0024          $	& $0.2549^{+0.0060}_{-0.0048}$	& $0.2577\pm 0.0034          $\\
  $r_*$ [Mpc] 	& $145.8\pm 2.0              $	& $143.0\pm 1.7              $	& $138.2^{+3.2}_{-4.3}       $	& $136.3\pm 2.4              $\\
  $100\theta_*               $ 	& $1.04626^{+0.00060}_{-0.00046}$	& $1.04629^{+0.00053}_{-0.00044}$	& $1.04604^{+0.00060}_{-0.00046}$	& $1.04604\pm 0.00056        $\\
  $D_{\rm{A}}$ [Gpc] 	& $13.93\pm 0.19             $	& $13.67\pm 0.16             $	& $13.21^{+0.30}_{-0.41}     $	& $13.03\pm 0.23             $\\
  $r_{\rm{drag}}$ [Mpc] 	& $148.5\pm 2.1              $	& $145.6\pm 1.8              $	& $140.8^{+3.3}_{-4.3}       $	& $138.8\pm 2.5              $\\
  \end{tabular}
  \end{ruledtabular}
 \end{table*}

\begin{table*}{}
  \caption{Moderately Interacting Neutrino Mode: Parameter $68\%$ Confidence Limits \label{tab:LCDM_nu_params}}
  \begin{ruledtabular}
  \begin{tabular}{ccccc}
  Parameter & TT,TE,EE & TT,TE,EE + lens + $H_0$ & TT + lens + BAO & TT + lens + BAO + $H_0$ \\
  \hline
  {$\Omega_\mathrm{b} h^2   $} 	& $0.02203\pm 0.00023        $	& $0.02246\pm 0.00018        $	& $0.02254^{+0.00030}_{-0.00035}$	& $0.02282\pm 0.00030        $\\
  {$\Omega_\mathrm{c} h^2   $} 	& $0.1191\pm 0.0031          $	& $0.1220\pm 0.0027          $	& $0.1220^{+0.0039}_{-0.0046}$	& $0.1256^{+0.0035}_{-0.0039}$\\
  {$100\theta_{\rm MC} $} 	& $1.04085\pm 0.00044        $	& $1.04063\pm 0.00040        $	& $1.04086\pm 0.00058        $	& $1.04062^{+0.00049}_{-0.00056}$\\
  {$\tau           $} 	& $0.0642^{+0.0095}_{-0.0082}$	& $0.0645^{+0.0090}_{-0.0073}$	& $0.108\pm 0.033            $	& $0.127^{+0.034}_{-0.029}   $\\
  {$\sum m_\nu   $} [eV] 	& $0.150^{+0.054}_{-0.16}    $	& $0.052^{+0.020}_{-0.052}   $	& $0.28^{+0.12}_{-0.23}      $	& $0.40^{+0.17}_{-0.23}      $\\
  {$N_\mathrm{eff}        $} 	& $2.95\pm 0.19              $	& $3.29\pm 0.16              $	& $3.44^{+0.30}_{-0.38}      $	& $3.79\pm 0.28              $\\
  {$\rm{log}_{10}(G_{\rm{eff}}{\rm MeV}^2)$} 	& $-4.44^{+0.58}_{-0.77}     $	& $-4.26\pm 0.69             $	& $-4.12\pm 0.77             $	& $-3.90^{+1.0}_{-0.93}      $\\
  {${\rm{ln}}(10^{10} A_s)$} 	& $3.059^{+0.022}_{-0.019}   $	& $3.067^{+0.019}_{-0.016}   $	& $3.150\pm 0.067            $	& $3.194^{+0.068}_{-0.056}   $\\
  {$n_\mathrm{s}            $} 	& $0.9548\pm 0.0089          $	& $0.9718\pm 0.0073          $	& $0.980^{+0.014}_{-0.015}   $	& $0.993^{+0.013}_{-0.012}   $\\
  \hline \\[-1.5ex]
  $H_0$ [km/s/Mpc] 	& $65.3^{+2.2}_{-1.7}        $	& $69.3\pm 1.2               $	& $69.3^{+1.7}_{-1.9}        $	& $71.2\pm 1.3               $\\
  $\Omega_{\rm m}                  $ 	& $0.335^{+0.012}_{-0.025}   $	& $0.3021^{+0.0077}_{-0.010} $	& $0.3075\pm 0.0092          $	& $0.3010\pm 0.0080          $\\
  $\sigma_8                  $ 	& $0.798^{+0.038}_{-0.016}   $	& $0.826^{+0.014}_{-0.011}   $	& $0.809^{+0.021}_{-0.018}   $	& $0.813^{+0.023}_{-0.020}   $\\
  $10^9 A_{\rm s}                  $ 	& $2.132\pm 0.043            $	& $2.148^{+0.039}_{-0.035}   $	& $2.34^{+0.14}_{-0.18}      $	& $2.44\pm 0.15              $\\
  $10^9 A_{\rm s} e^{-2\tau}       $ 	& $1.875\pm 0.018            $	& $1.888\pm 0.016            $	& $1.880\pm 0.021            $	& $1.892^{+0.019}_{-0.017}   $\\
  $Y_P                       $ 	& $0.2439\pm 0.0027          $	& $0.2486\pm 0.0022          $	& $0.2506^{+0.0041}_{-0.0048}$	& $0.2550\pm 0.0035          $\\
  $r_*$ [Mpc] 	& $145.5\pm 1.9              $	& $142.8\pm 1.5              $	& $141.9^{+3.0}_{-2.7}       $	& $139.1\pm 2.3              $\\
  $100\theta_*               $ 	& $1.04117\pm 0.00054        $	& $1.04066\pm 0.00047        $	& $1.04086\pm 0.00070        $	& $1.04041^{+0.00058}_{-0.00064}$\\
  $D_{\rm{A}}$ [Gpc] 	& $13.97\pm 0.17             $	& $13.72\pm 0.14             $	& $13.63^{+0.28}_{-0.25}     $	& $13.37\pm 0.21             $\\
  $r_{\rm{drag}}$ [Mpc] 	& $148.3\pm 1.9              $	& $145.4\pm 1.6              $	& $144.5^{+3.1}_{-2.8}       $	& $141.6\pm 2.3              $\\
\end{tabular}
\end{ruledtabular}
\end{table*}

\begin{widetext}
\section{Perturbation equations for interacting massive neutrinos}\label{app:Boltz_equns}
In this appendix, we derive the Boltzmann equation governing the evolution of the distribution function of massive self-interacting neutrinos which we denote by $\fn(\xx,{\bf P},\tau)$, where ${\bf P}$ is the canonical conjugate variable to the position $\xx$, and $\tau$ is the conformal time. In the scenario considered here, neutrinos can exchange energy and momentum via 2-to-2 scattering of the type $\nu_i+\nu_j\rightarrow\nu_k+\nu_l$. The Boltzmann equation of neutrino species $i$ can be written as
\be
\frac{df_{\nu_i}}{d\lambda} = \sum_{j,k,l=1}^3 C_{ \nu_i+\nu_j\rightarrow\nu_k+\nu_l}[f_{\nu_i},f_{\nu_j}f_{\nu_k},f_{\nu_l}]
\ee
where $\lambda$ is an affine parameter that described the trajectory of the observer (see below) and $C_{ \nu_i+\nu_j\rightarrow\nu_k+\nu_l}$ is the collision term for the process $\nu_i+\nu_j\rightarrow\nu_k+\nu_l$. In the conformal Newtonian gauge, the space-time metric takes the form
\be\label{metric}
ds^2 = a^2(\tau)[-(1+2\psi) d\tau^2+(1-2\phi) d\vec{x}^2],
\ee
where $a$ is the cosmological scale factor and $\phi$ and $\psi$ are the two gravitational potentials. We can define the affine parameter in terms of the four-momentum $P$ of an observer
\be
P^\mu \equiv \frac{d x^\mu}{d\lambda},
\ee
where $x=(\tau,\vec{x})$ is a four-vector parametrizing the trajectory of the observer. Using Eq.~(\ref{metric}), we can then write
 \be\label{eq:lambda}
 \frac{d}{d\lambda} = \frac{d\tau}{d\lambda}\frac{d}{d\tau}=P^0\frac{d}{d\tau}=\frac{E(1-\psi)}{a}\frac{d}{d\tau},
 \ee
 where we have used the dispersion relation $\ga P^\mu P^\nu = -m_\nu^2$. Here, we have defined $E = \sqrt{p^2+m_\nu^2}$, where $p=|\pp{}|$ is the proper momentum, which is related to the conjugate momentum ${\bf P}$ via the relation $p^2 = g_{ij}P^iP^j$. We note that Eq.~(\ref{eq:lambda}) is valid to first-order in perturbation theory. As in other work in the literature, we choose to write the distribution function in terms of the proper momentum $\pp{}$. This choice is valid as long as we also modify the phase-space volume element as $d^3 P \rightarrow a^3(1-3\phi)d^3p$. The left-hand side of the Boltzmann equation takes the form \cite{Ma:1995ey,Dodelson-Cosmology-2003}
 \be\label{eq:boltz_LHS}
 \frac{df_{\nu_i}}{d\tau} =\frac{\pa f_{\nu_i}}{\pa \tau} + \frac{\p}{E}\cdot\nabla f_{\nu_i}+p\frac{\pa f_{\nu_i}}{\pa p}\left[-\mathcal{H}+\frac{\pa \phi}{\pa \tau}-\frac{E}{p^2}\p\cdot\nabla\psi   \right],
\ee
where $\mathcal{H}\equiv\dot{a}/a$ is the conformal Hubble parameter, a overhead dot denoting a derivative with respect to conformal time. We expand the neutrino distribution function as
\be\label{eq:pert_exp}
f_{\nu_i}(\xx,\p,\tau)=f_{\nu_i}^{(0)}(\p,\tau)[1+\Theta_{\nu_i}(\xx,{\bf p},\tau)].
\ee
At early times, neutrinos form a relativistic tightly-coupled fluid with an equilibrium background distribution function $\fn^{(0)}(\p,\tau)$ given by the Fermi-Dirac distribution. If the interactions mediated by the Lagrangian in Eq.~(\ref{eq:int_lagragian}) go out of equilibrium while neutrinos are relativistic, the background distribution function would maintain this shape, with a temperature red shifting as $T_\nu\propto a^{-1}$. As mentioned in \autoref{sec:cosmo_pert}, we work under this approximation here and assume that the background distribution function maintains its equilibrium shape throughout the epoch of neutrino decoupling. In the absence of energy source or sink, and for the type of interaction we consider in this work, this is an excellent approximation \cite{oldengott17}.


Substituting Eq.~(\ref{eq:pert_exp}) in Eq.~(\ref{eq:boltz_LHS}) and keeping terms that are first order in the perturbation variables, we obtain
\begin{align}\label{eq:boltz_intermdiate1}
 f_{\nu_i}^{(0)}(p,\tau)\left[\frac{\pa\Theta_{\nu_i}}{\pa \tau} + \frac{\p}{E}\cdot\nabla \Theta_{\nu_i}\right] +  p\frac{\pa f_{\nu_i}^{(0)}(p,\tau)}{\pa p}\left[-\mathcal{H}\Theta_{\nu_i}+\frac{\pa \phi}{\pa \tau}-\frac{E}{p^2}\p\cdot\nabla\psi   \right] = \frac{a}{E}C^{(1)}_{ \nu_i}[\p],
\end{align}
where the superscript $C^{(1)}_{ \nu_i}$ denotes the part of the collision term that is first order in the perturbation variables $\Theta_{\nu_i}$. It is useful at this point to introduce the comoving momentum $\qq\equiv a \p$ and comoving energy $\epsilon \equiv a E$. Going to Fourier space, Eq.~(\ref{eq:boltz_intermdiate1}) becomes
\begin{equation}\label{eq:boltz_intermdiate2}
 f_{\nu_i}^{(0)}(q,\tau)\left[\frac{\pa\tilde{\Theta}_{\nu_i}}{\pa \tau} + i\frac{q}{\epsilon}k\mu \tilde{\Theta}_{\nu_i}\right]+ q\frac{\pa f_{\nu_i}^{(0)}(q,\tau)}{\pa q}\left[\frac{\pa \phi}{\pa \tau}-i\frac{\epsilon}{q}k\mu\psi\right] = \frac{a^2}{\epsilon}C^{(1)}_{ \nu_i}[\qq],
\end{equation}
where $\tilde{\Theta}_{\nu_i}$ is the Fourier transform of the perturbation variable $\Theta_{\nu_i}$, $\kk$ is the Fourier conjugate of $\xx$, $k=|\kk|$, $\mu\equiv\hat{q}\cdot\hat{k}$, and $\hat{k}=\kk/k$. In this work, we focus on (helicity) scalar perturbations and expand the angular dependence of the $\tilde{\Theta}_{\nu_i}$ variable in Legendre polynomials $P_l(\mu)$
\be\label{eq:Legendre_exp}
\tilde{\Theta}_{\nu_i}(\qq,\kk,\tau)=\sum_{l=0}^{\infty}(-i)^l(2l+1)\theta_l(k,q,\tau)P_l(\mu).
\ee
We note that this decomposition is always valid for scalar perturbations since they must be azimuthally symmetric with respect to $\kk$, independently of the structure of the collision term. Substituting the above expansion in the first-order Boltzmann equation and and integrating both sides with $\frac{1}{2(-i)^l}\int_{-1}^1d\mu P_l(\mu)$ yields the hierarchy of equations
\begin{align}\label{eq:boltz_before_coll}
f^{(0)}_{\nu_i}\left[\frac{\pa \theta_l}{\pa \tau} +k\frac{q}{\epsilon}\left(\frac{l+1}{2l+1}\theta_{l+1}-\frac{l}{2l+1}\theta_{l-1}\right)\right]+ q\frac{\pa f^{(0)}_{\nu_i}}{\pa q}\left[\frac{\pa \phi}{\pa\tau}\de_{l0}+\frac{k}{3}\frac{\epsilon}{q} \psi\de_{l1}\right]=\frac{a^2}{\epsilon}\frac{1}{2(-i)^l}\int_{-1}^1d\mu P_l(\mu)C^{(1)}_{\nu_i}[\qq],
\end{align}
where $\de_{ij}$ is the Kroenecker delta and where we have suppressed the arguments of $f^{(0)}_{\nu_i}$ and $\theta_l$ for succinctness. We now turn our attention to the collision integral.

\section{Collision Integrals}\label{app:coll_int}
We now compute the first-order collision term for neutrino scattering, $\nu_i(\pp{1}) +\nu_j(\pp{2})\leftrightarrow \nu_k(\pp{3}) + \nu_l(\pp{4})$. We start from the general expression \cite{Kolb:1990vq}
\be
C_{\nu}[\pp{1}]=\frac{1}{2}\int d\Pi_2 d\Pi_3 d\Pi_4  |\mathcal{M}|_{\nu}^2 (2\pi)^4\de^4(P_1+P_2-P_3-P_4)F(\pp{1},\pp{2},\pp{3},\pp{4}),
\ee
where $ |\mathcal{M}|_{\nu}^2$ here is the spin-summed (not averaged) matrix element for the scattering as defined in Eq.~(\ref{matrix_elem_majorana}), $\pp{i}$ denotes the $i^{th}$ three-momentum, $p_i = |\pp{i}|$, and where
\be
 d\Pi_i=\frac{d^3p_i}{(2\pi)^32E_i},
 \ee
and
\be
F(\pp{1},\pp{2},\pp{3},\pp{4})=\fn(\pp{4})\fn(\pp{3})(1-\fn(\pp{2}))(1-\fn(\pp{1}))-\fn(\pp{2})\fn(\pp{1})(1-\fn(\pp{4}))(1-\fn(\pp{3})).
\ee
Using Eq.~(\ref{eq:pert_exp}) and keeping only the first order term in the perturbation variable $\Theta_\nu$, we can rewrite the collision term as:
\ba\label{eq:initial_collision_term}
C^{(1)}_{\nu}&=&\frac{1}{2}\int d\Pi_2 d\Pi_3 d\Pi_4 |\mathcal{M}|_{\nu}^2\frac{(2\pi)^4\de^4(P_1+P_2-P_3-P_4)e^{(p_1+p_2)/T}}{(e^{p_1/T}+1)(e^{p_2/T}+1)(e^{p_3/T}+1)(e^{p_4/T}+1)}\en
&&\qquad\qquad\times\left(2(1+e^{-p_3/T})\Theta_\nu(\pp{3})-(1+e^{-p_2/T})\Theta_\nu(\pp{2})-(1+e^{-p_1/T})\Theta_\nu(\pp{1})\right),
\ea
where we have suppressed the ${\bf x}$ and $\tau$ dependence of the $\Theta_\nu$ variables to avoid clutter, and where we have used the symmetry $\pp{3}\leftrightarrow\pp{4}$ to simplify the integrand. Here, we have taken the background neutrino distribution function to have a relativistic Fermi-Dirac shape. As mentioned in \autoref{sec:cosmo_pert}, we assume neutrinos decouple in the relativistic regime and thus neglect the small neutrino mass in the computation of the collision integrals, $g_{\mu\nu}P^\mu P^\nu\approx0$ and $E\approx p$. We use the technique developed in Refs.~\cite{1976Ap&SS..39..429Y,Hannestad:1995rs,oldengott15} to perform the majority of the integrals. We first perform the $p_4$ integration using the identity
\be
\frac{d^3p_i}{2E_i}\equiv d^4P_i\de(P_i^2)H(P_i^0),
\ee
where $H(x)$ is the Heaviside step function. The collision term then reduces to:
\be
C^{(1)}_{\nu}=\pi\int d\Pi_2 d\Pi_3 |\mathcal{M}|_{\nu}^2\de(2(P_1\cdot P_2-P_1\cdot P_3-P_2\cdot P_3))H(p_1+p_2-p_3)\tilde{F}(p_1,p_2,p_3,p_1+p_2-p_3),
\ee
where we have gathered all the terms dependent on the distribution functions inside $\tilde{F}$. To make progress, we have to choose a coordinate system. We take $\pp{1}$ to point in the z-direction, and $\pp{3}$ to lie in the x-z plane. and define the following angles:
\be
\hat{p}_1\cdot\hat{k}=\mu\qquad\hat{p}_1\cdot\hat{p}_2=\cos{\alpha},\qquad\hat{p}_1\cdot\hat{p}_3=\cos{\theta},\qquad\hat{p}_2\cdot\hat{p}_3=\cos{\alpha}\cos{\theta}+\sin{\alpha}\sin{\theta}\cos{\beta},
\ee
where $\kk$ is the Fourier wavenumber of the perturbations and $\beta$ is the azimuthal angle for $\pp{2}$ to wrap around $\pp{1}$. The integration measure than takes the form
\be
d^3p_3=p_3^2dp_3d\cos{\theta}d\phi\qquad d^3p_2=p_2^2dp_2d\cos{\alpha}d\beta.
\ee
The $\phi$ angle is the azimuthal angle for $\pp{3}$ to wrap around $\pp{1}$. Since we are only dealing with scalar perturbations here, we are free to redefine this angle at will since no physical quantity depends on it. Within this coordinate system, we can write $P_1\cdot P_2 = -p_1p_2 + p_1p_2\cos{\alpha}$, $P_1\cdot P_3 = -p_1p_3 + p_1p_3\cos{\theta}$, and $P_2\cdot P_3 = -p_2p_3 + p_2p_3(\cos{\alpha}\cos{\theta}+\sin{\alpha}\sin{\theta}\cos{\beta})$. We can now use the delta function to do the $\beta$ integration. Setting the argument of the delta function to zero and solving for $\cos{\beta}$ yields
\be\label{eq:cos_beta_sol}
\cos{\beta} = -\frac{p_1p_2-p_1p_3-p_2p_3-p_1p_2\cos{\alpha}+p_1p_3\cos{\theta}+p_2p_3\cos{\alpha}\cos{\theta}}{p_2p_3\sin{\alpha}\sin{\theta}}.
\ee
Performing the integration introduces a Jacobian
\ba\label{eq:Theta_int_first}
C^{(1)}_\nu&=&\frac{1}{8(2\pi)^5}\int \frac{p_2dp_2p_3dp_3d(\cos{\alpha})d(\cos{\theta})d\phi }{\sqrt{a_\alpha \cos^2{\theta} + b_\alpha\cos{\theta}+c_\alpha}}|\mathcal{M}|_\nu^2\en
&&\qquad\times H(p_1+p_2-p_3)\tilde{F}(p_1,p_2,p_3,p_1+p_2-p_3)H(a_\alpha \cos^2{\theta} + b_\alpha\cos{\theta}+c_\alpha).
\ea
In the above, $a_\alpha$, $b_\alpha$, and $c_\alpha$ are
\be
a_\alpha=-p_3^2(p_1^2+p_2^2+2p_1p_2\cos{\alpha}),
\ee
\be
b_\alpha=2p_3(p_1+p_2\cos{\alpha})(p_2p_3+p_1(p_3-p_2)+p_1p_2\cos{\alpha}),
\ee
\be
c_\alpha=-(p_2p_3+p_1(p_3-p_2)+p_1p_2\cos{\alpha})^2+ p_2^2p_3^2(1 - \cos^2{\alpha}).
\ee
The above form of the collision term is useful when the $\theta$ integration needs to be performed first. In some instances, it will be easier to first perform the $\alpha$ integral first. In this latter case, the collision term can equivalently be written as:
\ba\label{eq:Alpha_int_first}
C^{(1)}_\nu&=&\frac{1}{8(2\pi)^5}\int \frac{p_2dp_2p_3dp_3d(\cos{\alpha})d(\cos{\theta})d\phi }{\sqrt{a_\theta \cos^2{\alpha} + b_\theta\cos{\alpha}+c_\theta}}|\mathcal{M}|_\nu^2\en
&&\qquad\times H(p_1+p_2-p_3)\tilde{F}(p_1,p_2,p_3,p_1+p_2-p_3)H(a_\theta \cos^2{\alpha} + b_\theta\cos{\alpha}+c_\theta).
\ea
In the above, $a_\theta$, $b_\theta$, and $c_\theta$ are
\be
a_\theta=-p_2^2(p_1^2+p_3^2-2p_1p_3\cos{\theta}),
\ee
\be
b_\theta=2p_2(p_1-p_3\cos{\theta})(p_1p_2-p_3(p_1+p_2)+p_1p_3\cos{\theta}),
\ee
\be
c_\theta=-(p_1p_2-p_3(p_1+p_2)+p_1p_3\cos{\theta})^2+ p_2^2p_3^2(1 - \cos^2{\theta}).
\ee
We remark that using Eq.~(\ref{eq:cos_beta_sol}) simplifies $P_2\cdot P_3$
\be
P_2\cdot P_3 \rightarrow p_1p_3-p_1p_2+p_1p_2\cos{\alpha}-p_1p_3\cos{\theta}=P_1\cdot P_2-P_1\cdot P_3.
\ee
We note that since the matrix element for the type of interaction of interest (see Eq.~\eqref{matrix_elem_majorana}) only depends on a sum of Mandelstam variables or squares of Mandelstam variables, we can write
\be
|\mathcal{M}|_\nu^2=16 G_{\rm eff}^2\left(\Delta_2(\theta)  \cos^2{\alpha}+\Delta_1(\theta)  \cos{\alpha}+\Delta_0(\theta)\right)\quad\text{or}\quad |\mathcal{M}|_\nu^2=16 G_{\rm eff}^2\left(\Delta_2(\alpha)  \cos^2{\theta}+\Delta_1(\alpha)  \cos{\theta}+\Delta_0(\alpha)\right),
\ee
depending on which of the $\theta$ or $\alpha$ integral we want to perform first. The coefficients are as follow:
\ba
\Delta_2(\theta) &=& p_1^2p_2^2\en
\Delta_1(\theta) &=&  p_1^2p_2\left(p_3-2p_2-p_3\cos{\theta}\right)\en
\Delta_0(\theta) &=& \left( p_1^2(p_2^2-p_2p_3 +p_3^2)+p_1p_3\cos{\theta}(p_1(p_2-2p_3)+p_1p_3\cos{\theta})  \right),
\ea
\ba
\Delta_2(\alpha) &=&  p_1^2p_3^2\en
\Delta_1(\alpha) &=& p_1p_3\left(p_1(p_2-2p_3)-p_1p_2\cos{\alpha}\right)\en
\Delta_0(\alpha) &=& \left( p_1^2(p_2^2-p_2p_3 +p_3^2)+p_1p_2\cos{\alpha}(p_1(-2p_2+p_3)+p_1p_2\cos{\alpha})  \right).
\ea
We note that we can perform the $\cos{\theta}$ or the $\cos{\alpha}$ integration using the following results:
\be\label{eq:form0}
\int_{-\infty}^\infty\frac{dx}{\sqrt{a x^2 + bx+c}}H(a x^2 + bx+c)=\frac{\pi}{\sqrt{-a}}H(b^2-4ac),
\ee
\be\label{eq:form1}
\int_{-\infty}^\infty\frac{xdx}{\sqrt{a x^2 + bx+c}}H(a x^2 + bx+c)=-\frac{b\pi}{2a\sqrt{-a}}H(b^2-4ac),
\ee
\be\label{eq:form2}
\int_{-\infty}^\infty\frac{x^2dx}{\sqrt{a x^2 + bx+c}}H(a x^2 + bx+c)=\frac{\pi(3b^2-4ac)}{8a^2\sqrt{-a}}H(b^2-4ac).
\ee
To make further progress in performing the angular integration, we need to specify the angular dependence of the $\Theta_\nu(\pp{i})$ variables. As in Eq.~\eqref{eq:Legendre_exp}, we expand their angular dependence in Legendre polynomials with respect to the angle between the vector $\kk$ and the $\pp{i}$ vectors. Within our coordinate system, these angles are
\be
\hat{p}_1\cdot \hat{k} = \cos{\gamma}\equiv \mu,\quad\hat{p}_2\cdot \hat{k} =\cos{\alpha}\cos{\gamma}+ \sin{\alpha}\sin{\gamma}\cos{(\phi-\beta)},\quad\hat{p}_3\cdot \hat{k} = \cos{\theta}\cos{\gamma}+\sin{\theta}\sin{\gamma}\cos{\phi}.
\ee
The following identity will be useful later in order to perform the remaining azymuthal integral (the ``$\phi$" integral)
\be\label{eq:usefull_identity_Leg}
\int_0^{2\pi}d\phi\, P_l(\cos{\theta}\cos{\gamma}+\sin{\theta}\sin{\gamma}\cos{\phi})=2\pi P_l(\cos{\theta})P_l(\cos{\gamma}).
\ee
We now consider separately the different terms in the perturbative expansion in $\Theta_\nu(\pp{i})$.
\subsection{Terms involving \texorpdfstring{$\Theta_\nu(\pp{1})$}{Thetanu}}
%
This is the simplest case since $\Theta_\nu(\pp{1})$ can be carried outside the integrals. The azymuthal $\phi$ integral is trivial and yield an extra factor of $2\pi$.

\ba
-\frac{\Theta_\nu(\pp{1})}{8(2\pi)^4}\int \frac{p_2dp_2p_3dp_3d(\cos{\alpha})d(\cos{\theta})}{\sqrt{a_\theta \cos^2{\alpha} + b_\theta\cos{\alpha}+c_\theta}}\langle|\mathcal{M}|_\nu^2\rangle H(p_1+p_2-p_3)H(a_\theta \cos^2{\alpha} + b_\theta\cos{\alpha}+c_\theta)\en
\qquad\times\frac{e^{p_2/T_\nu}}{(e^{p_2/T_\nu}+1)(e^{p_3/T_\nu}+1)(e^{(p_1+p_2-p_3)/T_\nu}+1)}
\ea
Performing the $\alpha$ integration first using Eqs.~(\ref{eq:form0})-(\ref{eq:form2}), we obtain
\ba
-\frac{16 G_{\rm eff}^2\Theta_\nu(\pp{1})}{128\pi^3}\int \frac{p_2dp_2p_3dp_3d(\cos{\theta})}{\sqrt{-a_\theta}}\left(\Delta_2(\theta)\frac{3b_\theta^2-4a_\theta c_\theta}{8 a_\theta^2}-\Delta_1(\theta)\frac{b_\theta}{2a_\theta}+\Delta_0(\theta)\right)H(p_1+p_2-p_3)H(b_\theta^2-4 a_\theta c_\theta)\en
\qquad\times\frac{e^{p_2/T_\nu}}{(e^{p_2/T_\nu}+1)(e^{p_3/T_\nu}+1)(e^{(p_1+p_2-p_3)/T_\nu}+1)}\qquad.
\ea
Writing $\eta\equiv\cos{\theta}$, we have
\ba
-\frac{G_{\rm eff}^2 \Theta_\nu(\pp{1})}{8\pi^3}\int \frac{dp_2p_3dp_3d\eta}{\sqrt{(p_1^2+p_3^2-2p_1p_3\eta)}} |\bar{\mathcal{M}}_{\eta}(p_1,p_2,p_3,\eta)|^2 H(p_1+p_2-p_3)H(b_\theta^2-4 a_\theta c_\theta)\en
\qquad\times\frac{e^{p_2/T_\nu}}{(e^{p_2/T_\nu}+1)(e^{p_3/T_\nu}+1)(e^{(p_1+p_2-p_3)/T_\nu}+1)},
\ea
where we have use the definition
\be
\left(\Delta_2(\theta)\frac{3b_\theta^2-4a_\theta c_\theta}{8 a_\theta^2}-\Delta_1(\theta)\frac{b_\theta}{2a_\theta}+\Delta_0(\theta)\right)\equiv |\bar{\mathcal{M}}_{\eta}(p_1,p_2,p_3,\eta)|^2 .
\ee
The Heaviside step function $H(b_\theta^2-4 a_\theta c_\theta)$ determines the range of integration of both $\eta$ and $p_3$. It yields
\be\label{eq:eta_limit}
{\rm Max}[\eta_-,-1]\leq\eta\leq 1 \quad\text{for}\quad 0\leq p_3\leq p_1+p_2,
\ee
where
\be
\eta_{-}=\frac{(p_1+2p_2)p_3 - 2p_2(p_1+p_2)}{p_1p_3}.
\ee
We can then write
\ba
-\frac{G_{\rm eff}^2\Theta_\nu(\pp{1})}{8\pi^3}\int_0^\infty dp_2 \frac{1}{(e^{-p_2/T_\nu}+1)}\int_{0}^{p_1+p_2} \frac{p_3dp_3}{(e^{p_3/T_\nu}+1)(e^{(p_1+p_2-p_3)/T_\nu}+1)}\en
\times\int_{{\rm Max}[\eta_-,-1]}^1 d\eta\frac{|\bar{\mathcal{M}}_\eta(p_1,p_2,p_3,\eta)|^2 }{\sqrt{(p_1^2+p_3^2-2p_1p_3\eta)}}.\qquad
\ea
Defining $x_i\equiv p_i/T_\nu$, we obtain
\ba
-\frac{G_{\rm eff}^2T_\nu^6\Theta_\nu(\pp{1})}{8\pi^3}\int_0^\infty dx_2 \frac{1}{(e^{-x_2}+1)}\int_{0}^{x_1+x_2} \frac{x_3dx_3}{(e^{x_3}+1)(e^{(x_1+x_2-x_3)}+1)}\qquad\en
\qquad\times\int_{{\rm Max}[\eta_-,-1]}^{1} d\eta\frac{|\bar{\mathcal{M}}_\eta(x_1,x_2,x_3,\eta)|^2 }{\sqrt{(x_1^2+x_3^2-2x_1x_3\eta)}}.
\ea
%
\subsection{Terms involving \texorpdfstring{$\Theta_\nu(\pp{2})$}{Thetanup2}}
For the term involving $\Theta_\nu(\pp{2})$, we start from Eq.~(\ref{eq:Theta_int_first}) and substitute the expansion from Eq.~(\ref{eq:Legendre_exp}).
\ba
-\sum_{l=0}^\infty\frac{(-i)^l(2l+1)}{8(2\pi)^5}\int \frac{p_2dp_2p_3dp_3d(\cos{\alpha})d(\cos{\theta})d\phi }{\sqrt{a_\alpha \cos^2{\theta} + b_\alpha\cos{\theta}+c_\alpha}}\langle|\mathcal{M}|_\nu^2\rangle H(p_1+p_2-p_3)H(a_\alpha \cos^2{\theta} + b_\alpha\cos{\theta}+c_\alpha)\en
\quad\times \frac{\theta_l(p_2)P_l(\cos{\alpha}\cos{\gamma}+ \sin{\alpha}\sin{\gamma}\cos{(\phi)})e^{p_1/T_\nu}}{(e^{p_1/T_\nu}+1)(e^{p_3/T_\nu}+1)(e^{(p_1+p_2-p_3)/T_\nu}+1)},
\ea
where we have used the available freedom to redefine the azymuthal angle $\phi$. We can now perform the $\phi$ integral using the identity given in Eq.~(\ref{eq:usefull_identity_Leg})
\ba
-\sum_{l=0}^\infty\frac{(-i)^l(2l+1)P_l(\mu)}{8(2\pi)^4}\int \frac{p_2dp_2p_3dp_3d(\cos{\alpha})d(\cos{\theta}) }{\sqrt{a_\alpha \cos^2{\theta} + b_\alpha\cos{\theta}+c_\alpha}}\langle|\mathcal{M}|_\nu^2\rangle H(p_1+p_2-p_3)H(a_\alpha \cos^2{\theta} + b_\alpha\cos{\theta}+c_\alpha)\en
\times \frac{\theta_l(p_2)P_l(\cos{\alpha})e^{p_1/T_\nu}}{(e^{p_1/T_\nu}+1)(e^{p_3/T_\nu}+1)(e^{(p_1+p_2-p_3-\mu_\nu)/T_\nu}+1)}.\,\,\,\,\,
\ea
Performing the $\cos{\theta}$ integral yields
\ba
-16 G_{\rm eff}^2 \sum_{l=0}^\infty\frac{(-i)^l(2l+1)P_l(\mu)}{128\pi^3(e^{-p_1/T_\nu}+1)}\int \frac{p_2dp_2p_3dp_3d(\cos{\alpha})}{\sqrt{-a_\alpha}}\left(\Delta_2(\alpha)\frac{3b^2_\alpha-4a_\alpha c_\alpha}{8a_\alpha^2} -\Delta_1(\alpha) \frac{b_\alpha}{2 a_\alpha}+\Delta_0(\alpha)\right)\en
\times H(p_1+p_2-p_3)H( b_\alpha^2-4a_\alpha c_\alpha)\frac{\theta_l(p_2)P_l(\cos{\alpha})}{(e^{p_3/T_\nu}+1)(e^{(p_1+p_2-p_3)/T_\nu}+1)}.
\ea
Similarly to the previous section, we define
\be
\left(\Delta_2(\alpha)\frac{3b^2_\alpha-4a_\alpha c_\alpha}{8a_\alpha^2} -\Delta_1(\alpha) \frac{b_\alpha}{2 a_\alpha}+\Delta_0(\alpha)\right)\equiv |\bar{\mathcal{M}}_\rho(p_1,p_2,p_3,\rho)|^2.
\ee
We use the Heaviside step function $H( b_\alpha^2-4a_\alpha c_\alpha)$ to determine the range of integration for $\rho\equiv\cos{\alpha}$ and $p_3$
\be
{\rm Max}[\rho_-,-1]\leq\rho\leq1\quad\text{for}\quad0\leq p_3\leq p_1+p_2,
\ee
where
\be
\rho_{-}=\frac{p_1p_2 - 2(p_1 + p_2)p_3 + 2p_3^2}{p_1p_2}.
\ee
We thus obtain
\ba\label{eq:theta_2_term_before_nu_l}
-\sum_{l=0}^\infty\frac{G_{\rm eff}^2(-i)^l(2l+1)P_l(\mu)}{8\pi^3(e^{-p_1/T_\nu}+1)}\int_0^\infty dp_2 \theta_l(p_2)p_2\int_0^{p_1+p_2}\frac{dp_3}{(e^{p_3/T_\nu}+1)(e^{(p_1+p_2-p_3)/T_\nu}+1)}\en
\times\int_{{\rm Max}[\rho_-,-1]}^1d\rho \frac{ |\bar{\mathcal{M}}_\rho(p_1,p_2,p_3,\rho)|^2P_l(\rho)}{\sqrt{(p_1^2+p_2^2+2p_1p_2\rho)}}.\qquad
\ea
The remaining difficulty is the $p_2$ dependence of $\theta_l(p_2)$. Since we are working in the thermal approximation in which the only possible neutrino perturbations are local temperature fluctuations, we note that the perturbation variable $\Theta(\pp{2})$ admits the form
\be
\Theta(\xx,\pp{2},\tau) = - \frac{d \ln{f_\nu^{(0)}}}{d\ln{p_2}} \frac{\de T_\nu(\xx,\tau)}{T^{(0)}_\nu(\tau)}.
\ee
It is therefore convenient to introduce the temperature fluctuation variables (see Eq.~\eqref{eq:temp_fluct_def} in main text) $\nu_l$
\be
\nu_l \equiv \frac{-4 \theta_l(p_2)}{\frac{d \ln{f_\nu^{(0)}}}{d\ln{p_2}}},
\ee
which are independent of $\pp{2}$ in the ultra-relativistic limit. We note that as the neutrinos transition to the non-relativistic regime, the $\nu_l$ variables will develop a momentum dependence due to the presence of the mass term in the left-hand side of the Boltzmann equations. However, since we expect the neutrinos to self-decoupled in the relativistic regime, we can safely assume that $\nu_l$ is independent of the neutrino momentum. Substituting
\be
\theta_l(p_2) =\frac{1}{4} \frac{e^{p_2/T_\nu}}{1+e^{p_2/T_\nu}} \frac{p_2}{T_\nu} \nu_l
\ee
in Eq.~\eqref{eq:theta_2_term_before_nu_l} and writing down the answer in terms of $x_i$, we obtain
\ba
-\sum_{l=0}^\infty\frac{G_{\rm eff}^2 T_\nu^6(-i)^l(2l+1)\nu_lP_l(\mu)}{32\pi^3(e^{-x_1}+1)}\int_0^\infty dx_2 x_2^2\frac{e^{x_2}}{1+e^{x_2}}\int_0^{x_1+x_2}\frac{dx_3}{(e^{x_3}+1)(e^{(x_1+x_2-x_3)}+1)}\en
\times\int_{{\rm Max}[\rho_-,-1]}^{1}d\rho\frac{|\bar{\mathcal{M}}_\rho(x_1,x_2,x_3,\rho)|^2P_l(\rho)}{\sqrt{(x_1^2+x_2^2+2x_1x_2\rho)}}.
\ea
%
\subsection{Terms involving \texorpdfstring{$\Theta_\nu(\pp{3})$}{Thetanup3}}
For the term involving $\Theta_\nu(\pp{3})$, we begin from Eq.~(\ref{eq:Alpha_int_first}) and substitute the expansion from Eq.~(\ref{eq:Legendre_exp})
\ba
\sum_{l=0}^\infty\frac{(-i)^l(2l+1)}{8(2\pi)^5}\int \frac{p_2dp_2p_3dp_3d(\cos{\alpha})d(\cos{\theta})d\phi }{\sqrt{a_\theta \cos^2{\alpha} + b_\theta\cos{\alpha}+c_\theta}}\langle|\mathcal{M}|_\nu^2\rangle H(p_1+p_2-p_3)H(a_\theta \cos^2{\alpha} + b_\theta\cos{\alpha}+c_\theta)\en
\qquad\times \frac{e^{(p_1+p_2-p_3)/T_\nu}\theta_l(p_3)P_l(\cos{\theta}\cos{\gamma}+\sin{\theta}\sin{\gamma}\cos{\phi})}{(e^{p_1/T_\nu}+1)(e^{p_2/T_\nu}+1)(e^{(p_1+p_2-p_3)/T_\nu}+1)}.
\ea
We can now perform the $\phi$ integral using the identity given in Eq.~(\ref{eq:usefull_identity_Leg})
\ba
\sum_{l=0}^\infty\frac{(-i)^l(2l+1)P_l(\mu)}{8(2\pi)^4}\int \frac{p_2dp_2p_3dp_3d(\cos{\alpha})d(\cos{\theta}) }{\sqrt{a_\theta \cos^2{\alpha} + b_\theta\cos{\alpha}+c_\theta}}\langle|\mathcal{M}|_\nu^2\rangle H(p_1+p_2-p_3)H(a_\theta \cos^2{\alpha} + b_\theta\cos{\alpha}+c_\theta)\en
\qquad\times \frac{e^{(p_1+p_2-p_3)/T_\nu}\theta_l(p_3)P_l(\cos{\theta})}{(e^{p_1/T_\nu}+1)(e^{p_2/T_\nu}+1)(e^{(p_1+p_2-p_3)/T_\nu}+1)}.\,\,
\ea
Performing the $\cos{\alpha}$ integral yields
\ba
16 G_{\rm eff}^2\sum_{l=0}^\infty\frac{(-i)^l(2l+1)P_l(\mu)}{128\pi^3}\int \frac{p_2dp_2p_3dp_3d(\cos{\theta}) }{\sqrt{-a_\theta}}\left(\Delta_2(\theta)\frac{3b_\theta^2-4a_\theta c_\theta}{8 a_\theta^2}-\Delta_1(\theta)\frac{b_\theta}{2a_\theta}+\Delta_0(\theta)\right)\en
\qquad\times H(p_1+p_2-p_3)H( b_\theta^2-4 a_\theta c_\theta)\frac{e^{(p_1+p_2-p_3)/T_\nu}\theta_l(p_3)P_l(\cos{\theta})}{(e^{p_1/T_\nu}+1)(e^{p_2/T_\nu}+1)(e^{(p_1+p_2-p_3)/T_\nu}+1)}.
\ea
Writing $\eta\equiv\cos{\theta}$ and using the same integration limits as in Eq.~(\ref{eq:eta_limit}), we obtain
\ba
\sum_{l=0}^\infty\frac{G_{\rm eff}^2(-i)^l(2l+1)P_l(\mu)}{8\pi^3(e^{p_1/T_\nu}+1)}\int_0^\infty \frac{dp_2}{e^{p_2/T}+1}\int_0^{p_1+p_2} dp_3 \frac{p_3 \theta_l(p_3)}{e^{-(p_1+p_2-p_3)/T_\nu}+1}\qquad\en
\qquad\times\int_{{\rm Max}[\eta_-,-1]}^{1} d\eta \frac{|\bar{\mathcal{M}}_\eta(p_1,p_2,p_3,\eta)|^2P_l(\eta)}{\sqrt{p_1^2+p_3^2-2p_1p_3\eta}} .\,\,\,\,\,\,
\ea
Again, writing down the $p_3$ dependence of $\theta_l(p_3)$ in terms of the $\nu_l$ variables and writing the integrals in terms of the dimensionless variables $x_i$, we get
\ba
\sum_{l=0}^\infty\frac{G_{\rm eff}^2T_\nu^6(-i)^l(2l+1)\nu_lP_l(\mu)}{32\pi^3(e^{x_1}+1)}\int_0^\infty \frac{dx_2}{e^{x_2}+1}\int_0^{x_1+x_2} dx_3\ \frac{x_3^2e^{x_3}}{(1+e^{x_3})(e^{-(x_1+x_2-x_3)}+1)}\qquad\en
\qquad\times\int_{{\rm Max}[\eta_-,-1]}^{1} d\eta \frac{|\bar{\mathcal{M}}_\eta(x_1,x_2,x_3,\eta)|^2P_l(\eta)}{\sqrt{x_1^2+x_3^2-2x_1x_3\eta}}.
\ea
\subsection{Total Collision Term}
The complete collision term can then be written as
\be
C_\nu^{(1)}[\pp{1}] = \frac{G_{\rm eff}^2T_\nu^6}{4}\frac{ \pa \ln f^{(0)}(p_1)}{\pa \ln p_1} \sum_{l=0}^\infty(-i)^l(2l+1)\nu_lP_l(\mu)\left(A\left(\frac{p_1}{T_\nu}\right) +  B_l\left(\frac{p_1}{T_\nu}\right)-2 D_l\left(\frac{p_1}{T_\nu}\right)\right),
\ee
where
\ba\label{eq:A_coll}
A(x_1) = \frac{1}{8\pi^3}\int_0^\infty  \frac{e^{x_2}dx_2}{e^{x_2}+1}\int_{0}^{x_1+x_2} \frac{x_3\,dx_3}{(e^{x_3}+1)(e^{(x_1+x_2-x_3)}+1)}\int_{{\rm Max}[\eta_-,-1]}^{1} d\eta\frac{|\bar{\mathcal{M}}_\eta(x_1,x_2,x_3,\eta)|^2 }{\sqrt{(x_1^2+x_3^2-2x_1x_3\eta)}},
\ea
\ba\label{eq:Bl_coll}
B_l(x_1)=\frac{1}{8\pi^3x_1}\int_0^\infty \frac{e^{x_2}x_2^2dx_2}{e^{x_2}+1}\int_0^{x_1+x_2}\frac{dx_3}{(e^{x_3}+1)(e^{(x_1+x_2-x_3)}+1)}\int_{{\rm Max}[\rho_-,-1]}^{1}d\rho\frac{|\bar{\mathcal{M}}_\rho(x_1,x_2,x_3,\rho)|^2P_l(\rho)}{\sqrt{(x_1^2+x_2^2+2x_1x_2\rho)}},
\ea
\ba\label{eq:Dl_coll}
D_l(x_1)= \frac{e^{-x_1}}{8\pi^3 x_1} \int_0^\infty \frac{dx_2}{e^{x_2}+1}\int_0^{x_1+x_2} \frac{e^{x_3}x_3^2dx_3}{(e^{x_3}+1)(e^{-(x_1+x_2-x_3)}+1)}\int_{{\rm Max}[\eta_-,-1]}^{1} d\eta \frac{|\bar{\mathcal{M}}_\eta(x_1,x_2,x_3,\eta)|^2P_l(\eta)}{\sqrt{x_1^2+x_3^2-2x_1x_3\eta}},
\ea
where $x_i = p_i/T_\nu$.

\end{widetext}

\newpage

\bibliography{Interacting_neutrinos}

\end{document}